\documentclass[useAMS,usenatbib]{mn2e}

\usepackage{graphicx}
\usepackage{rotating}
\usepackage{caption}

\title[Cavity dynamics of a sample of galaxy groups and clusters]{A volume-limited sample of X-ray galaxy groups and clusters - II. X-ray cavity dynamics}
\author[E. K. Panagoulia et~al.]{E. K. Panagoulia$^{1}$\thanks{E-mail:
epanagoulia@ast.cam.ac.uk}, A. C. Fabian$^{1}$, J. S. Sanders$^{2}$ and J. Hlavacek-Larrondo$^{3, 4}$\\
$^{1}$Institute of Astronomy, Madingley Road, Cambridge CB3 0HA\\
$^{2}$Max-Planck-Institute f\"{u}r extraterrestrische Physik, 85748, Garching, Germany\\
$^{3}$Kavli Institute for Particle Astrophysics and Cosmology, Stanford University, 452 Lomita Mall, Stanford, CA 94305-4085, USA\\
$^{4}$Department of Physics, Stanford University, 452 Lomita Mall, Stanford, CA 94305-4085, USA.}
\begin{document}

\date{Accepted . Received ; in original form }
\pagerange{\pageref{firstpage}--\pageref{lastpage}} \pubyear{2014}

\maketitle

\label{firstpage}

\begin{abstract}
We present the results of our study of a volume-limited sample ($z$ $\leq$ 0.071) of 101 X-ray galaxy groups and clusters, in which we explore the X-ray cavity energetics. Out of the 101 sources in our parent sample, X-ray cavities are found in 30 of them, all of which have a central cooling time of $\leq$3 Gyr. New X-ray cavities are detected in three sources. We focus on the subset of sources that have a central cooling time of $\leq$3 Gyr, whose active galactic nucleus (AGN) duty cycle is $\simeq$61 percent (30/49). This rises to $>$80 percent for a central cooling time of $\leq$0.5 Gyr. 
When projection effects and central radio source detection rates are considered, the actual duty cycle is probably much higher. In addition, we show that data quality strongly affects the detection rates of X-ray cavities. 
After calculating the cooling luminosity and cavity powers of each source with cavities, it is evident that the bubbling process induced by the central AGN has to be, on average, continuous, to offset cooling. We find that the radius of the cavities, $r$, loosely depends on the ambient gas temperature as $r \propto T^{0.5}$, above about 1.5 keV, with much more scatter below that temperature. Finally, we show that, at a given location in a group or cluster, larger bubbles travel faster than smaller ones. This means that the bubbles seen at larger distances from cluster cores could be the result of the merging of several smaller bubbles, produced in separate AGN cycles.  
\end{abstract}


\section{Introduction}
The central cooling time in many groups and clusters of galaxies is $\leq$1 Gyr. If no heating is present, a cooling flow \citep{Fabian94} would be expected to form, wherein the gas cools out of the X-ray phase, and accretes onto the central galaxy. This gas would eventually form molecular clouds, and form stars. Studies of cooling flows with {\it XMM-Newton} and {\it Chandra} have confirmed the presence of positive temperature gradients and short cooling times in the cores of clusters, which are the fingerprint of a cooling flow. However, the amount of gas observed to be cooling at temperatures 2--3 times lower than the virial temperature, is overpredicted by the cooling flow model \citep{McNamara07, McNamara12, Peterson06}. In addition, UV and optical observations have shown that star formation rates in cluster cores are about an order of magnitude lower than predicted \citep{Johnstone87, Nulsen87}, as is the amount of cold gas \citep[e.g.][]{Edge01}. 

Several mechanisms have been proposed over the years, which could suppress the cooling flux at lower energies, while maintaining the high cooling rates. The most prevalent solution to the cooling flow problem is that cooling is balanced, or almost so, by heating. This almost definitely requires the existence of one or more heating mechanisms, which work in a self-regulating feedback loop. Thermal conduction from the hot outer layers of clusters is one such mechanism that has been suggested. In some cases, it may be capable of balancing a cooling flow \citep[e.g][]{Tucker83, Voigt02}, though it can be unstable and needs fine tuning \citep{Soker03}. Furthermore, thermal conduction cannot offset cooling in all clusters \citep{Voigt02}, and is not considered a general solution to the cooling flow problem. Other heating mechanisms, such as cosmic rays \citep{Loewenstein91} and supernovae \citep{McNamara04}, may provide additional heating, but overall these mechanisms are unable to balance radiative losses. 

By far the most favoured heating mechanism is feedback from the central active galactic nuclei (AGNs) in clusters. Outbursts deposit large amounts of energy into the surrounding ICM, which are sufficient to reduce, or even quench, cooling flows \citep[for reviews, see][]{McNamara07, McNamara12, Fabian12, Gitti12}. Many studies now support the fact that AGN affect their surrounding intracluster medium (ICM) \citep[e.g.][]{Birzan04, Dunn08}, though the details of this process are still poorly defined. The detection of X-ray cavities in {\it Chandra} images of many clusters gave credence to the idea that it is indeed AGN feedback that primarily heats up the ICM. These images show radio lobes displacing the ICM, leaving behind X-ray surface brightness (SB) depressions. These depressions are cospatial with radio emission seen from the radio lobes, as seen in e.g. Perseus \citep{Fabian00, Bohringer93} and Hydra~A \citep{McNamara00, Nulsen02}. The displacement of the hot ICM gas creates a low-density, buoyantly rising bubble in the ICM, which is in pressure equilibrium with its surrounding gas. Observations also show X-ray SB depressions coincident with radio emission of a very low frequency. These have been dubbed ``ghost cavities'', and are believed to originate from past AGN outbursts, and whose high-frequence radio emission has since faded \citep[e.g. the ghost cavities in Abell~2597 and the Hydra A cluster,][respectively]{McNamara01, Wise07}.

X-ray cavities provide a reliable method of obtaining lower limits on the energy output of AGN outbursts, through measuring the enthalpy of the X-ray cavities, under the assumption that they are in pressure equilibrium with the surrounding gas. In addition, it is possible to measure the power associated with AGN feedback processes, by calculating either the time it would take the bubble to buoyantly rise to its present position, or the time required to inflate the bubble at the local sound speed \citep[e.g.][]{Birzan04}. The latter assumes that there are no strong shocks driven into the ICM by the bubbles, and so the bubbles are being inflated at a speed similar to the local sound speed. 

X-ray cavities also provide a unique insight into how the AGN feedback cycle works. 60--70 per cent of cool core (CC) clusters have radio sources \citep{Burns90, Mittal09, Birzan12}, and, of the clusters that require heating (i.e. are likely to be CC clusters), a similar fraction display bubbles \citep{Dunn05, Dunn06}. Futhermore, the process of creating bubbles in cluster cores is thought to be the main method of transferring energy from the AGN to the cluster core, and so preventing the occurence of runaway cooling. This in turn generates pressure (i.e. sound) waves in the ICM, which can carry energy out to large distances from the cluster centre, in the form of continuous heating, which is distributed as is necessary \citep{Fabian03a, Voigt04}. 

Several extensive studies on X-ray cavity dynamics have been done before, by a number of authors \citep[e.g.][]{Dunn06, Rafferty06, Birzan12, Hlavacek12}. However, these studies have primarily focused on the brightest sources in the sky, or on a particular type of sources. For example, \cite{Birzan12} study the Brightest 55 clusters of galaxies (B55) and the HIghest X-ray FLUx Galaxy Cluster Sample (HIFLUGCS), \cite{Dunn06} analyse a subsample of the B55 sample, while \cite{Rafferty06} examine a sample of 33 galaxies that are located at the centre of a cooling flow. This means that quite a lot of past studies are flux-limited, and focus predominantly on galaxy clusters. 

In this paper, we examine a distance- and X-ray luminosity-limited parent sample of 101 clusters and groups. We use unsharp-masking to examine whether individual sources in our parent sample harbour X-ray cavities. We then focus on a subsample of 49 clusters and groups, all of which have central cooling times $\leq$3 Gyr, and study their X-ray cavity ``demographics''. The larger parent sample is discussed in detail in \cite{Panagoulia14} (hereafter referred to as Paper I). The parent sample and subsample selection are discussed in Section 2. The data preparation is briefly summarised in Section 3. The extraction of cooling time profiles is described in Section 4, and the imaging analysis is outlined in Section 5. The results of our analysis are presented and discussed in Section 6.

 In this paper, we adopt a flat ${\rm \Lambda}$CDM cosmology, with H$_{0}$ = 71 km s$^{-1}$ Mpc$^{-1}$, $\Omega_{\rm m}$ = 0.27 and $\Omega_{\Lambda}$ = 0.73. All abundances in this paper are relative to solar, as defined in \cite{Anders89}. In all the images shown in this paper, north is to the top and east is to the left. The errorbars presented are at 90 percent confidence, unless otherwise stated.  

\section{A volume-limited sample of groups and clusters of galaxies}

\begin{table*}
\begin{center}
\footnotesize{ 
\begin{tabular}{cccccccc} 
    \multicolumn{1}{c}{Source name}&\multicolumn{1}{c}{Alt. source name}&\multicolumn{1}{c}{RA}&\multicolumn{1}{c}{DEC}&\multicolumn{1}{c}{Redshift}&\multicolumn{1}{c}{$L_{\rm {X}}$}&\multicolumn{1}{c}{Radio}&\multicolumn{1}{c}{Central $t_{\rm {cool}}$}\\
    \multicolumn{1}{c}{}&\multicolumn{1}{c}{}&\multicolumn{1}{c}{(2000)}&\multicolumn{1}{c}{(2000)}&\multicolumn{1}{c}{}&\multicolumn{1}{c}{($\times$10$^{44}$ erg s$^{-1}$)}&\multicolumn{1}{c}{}&\multicolumn{1}{c}{(Gyr)}\\
    \multicolumn{1}{c}{(1)}&\multicolumn{1}{c}{(2)}&\multicolumn{1}{c}{(3)}&\multicolumn{1}{c}{(4)}&\multicolumn{1}{c}{(5)}&\multicolumn{1}{c}{(6)}&\multicolumn{1}{c}{(7)}&\multicolumn{1}{c}{(8)}\\
    \hline
RXCJ1242.8+0241 & NGC~4636 & 190.7063 & 2.6856 & 0.0031 & 0.01 &  yes (1) & 0.19$_{-0.04}^{+0.17}$ \\
RXCJ1229.7+0759 & M~49-Virgo-NGC~4472 & 187.4403 & 7.9870 & 0.0033 & 0.01 & yes (2) & 0.04$_{-0.003}^{+0.002}$ \\
RXCJ0338.4-3526 & FORNAX-NGC~1399 & 54.6167 & -35.4483 & 0.0051 & 0.012 & yes (1) & 0.06$_{-0.003}^{+0.004}$ \\
RXCJ1501.2+0141 & NGC~5813 & 225.3016 & 1.6960 & 0.0065 & 0.02 & yes (3) & 0.18$\pm$0.01\\
RXCJ1506.4+0136 & NGC~5846 & 226.6250 & 1.6022 & 0.0066 & 0.008 & yes (2) & 0.15$_{-0.01}^{+0.02}$\\
RXCJ1315.3-1623 & NGC~5044 & 198.8500 & -16.3897 & 0.0087 & 0.097 & yes (2) & 0.22$_{-0.02}^{+0.03}$\\
RXCJ1248.7-4118 & A~3526-NGC~4696 & 192.20 & -41.3078 & 0.0114 & 0.721 & yes (1, 2) & 0.09$_{-0.003}^{+0.004}$\\
RXCJ1036.6-2731* & A~1060 & 159.1750 & -27.5244 & 0.0126 & 0.297 & yes (1) & 0.32$\pm$0.04\\
RXCJ0419.6+0224 & NGC~1550 & 64.9083 & 2.4139 & 0.0131 & 0.153 & yes (1) & 0.19$_{-0.03}^{+0.02}$\\
RXCJ1751.7+2304* & NGC~6482 &  267.9480 & 23.0705 & 0.0132 & 0.02 &  yes (13) & 0.15$_{-0.006}^{+0.035}$\\
RXCJ1253.0-0912 & HCG~62 & 193.2750 & -9.2003 & 0.0146 & 0.037 & yes (4) & 0.19$_{-0.01}^{+0.02}$\\
RXCJ1050.4-1250 & NGC~3402 & 162.6083 & -12.8464 & 0.0155 & 0.059 & yes (14) & 0.098$_{-0.003}^{+0.014}$\\
RXCJ0152.7+3609 & A~0262 & 28.1948 & 36.1513 & 0.0163 & 0.81 & yes (1) & 0.13$\pm$0.02\\
RXCJ0200.2+3126 & NGC~0777 & 30.0687 & 31.4365 & 0.0168 & 0.04 & yes (5) & 0.325$_{-0.004}^{+0.024}$\\
RXCJ0125.5+0145* & NGC~533 & 21.3750 & 1.7622 & 0.0174 & 0.032 & yes (2) & 0.10$_{-0.01}^{+0.02}$\\
RXCJ1204.4+0154* & MKW~4 & 181.1065 & 1.9010 & 0.0195 & 0.28 & yes (1) & 0.24$_{-0.02}^{+0.07}$ \\
RXCJ1407.4-2700 & A~3581 & 211.8667 & -27.0153 & 0.0230 & 0.316 & yes (6) & 0.29$_{-0.04}^{+0.02}$ \\
RXCJ1223.1+1037 & NGC~4325 & 185.7772 & 10.6240 & 0.0258 & 0.20 &  yes (15) & 0.27$_{-0.03}^{+0.06}$\\
RXCJ1715.3+5724 & NGC~6338 & 258.8414 & 57.4074 & 0.0276 & 0.49 & yes (7) & 0.22$_{-0.02}^{+0.05}$ \\
RXCJ1628.6+3932 & A~2199 & 247.1582 & 39.5487 & 0.0299 & 3.77 & yes (1) & 0.48$_{-0.09}^{+0.06}$ \\
RXCJ0433.6-1315 & A~0496 & 68.4083 & -13.2592 & 0.0326 & 1.746 & yes (1) & 0.33$_{-0.09}^{+0.04}$\\
RXCJ0338.6+0958 & 2A0335+096 & 54.6699 & 9.9745 & 0.0347 & 4.21 & yes (1) & 0.16$\pm$0.02  \\
RXCJ1516.7+0701 & A~2052 & 229.1834 & 7.0185 & 0.0353 & 2.58 & yes (1) & 0.21$_{-0.11}^{+0.06}$ \\
RXCJ0425.8-0833* & RBS~0540 & 66.4625 & -8.5592 & 0.0397 & 1.008 & yes (1) & 0.33$_{-0.05}^{+0.14}$\\
RXCJ1521.8+0742 & MKW3s & 230.4582 & 7.7088 & 0.0442 & 2.70 & yes (1) & 0.09$\pm$0.03 \\
RXCJ1252.5-3116* & - & 193.1417 & -31.2678 & 0.0535 & 0.861 &  yes (16) & 0.32$_{-0.04}^{+0.12}$\\
RXCJ0041.8-0918 & A~85 & 10.4583 & -9.2019 & 0.0555 & 5.293 & yes (1) & 0.40$_{-0.03}^{+0.11}$\\
RXCJ2313.9-4244 & A~S1101 & 348.4958 & -42.7339 & 0.0564 & 1.738 & yes (1) & 0.34$_{-0.05}^{+0.08}$\\
RXCJ0102.7-2152 & A~0133 & 15.6750 & -21.8736 & 0.0569 & 1.439 & yes (1) & 0.18$\pm$0.03\\
RXCJ2205.6-0535* & A~2415 & 331.4417 & -5.5933 & 0.0582 & 1.135 & yes (8) & 0.48$_{-0.04}^{+0.13}$\\
RXCJ1454.5+1838 & A~1991 & 223.6309 & 18.6420 & 0.0586 & 1.46 & yes (8) & 0.23$\pm$0.01\\
RXCJ1348.8+2635 & A~1795 & 207.2207 & 26.5956 & 0.0622 & 9.93 & yes (1, 8) & 0.27$_{-0.06}^{+0.12}$\\
\hline
RXCJ0123.1+3327 & NGC~499 & 20.7970 & 33.4620 & 0.0153 & 0.04 & no (2) & 0.68$_{-0.03}^{+0.12}$ \\
\underline{RXCJ1627.3+4240}* & A~2192 & 246.8482  & 42.6784 & 0.0317 &  0.12 &  yes (18) & 0.65$_{-0.01}^{+0.45}$ \\
RXCJ1604.9+2355* & AWM4 & 241.2377 & 23.9206 & 0.0326 & 0.55 & yes (9) & 0.70$_{-0.06}^{+0.16}$\\
RXCJ2357.0-3445 & A~4059 & 359.2583 & -34.7606 & 0.0475 & 1.698 & yes (10) & 0.65$_{-0.06}^{+0.05}$\\
RXCJ0918.1-1205 & A~780 - Hydra A & 139.5292 & -12.0933 & 0.0539 & 2.659 & yes (1) & 0.53$_{-0.10}^{+0.06}$\\
RXCJ2336.5+2108* & A~2626 & 354.1262 & 21.1424 & 0.0565 & 1.55 & yes (8) & 0.60$_{-0.27}^{+0.44}$\\
RXCJ1303.7+1916* & A~1668 & 195.9398 & 19.2715 & 0.0643 & 1.79 & yes (8) & 0.80$_{-0.10}^{+0.14}$\\
\hline
RXCJ2347.7-2808* & A~4038 & 356.9292 & -28.1414 & 0.0300 & 1.014 & yes (1) & 1.07$_{-0.47}^{+0.45}$\\
RXCJ1347.4-3250* & A3571 & 206.8667 & -32.8497 & 0.0391 & 3.996 & yes (1) & 1.07$_{-0.37}^{+3.01}$ \\
\hline
RXCJ2009.9-4823* & S~0851-NGC~6868 & 302.4750 & -48.3931 & 0.0097 & 0.007 &  yes (2) & 1.65$_{-0.14}^{+0.34}$\\
\underline{RXCJ1840.6-7709}* & - & 280.1542 & -77.1556 & 0.0194 & 0.087 &  yes (17) & 1.59$_{-0.21}^{+0.18}$\\
RXCJ1523.0+0836* & A2063 & 230.7724 & 8.6025 & 0.0355 & 1.94 & yes (1) & 1.88$_{-0.77}^{+2.13}$\\
RXCJ0721.3+5547* & A0576 & 110.3426 & 55.7864 & 0.0381 & 1.41 & yes (1) & 1.51$_{-0.57}^{+3.40}$\\
RXCJ1257.1-1724 & A1644 & 194.2917 & -17.4003 & 0.0473 & 1.952 & yes (1) & 1.58$_{-0.20}^{+0.21}$\\
    \hline
\end{tabular}
}
\end{center}
\caption{List of the 49 sources in our subsample and their properties.}
\label{tab:properties}
\end{table*}

\begin{table*}
\begin{center}
  \contcaption{}
  \begin{tabular}{cccccccc}
     \multicolumn{1}{c}{Source name}&\multicolumn{1}{c}{Alt. source name}&\multicolumn{1}{c}{RA}&\multicolumn{1}{c}{DEC}&\multicolumn{1}{c}{Redshift}&\multicolumn{1}{c}{$L_{\rm {X}}$}&\multicolumn{1}{c}{Radio}&\multicolumn{1}{c}{Central $t_{\rm {cool}}$}\\
    \multicolumn{1}{c}{}&\multicolumn{1}{c}{}&\multicolumn{1}{c}{(2000)}&\multicolumn{1}{c}{(2000)}&\multicolumn{1}{c}{}&\multicolumn{1}{c}{($\times$10$^{44}$ erg s$^{-1}$)}&\multicolumn{1}{c}{}&\multicolumn{1}{c}{(Gyr)}\\
    \multicolumn{1}{c}{(1)}&\multicolumn{1}{c}{(2)}&\multicolumn{1}{c}{(3)}&\multicolumn{1}{c}{(4)}&\multicolumn{1}{c}{(5)}&\multicolumn{1}{c}{(6)}&\multicolumn{1}{c}{(7)}&\multicolumn{1}{c}{(8)} \\
    \hline
RXCJ1733.0+4345 & IC~1262 & 263.2607 & 43.7629 & 0.0307 & 0.47 & yes (11) &  2.10$_{-0.28}^{+0.44}$\\
\hline
\underline{RXCJ2315.7-0222}* & - & 348.9375 & -2.3769 & 0.0267 & 0.134 &  yes (16) & 2.76$_{-0.23}^{+0.25}$ \\
RXCJ1254.6-2913* & A3528S & 193.6708 & -29.2233 & 0.0544 & 1.064 & yes (12) & 2.68$_{-0.56}^{+0.81}$\\
\hline
\end{tabular}
\end{center}
\raggedright{List of the 49 sources in our short central cooling time subsample and their properties, adapted from Paper I. (1) source name as in the REFLEX or NORAS catalogue, (2) alternative source name (sometimes referring to the central dominant galaxy), (3) and (4) source right ascension and declination in epoch 2000 coordinates, (5) source redshift, (6) source X-ray luminosity in the rest frame 0.1--2.4 keV band, (7) existence or absence of a central radio source, and (8) central cooling time in Gyr. The starred sources show no sign of X-ray cavities, and the underlined sources are the ones for which {\it XMM-Newton} data were used. The sources are split into 0.5 Gyr central cooling time bins (i.e. the top group have a central cooling time $\leq$0.5 Gyr, the second group 0.5--1.0 Gyr and so on). REFERENCES.- (1) \cite{Birzan12} and references therein, (2) \cite{Dunn10}, (3) \cite{Randall11}, (4) \cite{Gitti10}, (5) \cite{Ho01}, (6) \cite{Canning13}, (7) \cite{Pandge12}, (8) \cite{Owen97}, (9) \cite{OSullivan10}, (10) \cite{Choi04}, (11) \cite{Trinchieri07}, (12) \cite{Venturi01}, (13) \cite{Goudfrooij94}, (14) \cite{OSullivan07}, (15) \cite{Dong10}, (16) \cite{Magliocchetti07}, (17) Sydney University Molonglo Sky Survey image \citep[SUMSS;][]{Mauch03}, (18) NRAO VLA Sky Survey image \citep[NVSS;][]{Condon98}.}
\end{table*}

The motivation behind the work in both this paper and Paper I, is the study of the radial properties of the ICM in X-ray groups and clusters of galaxies. Ultimately, we aim to determine the importance and the effect of non-gravitational processes, such as AGN feedback, on the ICM. We are specifically interested in cluster cores (the central $\sim$10 kpc), for which we need high spatial resolution data. For this reason, we chose to look at a sample of nearby groups and clusters of galaxies, for which {\it Chandra} and/or {\it XMM-Newton} data are available. 

For details on the sample selection, we refer readers to Section 2 of Paper I. We briefly summarise the selection process below:
\begin{itemize}
  \item{Using the Northern {\it ROSAT} All Sky catalogue \citep[NORAS;][]{Bohringer00} and the {\it ROSAT-ESO} Flux Limited X-ray galaxy cluster survey \citep[REFLEX;][]{Bohringer04}, we constructed a volume-limited sample of sources. These were chosen to lie at a distance of $\leq$300 Mpc. In total, 289 sources from these catalogues meet this criterion.}
      \item{We require a statistically complete sample, so we make cuts in the X-ray luminosity, $L_{\rm {X}}$, as well as distance, to avoid groups of sources that have no data. For details of the cuts in distance and $L_{\rm {X}}$, see section 2 of Paper I.} 
          \item{After the cuts in both $L_{\rm {X}}$ and distance, we end up with a final sample, containing 101 sources. Of these 101 sources, all but four of them have {\it Chandra} and/or {\it XMM-Newton} data. Where possible, we use {\it Chandra} data, to benefit from the higher spatial resolution and lower background levels of the Advanced CCD Imaging Spectrometer (ACIS) detectors. The details of all the sources in our sample, and the observations used in subsequent data analysis, are given in Tables 1--4 in Paper I.} 
\end{itemize}

We note that the Perseus cluster, and the central galaxies of the Virgo cluster, M86 and M87, are not included in the NORAS and REFLEX catalogues. The reason given for the exclusion of M86 and M87  is the uncertainty in making individual flux measurements for these sources, given that they are surrounded by diffuse emission from the Virgo cluster. The Perseus cluster (NGC~1275) is excluded as it lies within the band of the Milky Way, which is defined as the region of the sky with galactic latitude $|b_{\rm {II}}|$ $\leq$ 20$^{\circ}$ in both surveys (the Perseus cluster has a galactic latitude of -13.26$^{\circ}$). This Milky Way band was excluded from both the NORAS and REFLEX surveys.

There is some overlap between our sample and those of \cite{Reiprich02} and \cite{Edge90}. We cross-checked the sources in these two samples with those in our sample, and find that the sources that appear in the \cite{Reiprich02} and \cite{Edge90} samples but not ours, are either too distant, or were not included in the NORAS and REFLEX catalogues, or did not meet our sample selection criteria. 

\subsection{Short central $\mathbf{{t_{cool}}}$ subsample}
\begin{figure*}
\begin{center}
\includegraphics[trim = 0cm 14cm 4.5cm 0cm, clip, height=3.4in, width=3.4in]{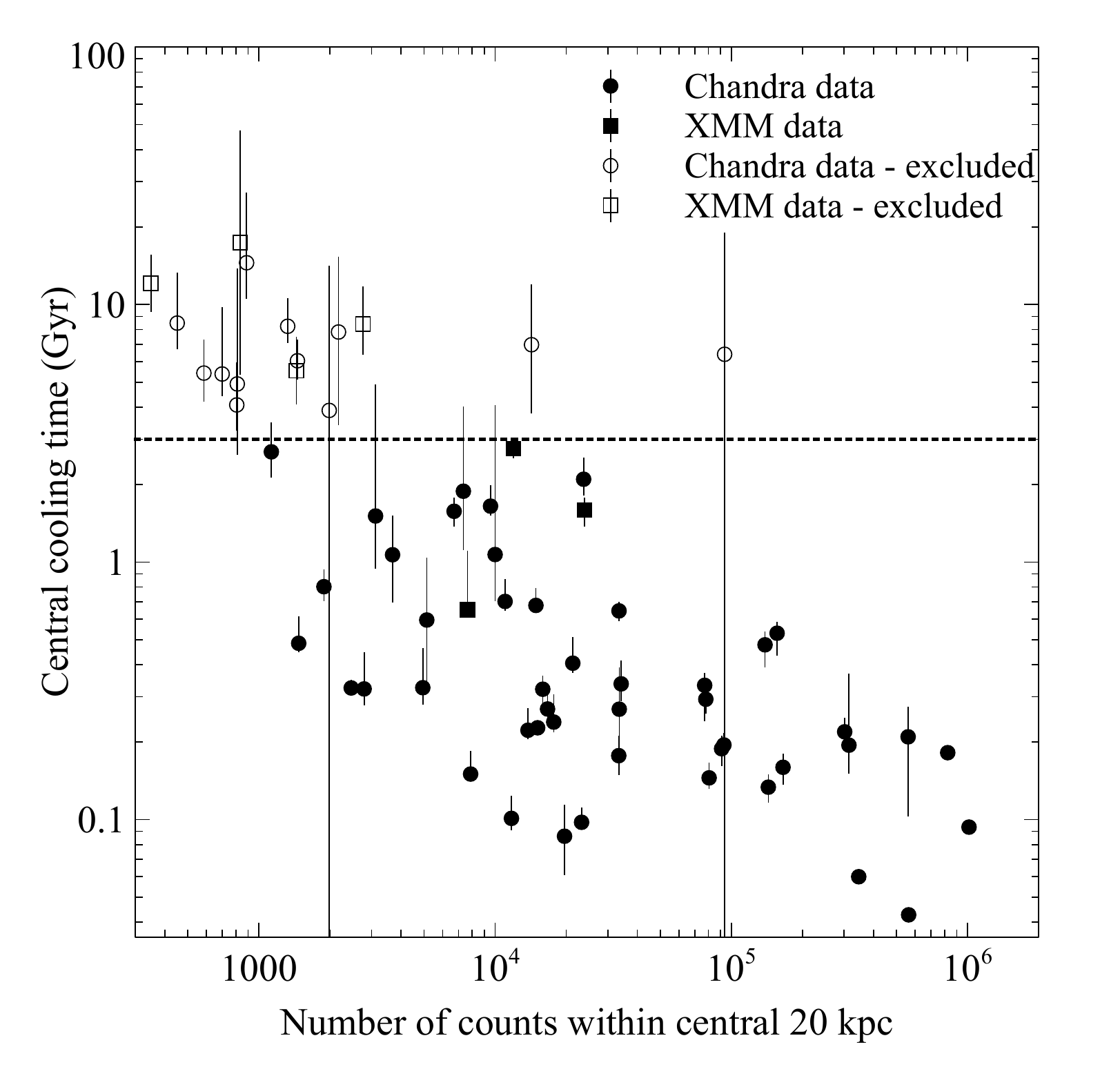}
\includegraphics[trim = 0cm 14cm 4.5cm 0cm, clip, height=3.4in, width=3.4in]{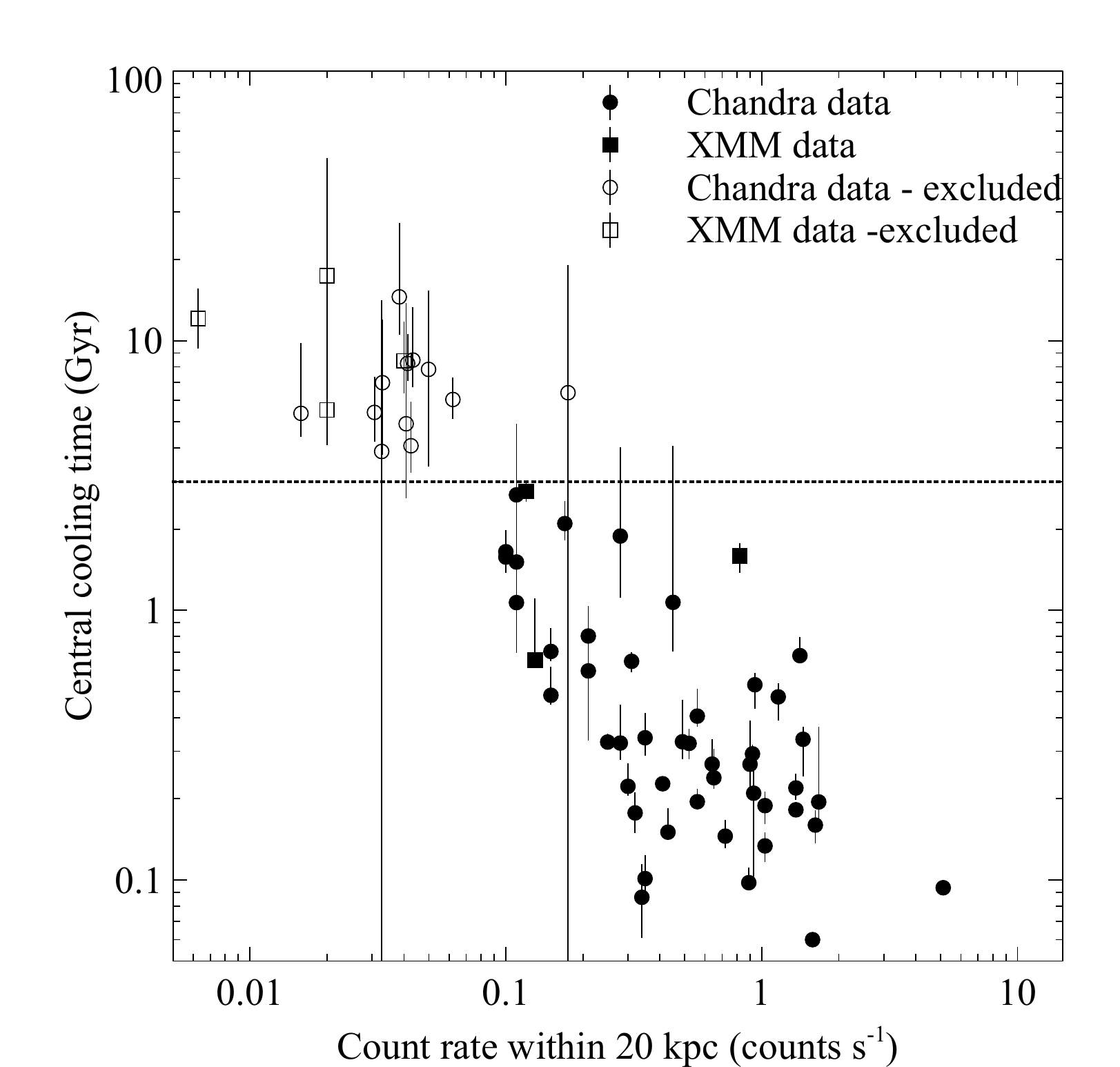}
\caption[]{{\it Left:} Central cooling time vs. counts within the central 20 kpc for the 65 sources for which we have a central cooling time measurement. {\it Right:} Central cooling time vs. count rate within the central 20 kpc for the 65 sources for which we have a central cooling time measurement. In both plots, the circles and squares represent sources for which {\it Chandra} and {\it XMM-Newton} data were used in the analysis, respectively. The filled symbols indicate the sources with a central cooling time $\leq$3 Gyr, which formed our short central cooling time subsample, and the empty symbols the sources that have longer central cooling times. The rightmost empty circle in both plots is the Coma cluster. The dashed line in both plots simply represents the 3 Gyr cut-off in the central cooling time.}
\label{fig:tcoolselection}
\end{center}
\end{figure*}

For this work, we are particularly interested in investigating the connection between central cooling times, and the presence, or absence, of X-ray cavities. Here, we define the central cooling time as the cooling time calculated for the gas in the innermost spectral bin for each source (for details of the creation of the spectral extraction regions, see Section 4.1), which we were able to reliably calculate for 65 out of the 101 sources in our parent sample, using  Equation \ref{eq:tcoolspec2} (see Section 4.2) (these are the starred sources in tables 1 and 3 in Paper I). In order to carry out a detailed analysis of the central regions of groups and clusters, a large number of counts is required there. To select the sources with the highest number of counts in their central region, we calculated the total number of counts within the central 20 kpc for all 65 groups and clusters for which we had a central cooling time. We plot the central cooling time of each source vs. the total counts in the central 20 kpc in the left-hand panel of Fig. \ref{fig:tcoolselection}. The black circles represent sources for which {\it Chandra} data were used in the analysis, and the black squares indicate sources for which {\it XMM-Newton} data were used. The filled and empty circles illustrate sources with a central cooling time of $\leq$3 Gyr and $\geq$3 Gyr, respectively. The two rightmost empty circles are Abell~3667 and the Coma cluster, both of which are well-studied major mergers. The dashed line represents the 3 Gyr cut-off in the central cooling time, in both figures. As can be seen, a cut-off of the central cooling time at $\leq$3 Gyr excludes most sources that do not have a large number of counts in their central 20 kpc. Our $\leq$3 Gyr cut-off is also in line with previous studies of relatively nearby groups and clusters of galaxies \citep[e.g.][]{Dunn06, Dunn10}. In total, 49 sources out of the 65, that have a reliable central cooling time value, meet this requirement. Of these sources, 15 are groups, while the rest are clusters, where we define a source as a group or cluster based on the currently available literature. As we used different, and sometimes multiple, literary sources to define a source as a group or cluster, this classification will be based on different classification criteria, such as optical richness and X-ray luminosity. Despite this variety of classification methods used for each source, the separate classifications in the literature of a source as a group or cluster are in agreement for the vast majority of the sources in our subsample. We consider all 49 sources with a central cooling time $\leq$ 3 Gyr to be cool cores. The 16 sources that were not included in our subsample are: IIIZW54, Abell~1736, Abell~2665, Abell~2734, Abell~3158, Abell~3376, Abell~3391, Abell~3395, Abell~3562, Abell~3667, Abell~376, Abell~754, the Coma cluster, MKW~8, NGC~6099, and RXCJ1109.7+2145. Hereafter, we refer to these 16 sources as the excluded sources. We note that many of these sources are well-studied merging clusters. 

To test the validity of our selection method, we also plot the central cooling time of the 65 sources, for which we have a central cooling time measurement, against the respective count rate within the central 20 kpc. The plot is shown in the right-hand panel of Fig. \ref{fig:tcoolselection}, where the symbols are the same as in the left-hand panel of the same figure, and the rightmost empty circle is the Coma cluster. The separation between the sources with a cooling time of less than and more than 3 Gyr is now even more pronounced. In addition, we plotted the central cooling time of each of the 65 sources against the count rate within the central 20 kpc, multiplied by (1+$z$)$^{2}$, where $z$ is the redshift of a source, to get rid of any dependence of the count rate on the redshift of the source. The result is very similar to that in the left-hand panel of Fig. \ref{fig:tcoolselection}, but with a slightly wider separation between the sources that have a central cooling time of $\geq$3 Gyr, and those with a central cooling time $\leq$3 Gyr. This indicates that there is a separation between the groups and clusters that have a central cooling time larger or smaller than 3 Gyr, that is intrinsic rather than affected by data quality. 

Note that we have selected to plot the cooling time in the innermost central bin vs. the total counts or count rate within the central 20 kpc of each source, as we are mainly interested in studying the cavity dynamics of this short central cooling time subsample. In sources within the redshift range we are studying, X-ray cavities are expected to lie within this radius, and not just in the region covered by the central spectral bin.

The properties of the sources in our short central cooling time subsample, as listed in the NORAS or REFLEX catalogues, and their calculated central cooling times, are given in columns 1--6 and 8 in Table \ref{tab:properties}, respectively. The sources are split into 0.5 Gyr central cooling time bins, i.e. the top group sources have a central cooling time $\leq$0.5 Gyr, the second group 0.5--1.0 Gyr, and so on. The underlined sources are the ones for whose analysis {\it XMM-Newton} data were used. The starred sources are the ones which show no evidence of X-ray cavities. 


\section{Observations and data analysis}
\subsection{{\it\textbf{Chandra}} data}
The observation IDs, as well as the clean exposure times, for each source are given in Tables 2 and 4 in Paper I. We refer the reader to Section 3.1 of Paper I for details of the data analysis. To summarise, new Level 2 events files were created from the Level 1 event files, using the {\sc ciao acis\_reprocess\_events} pipeline. Background lightcurves were generated and examined, in order to excise periods of X-ray background flaring. Blank-sky observations were selected and adjusted to match individual observations, and were used to create background images and spectra. If more than one dataset was available for each source, the datasets were reprojected onto the same set of coordinates. Finally, a 0.5--7.0 keV background-subtracted and exposure-corrected image was made for each source, and examined to identify contaminating point sources. These were then excluded from subsequent spectral analysis. 

\subsection{{\it\textbf{XMM-Newton}} data}
For details of the analysis of the {\it XMM-Newton} data, see Section 3.2 in Paper I. We used data from all three EPIC detectors, and these were reprocessed using {\sc emchain} and {\sc epchain} for the MOS and pn, respectively. Lightcurves were extracted in the appropriate energy bands for each of the three detectors, to filter out periods of X-ray background flaring. 0.5--7.0 keV composite images were then created for each source, and were visually examined for point sources. These were excised from further spectral analysis. 

\section{Cooling time profiles}
\subsection{Spectral analysis}
In order to obtain radial cooling time profiles, we first need radial temperature and electron number density profiles of the individual groups and clusters of our parent sample. In order to extract spectra, we generated a series of concentric annuli for each source. All annuli for the same source were defined in such a way to ensure that they all had the same number of counts in them, or signal-to-noise ratio. The signal-to-noise ratio of, or number of counts in, the annuli of a source was defined by the quality of the available data for the source in question. We followed the standard {\sc ciao} and {\sc sas} routines for the generation of source and background spectra, ancillary region files (ARFs) and redistribution matrix files (RMFs) for sources with {\it Chandra} and {\it XMM-Newton} data, respectively. To take projection effects into account, the spectra were deprojected using the {\sc dsdeproj} code \citep{Sanders07}. All spectral fits were performed in {\sc xspec} \citep{Arnaud96} v12.7.1b. For further details on our spectral analysis and fitting, we refer the reader to Section 4.1 of Paper I.

We note that we were not able to extract deprojected profiles from all the sources in our parent sample. This is because the data quality was not always sufficient to give reliable deprojected temperature and electron number density profiles.  For this reason, we did not perform any further spectral analysis, such as calculation of a cooling time profile, on these sources. We were able to extract reliable deprojected profiles for 65 sources out of the 101 sources in our parent sample, and these sources are listed in tables 1 and 3 of Paper I (indicated by an asterisk next to their name). We did however search for X-ray cavities in all the sources in our parent sample (see Section 5), and list the sources that do or might harbour cavities in Table \ref{tab:cavities}. We point out that Table \ref{tab:properties} only contains sources for which we were able to reliably obtain a deprojected number density and temperature profile, and hence calculate a central cooling time.

\subsection{Deprojected cooling time profiles}
Having obtained deprojected radial temperature and electron number density profiles for 65 out of 101 of the groups and clusters in our parent sample, as previously mentioned in Section 2.1, we were able to calculate individual cooling time profiles. These were calculated using the method described in Section 4.2 of Paper I, and are included in the overall cooling time profile in the right-hand panel of figure 2 in the same paper. These individual cooling time profiles were calculated in order to obtain a rough idea of the cooling behaviour of the clusters and groups in our parent sample. However, the equations used in that paper provide only a crude approximation of the cooling behaviour of the sources in our subsample, as they assume solar abundances. In order to calculate the central cooling times of each source in our subsample, we follow the method described below. 

The cooling time, $t_{\rm {cool}}$, of a gas parcel with a total number density $n_{\rm {t}}$, temperature $T$ and emissivity $\epsilon$ is given by 
\begin{equation}
t_{\rm {cool}} = \frac{\frac{5}{2} n_{\rm {t}} k_{\rm {B}} T}{\epsilon}.
\label{eq:tcoolspec1}
\end{equation}
Using the definition of the normalisation, $A$, of the {\sc apec} or {\sc vapec} component from {\sc xspec},
\begin{equation}
A = \frac{10^{-14}}{4 \pi [D_{\rm {A}}(1+z)]^{2}} \int{n_{\rm {e}} n_{\rm {H}} dV},
\label{eq:norm}
\end{equation}
Equation \ref{eq:tcoolspec1} can be rewritten as
\begin{equation}
t_{\rm {cool}} = \frac{5}{2}\times\frac{27}{14}\times\sqrt{\frac{14\pi10^{14}}{3}}\times D_{\rm {A}}(1+z)\frac{\sqrt{A V} k_{\rm {B}} T}{L_{\rm {X}}},
\label{eq:tcoolspec2}
\end{equation}
where $D_{\rm {A}}$ is the angular diameter distance to the source, $z$ is the source redshift, $n_{\rm {e}}$ and $n_{\rm {H}}$ are the electron and hydrogen number densities respectively, $V$ is the volume of the gas parcel and $L_{\rm {X}}$ is its luminosity. The temperature $T$, luminosity $L_{\rm {X}}$ and normalisation $A$ were obtained from the spectral fits. Since we obtain the aforementioned values from the spectral fitting, no assumptions about the gas abundance were necessary. We have used 
\begin{equation}
n_{\rm {H}} = \frac{6}{7}n_{\rm {e}},
\label{eqn:ne}
\end{equation}
and 
\begin{equation}
n_{\rm {t}} = \frac{27}{14}n_{\rm {e}}.
\end{equation}
The two above equations were calculated by ignoring contributions from all elements other than hydrogen and helium, and assuming mass fractions of 75 and 25 percent, respectively. We have also assumed that helium and hydrogen are both fully ionised.

\begin{figure}
\begin{center}
\includegraphics[trim = 0cm 14cm 4.5cm 0cm, clip, height=3.4in, width=3.4in]{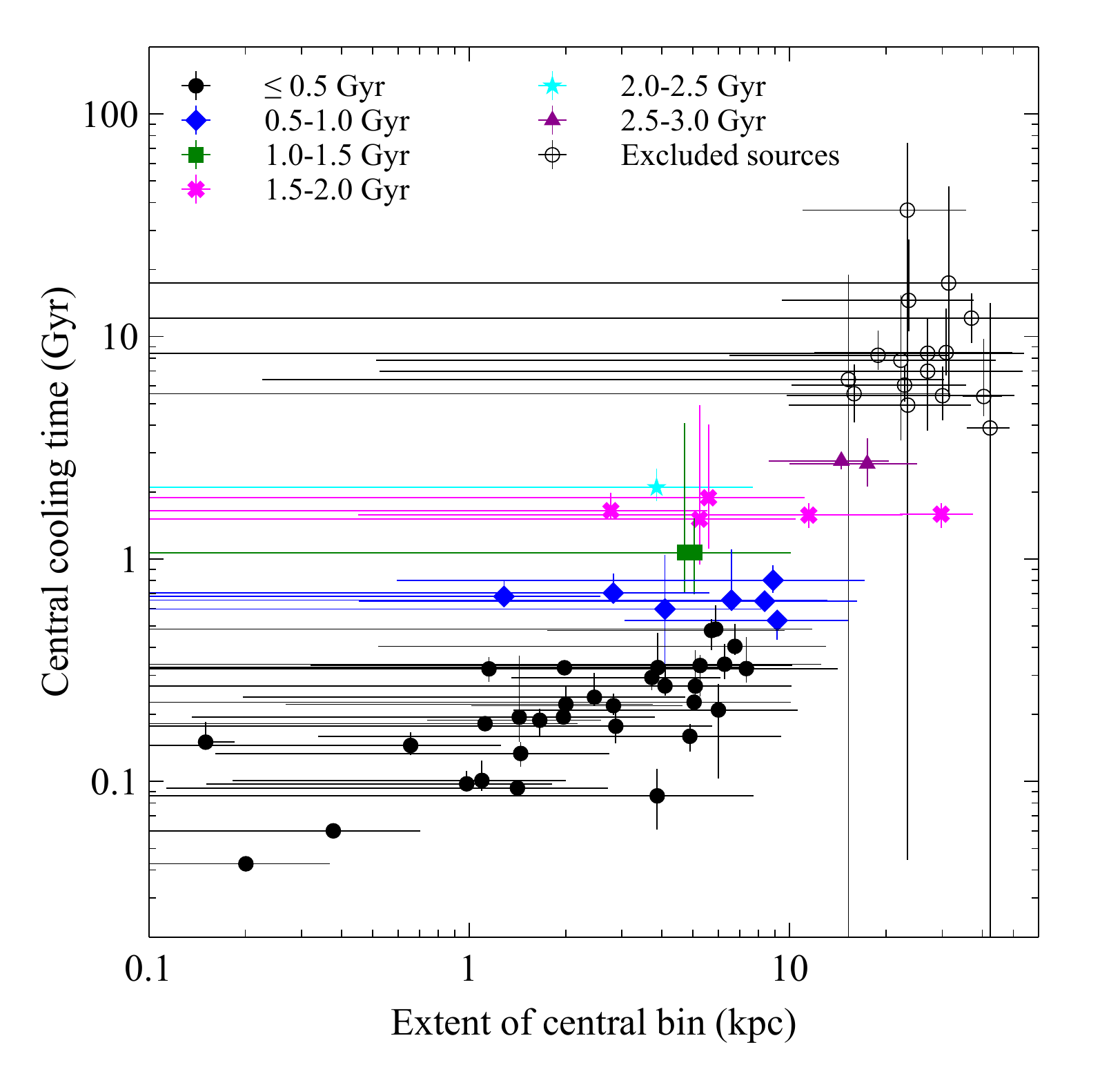}
\caption[]{Central cooling time vs. extent of the central spectral bin for the 65 sources in our parent sample, for which we have obtained central cooling time measurements. The different filled symbols indicate sources belonging to different central cooling time bins, while the empty circles indicate the 16 excluded sources.}
\label{fig:cooltimeprofs}
\end{center}
\end{figure}

For the sources in our short central cooling time subsample, the resulting central cooling times are shown in column 8 of Table \ref{tab:properties}. The central cooling times are plotted against the extent of the corresponding central spectral bin in Fig. \ref{fig:cooltimeprofs}, for the 65 sources for which we were able to obtain reliable central cooling time estimates. The different symbols represent systems in different central cooling time bins (see Table \ref{tab:properties}), and the empty symbols indicate the 16 excluded sources. 
We point out that the x-axis of Fig. \ref{fig:cooltimeprofs} is effectively the mean radius of each spectral bin, and the horizontal error bars are its  radial extent. 

It is clear that there is a weak trend between the central cooling time calculated here, and the radial extent covered by the central spectral bin. However, there are sources with central spectral bins of a similar size, that have significantly different central cooling times. There are a few exceptions to this, notably the two sources indicated by the purple triangles, which are included in our short central cooling time subsample, and the excluded sources. Therefore, we do not expect our estimates of the central cooling time to be noticeably affected by our choice of spectral bin size. Rather, it seems that calculations of the central cooling time are more strongly affected by the count rate within the central 20 kpc of each source (see Section 2.1 and Fig. \ref{fig:tcoolselection}), which is an intrinsic property of each source. From the right hand panel of Fig. \ref{fig:tcoolselection}, we can see that there is a spread of more than an order of magnitude in the count rate, of the sources that have a central cooling time $\leq$ 3 Gyr. In fact, if the central cooling time was defined as the cooling time of the gas at a certain radius from the source centre, for example at 1 kpc as is done in \cite{Birzan12}, the central cooling time of sources whose innermost bins are centered at a distance greater than 1 kpc, is likely to have stayed the same, or even increased due to the poorer data quality, which leads to noisier deprojected spectra. On the other hand, for sources with better resolved central regions, again the central cooling time would either stay the same, or increase due to the inclusion of regions at distances larger than those covered by the central spectral bin. This is based on the shape of the overall cooling time profile, shown in fig. 2 of Paper I, which gives an indication of the general cooling time behaviour of the sources in the parent sample. 

We note that, in Fig. \ref{fig:cooltimeprofs}, there are some sources whose central spectral bin does not cover the innermost regions of the core (i.e. the spectral bin does not go down to, in this case, 0.1 kpc or less). This is either due to the presence of a central point source, which was excised during the spectral analysis, or, less often, because the deprojection of the inner core regions did not give reliable temperature and electron number density estimates. We point out that the longer central cooling time calculated for the 16 excluded sources is not entirely due to resolution effects, as there are sources in our short central cooling time subsample which have central bins of a similar size to that of the 16 excluded sources. 


\section{Imaging analysis}

\begin{table*}
\begin{center}
\footnotesize{ 
\begin{tabular}{cccccccc} 
    \multicolumn{1}{c}{Source name}&\multicolumn{1}{c}{Group/cluster}&\multicolumn{1}{c}{Cavity location}&\multicolumn{1}{c}{$r_{\rm {a}}$ (kpc)}&\multicolumn{1}{c}{$r_{\rm {b}}$ (kpc)}&\multicolumn{1}{c}{Distance from core (kpc)}&\multicolumn{1}{c}{Cavity classification}&\multicolumn{1}{c}{H$\alpha$ filaments} \\ 
    \multicolumn{1}{c}{(1)}&\multicolumn{1}{c}{(2)}&\multicolumn{1}{c}{(3)}&\multicolumn{1}{c}{(4)}&\multicolumn{1}{c}{(5)}&\multicolumn{1}{c}{(6)}&\multicolumn{1}{c}{(7)}&\multicolumn{1}{c}{(8)}\\
    \hline
2A0335+096 & Cluster & NW &  15.0 & 6.7 & 43.9 &  C & yes (a)\\	 
 & & NW & 11.6 &  9.1 & 35.9 & C &  ''\\
 & & N & 6.5 & 13.0 & 66.3 & C & '' \\
NGC~4472 & Group & E & 1.3 & 0.8 & 3.2 & C & no (c, i) \\				
 & & W & 1.3 & 0.9 & 3.9 & C & ''\\
NGC~4696 & Cluster & E & 1.7 & 3.0 & 3.0 & C & yes (d)\\
 & & W & 1.6 & 2.8 & 3.1 & C & '' \\
A~262 &  Cluster & E & 3.5 & 3.4 & 7.5 & C & yes (f) \\				
 & & W & 4.1 & 3.4 & 6.5 & C & '' \\
HCG~62 &  Group & N & 2.1 & 2.8 & 7.5 & C & no (g) \\
 & & S & 2.7 & 2.4 & 5.2 & C & ''\\
NGC~1399  & Cluster & N & 2.1 & 1.3 & 5.0 & C & no (h, i)\\
 & & S & 2.2 & 1.7 & 6.4 & C & ''\\
NGC~4636  & Group & E & 1.5 & 1.4 & 3.4 & C & yes (i)\\		
 & & NW & 1.2 & 1.8 & 3.6 & C & '' \\
 & & SW & 1.8 & 1.9 & 4.9 & C & '' \\
NGC~5044 & Group & N & 3.8 & 3.5 & 5.7 & C & yes (h) \\
 & & S & 3.3 & 4.4 & 8.3 & C & '' \\	
 & & E & 2.8 & 3.5 & 7.5 & C & ''\\
 & & W & 2.7 & 1.3 & 4.1 & C & ''\\
NGC~5813 & Group & NE & 0.8 & 0.4 & 1.4 & C & yes (h)\\
 & & NE & 1.6 & 1.3 & 4.0 & C & ''\\
 & & NE & 4.2 & 6.2 & 14.2 & C & ''\\
 & & SW & 0.6 & 0.7 & 1.5 & C & ''\\
 & & SW & 3.1 & 3.1 & 6.5 & C & ''\\
NGC~6338 & Cluster & SW & 3.1 & 3.6 & 6.3 & C & yes (j)\\
 & & NE & 3.6 & 2.8 & 5.5 & C & '' \\
A~1991 & Cluster & N & 5.0 & 15.5 & 12.9 & C & yes (k) \\
 & & S & 4.4 & 6.7 & 10.7 & C & '' \\
IC~1262 & Group & N & 4.0 & 7.3 & 8.3 & C & yes (l) \\
A~S1101 & Cluster & SE & 4.7 & 7.4 & 14.8 & C & yes (m) \\
A~133 & Cluster & NW & 7.0 & 8.4 & 19.4 & C & yes (n) \\
 & & SW & 7.5 & 9.9 & 24.3 & C & ''\\
A~85 & Cluster & S & 4.6 & 7.9 & 13.9 & C & no (n) \\
A~1795 & Cluster & N & 5.8 & 16.7 & 10.6 & C & yes (o)\\ 
A~2052 & Cluster & N & 4.7 & 7.1 & 8.3 & C & yes (o) \\
 & & N & 3.2 & 10.3 & 18.2 & C & ''\\
 & & S & 4.9 & 13.1 & 8.1 & C & ''\\		
 & & S	& 4.3 & 7.0 & 19.9 & C & ''\\
A~2199 & Cluster & E & 6.9 & 5.3 & 16.8 & C & no (p)\\
 & & W & 8.9 & 6.9 & 23.4 & C & '' \\
A~4059 & Cluster & N & 14.2 & 17.9 & 22.6 & C & yes (n) \\
NGC~1550 & Group & SE & 2.1 & 3.0 & 8.5 & C & no (l)\\
NGC~3402 & Group & SW & 1.6 & 1.4 & 3.6 & C & no (k)\\ 
Hydra~A & Cluster & NE & 11.8 & 11.3 & 25.9 & C & yes (n) \\				
 & & NE & 35.6 & 39.7 & 83.1 & C & ''\\
 & & SW & 13.4 & 12.7 & 26.9 & C & '' \\
 & & SW & 34.6 & 36.6 & 109.2 & P & '' \\
A~3581 & Cluster & E & 2.4 & 2.5 & 4.9 & C & yes (b) \\
 & & E & 2.2 & 4.5 & 12.9 & C & '' \\
 & & W & 3.1 & 3.3 & 4.3 & C & ''\\
 & & W & 3.3 & 7.6 & 22.8 & P & ''\\
MKW3s & Cluster & SW & 13.5 & 6.7 & 58.6 & C & yes (q)\\
 & & S & 16.9 & 15.4 & 68.2 & P & '' \\
NGC~5846 & Group & S & 0.4 & 0.9 & 0.7 & C & yes (e) \\
 & & N & 0.4 & 0.8 & 1.0 & C & ''\\
 & & E & 1.2 & 1.3 & 5.0 & P & ''\\
 & & W & 1.7 & 1.5 & 5.5 & P & ''\\
NGC~4325 & Group & E & 2.4 & 4.6 & 11.8 & P & yes (o)\\      
 & & W & 2.6 & 4.3 & 7.7 & P & ''\\
A~1644 & Cluster & S & 8.3 & 6.1 & 15.0 & P & yes (n)\\
A~496 & Cluster & N & 5.6 & 4.9 & 14.2 & P & yes (n) \\
& & S & 2.8 & 7.9 & 7.1 & P & ''\\
NGC~499 & Group & SE & 1.4 & 2.0 & 4.5 & P & ?\\
 & & SW & 1.1 & 1.3 & 3.0 & P & ''\\
    \hline
\end{tabular}
}
\end{center}
\caption{List of the properties of the X-ray cavities of the groups and clusters in our subsample.}
\label{tab:cavities}
\end{table*}


\begin{table*}
\begin{center}
 \contcaption{}
\begin{tabular}{cccccccc}
\multicolumn{1}{c}{Source name}&\multicolumn{1}{c}{Group/cluster}&\multicolumn{1}{c}{Cavity location}&\multicolumn{1}{c}{$r_{a}$ (kpc)}&\multicolumn{1}{c}{$r_{b}$ (kpc)}&\multicolumn{1}{c}{Distance from core (kpc)}&\multicolumn{1}{c}{Cavity classification}&\multicolumn{1}{c}{H$\alpha$ filaments} \\ 
    \multicolumn{1}{c}{(1)}&\multicolumn{1}{c}{(2)}&\multicolumn{1}{c}{(3)}&\multicolumn{1}{c}{(4)}&\multicolumn{1}{c}{(5)}&\multicolumn{1}{c}{(6)}&\multicolumn{1}{c}{(7)}&\multicolumn{1}{c}{(8)}\\
\hline
 NGC~777 &  Group & E & 1.9 & 2.3 & 4.6 & P & ?\\
 & & W & 2.1 & 2.4 & 4.0 & P & ''\\ 
\hline
\end{tabular}
\end{center}
\raggedright{List of the properties of the X-ray cavities of the groups and clusters in our sample, all of which are also part of the short central cooling time subsample. All the sources are also part of the short central cooling time subsample. ``C'' denotes sources with certain cavities, while ``P'' indicates sources with possible cavities. We were unable to find evidence in the literature supporting either the presence or absence of H$\alpha$ filaments in NGC~499 and NGC~777, which we indicate with ``?'' in the relevant column. (1) source name, (2) group/cluster classification of source, (3) location of X-ray cavity, (4) length of axis along the direction of the jet in kpc, (5) length of axis perpendicular to the jet in kpc (6) distance of cavity from source centre, (7) classification of X-ray cavities as certain (C) or possible (P) (for more details, see Section 5), and (8) presence/absence of H$\alpha$ filaments and corresponding reference. REFERENCES.- (a) \cite{Romanishin88}, (b) \cite{Canning13}, (c) \cite{Battaia12}, (d) \cite{Crawford05b}, (e) \cite{Goudfrooij98}, (f) \cite{Plana98}, (g) \cite{Valluri96}, (h) \cite{Goudfrooij94}, (i) \cite{Werner13}, (j) \cite{Martel04}, (k) \cite{McDonald11}, (l) \cite{Crawford99}, (m) \cite{Jaffe05}, (n) \cite{McDonald10}, (o) \cite{McDonald12}, (p) \cite{Godon94}, (q) \cite{Edwards09}.} 
\end{table*}

\subsection{X-ray cavity detection}
As previously mentioned, the main aim of this paper is the study of the cavity dynamics of a sample of X-ray groups and clusters. We thus employ unsharp-masking to detect cavities in all the sources of our parent sample, not just those in our short central cooling time subsample. The way unsharp-masking works is the same image is smoothed twice, once with a wider smoothing kernel, and then with a less wide kernel. The more heavily-smoothed image is then subtracted from the less-smoothed image, thus revealing small-scale inhomogeneities in the image, such as X-ray cavities. In our analysis, we smoothed 0.5--7.0 keV background-subtracted, exposure-corrected {\it Chandra} images with 1-, 2-, 8- and 10-pixel Gaussians. We then subtracted each of the two more heavily-smoothed images from each of the less-smoothed images. The 0.5--7.0 keV background-subtracted, exposure-corrected images and the unsharp-masked images are presented in Fig. A1 of Appendix A. We only show the unsharp-masked image that best highlighted the presence of X-ray cavities, which for most sources, is the subtraction of the 8-pixel Gaussian smoothed image from the 2-pixel Gaussian smoothed image. 

We classify sources as having certain (``C''), possible (``P'') or no cavities (``N''), based on the following classification scheme:
\begin{itemize}
\item{Sources with {\bf certain} cavities. The cavities are (a) clearly visible in the original image (i.e. background-subtracted and exposure-corrected image), as well as the unsharp-masked image, or (b) are unambiguously visible in the unsharp-masked image.}
  \item{Sources with {\bf possible} cavities. These cavities are either (a) hinted at in the original image, but not clearly visible in the unsharp-masked image, or (b) visible in the unsharp-masked image, but the data are too noisy for the presence of a cavity to be certain. Cavities visible only in the unsharp-masked image are listed as only possible because, what might appear like a cavity, may be a conveniently located and shaped surface brightness depression, which is actually an artefact of the unsharp-masking.}
  \item{Sources with {\bf no} cavities. In this case, both the original and unsharp-masked image show no visual sign of an X-ray cavity. Included in this category are sources in whose unsharp-masked image, the surface brightness depression forms a dark annulus about the source centre. This dark annulus is most likely an artefact introduced by the unsharp-masking, through the presence of a sharp central brightness excess.} 
\end{itemize}
Table \ref{tab:cavities} lists all the sources which display certain or possible cavities, along with the size of the cavities and the distance from their centre to the centre of their host group or cluster, both of which were calculated from the unsharp masked images. Here, we have assumed that the cavities are prolate ellipsoids, with a semimajor axis along the direction of the jet $r_{\rm {a}}$, and a semimajor axis perpendicular to the jet direction, $r_{\rm {b}}$. We also indicate whether there is extended H$\alpha$ filamentary structure in the core of each source, along with the relevant reference. The ``?'' in the cases of NGC~499 and NGC~777 indicates that we were unable to find any evidence from the literature, which supported the presence or absence of H$\alpha$ filaments in these sources. All the sources that have certain or possible X-ray cavities also have a central cooling time of $\leq$3 Gyr. Hence, out of the 49 sources in our subsample, a maximum of 30 of them display X-ray cavities ($\sim$61 percent). We note that there are a few sources with short central cooling times that do not show any signs of X-ray cavities. These sources are the starred sources in Table \ref{tab:properties}.

Through our search for X-ray cavities using unsharp-masking, we have uncovered some previously undiscovered  cavities in some sources in our sample. These newly-discovered cavities are the cavity in NGC~3402, the ``possible'' cavity in Abell~1644 and the eastern ghost cavity in NGC~5846. A ghost cavity to the west of the centre of NGC~5846 has been reported previously by \cite{Machacek11}, though the same authors do not detect a cavity to the east. A cavity has not been previously detected in NGC~3402 or Abell~1644. 

\subsubsection{{\it\textbf{XMM-Newton}} images}
We note that there are some sources in our sample, for whose analysis {\it XMM-Newton} data were used. These are Abell~1736, Abell~2192, Abell~2197, Abell~3376, Abell~3390, Abell~3391, Abell~3395, Abell~S0405, Abell~S0805, IC~4296, IC~4329, NGC~410, RXCJ0920.0+0102, RXCJ1840.6-7709, RXCJ2314.7-0222 and UGC~4052. To search these sources for cavities, we used only the cleaned (i.e. periods of background flaring have been removed) EPIC MOS data. We excluded the pn detector from our imaging analysis as it has numerous detector gaps in its centre, which is where we are most likely to see X-ray cavities. We generated 0.5--7.0 keV images for each MOS detector seperately, and then added these together. The total image was then smoothed with 1-, 2-, 8- and 10-pixel Gaussians, as was done with the {\it Chandra} data. Each of the two more heavily smoothed images were then subtracted from each of the less smoothed images. The total image of each source, and one of the unsharp-masked images, are shown in Fig. A2 of Appendix A. We did not discover X-ray cavities in any of the aforementioned sources, which may be in part due to the poorer angular resolution and larger point spread function (PSF) of the EPIC instruments. We note that the data for Abell~2197 were so heavily contaminated by background flaring, that no usable exposure time was left after removing periods of background flaring. We show no combined MOS image for IC~4329, as the object in the centre of the detectors is a very bright FR-I source, and so the MOS detectors were operated in large window mode. This means that the diffuse emission from the cluster is not visible in the images.

\subsection{Sources without X-ray cavities}
To verify that we are not ``missing'' any X-ray cavities, that happen to be too small or faint to be readily visible in the currently available data, we estimated the size of X-ray cavities necessary to offset cooling in the cores of some of the sources in our sample. We performed this calculation for the three sources in our sample which have a central cooling time $\leq$0.25 Gyr, and which do not have cavities, namely MKW4, NGC~533 and NGC~6482. We assumed that the cavities are spherical in shape, that they are produced in pairs and that they move and expand at the local sound speed. We determined the size of the bubbles needed to offset the cooling within the cooling radius, $r_{\rm {cool}}$ (the radius within which the cooling time is $\leq$3 Gyr), and we used the results from the spectral fitting to estimate the local sound speed. 

We find that if the X-ray cavities for the three sources were located close to the centre of the host source, only the cavities in MKW4 would be just about large enough to be resolved in the currently available observations. However, not knowing the size or location of these bubbles significantly decreases the possibility of their detection, as does the fact that they may not have bright rims. We note that previous searches for X-ray cavities in these sources have not found them either \citep{Dong10, Khosroshahi04, Gu12}.     


\begin{figure*}
\begin{center}
\includegraphics[trim =  0cm 14cm 4.5cm 0cm, clip, height=3.3in, width=3.3in]{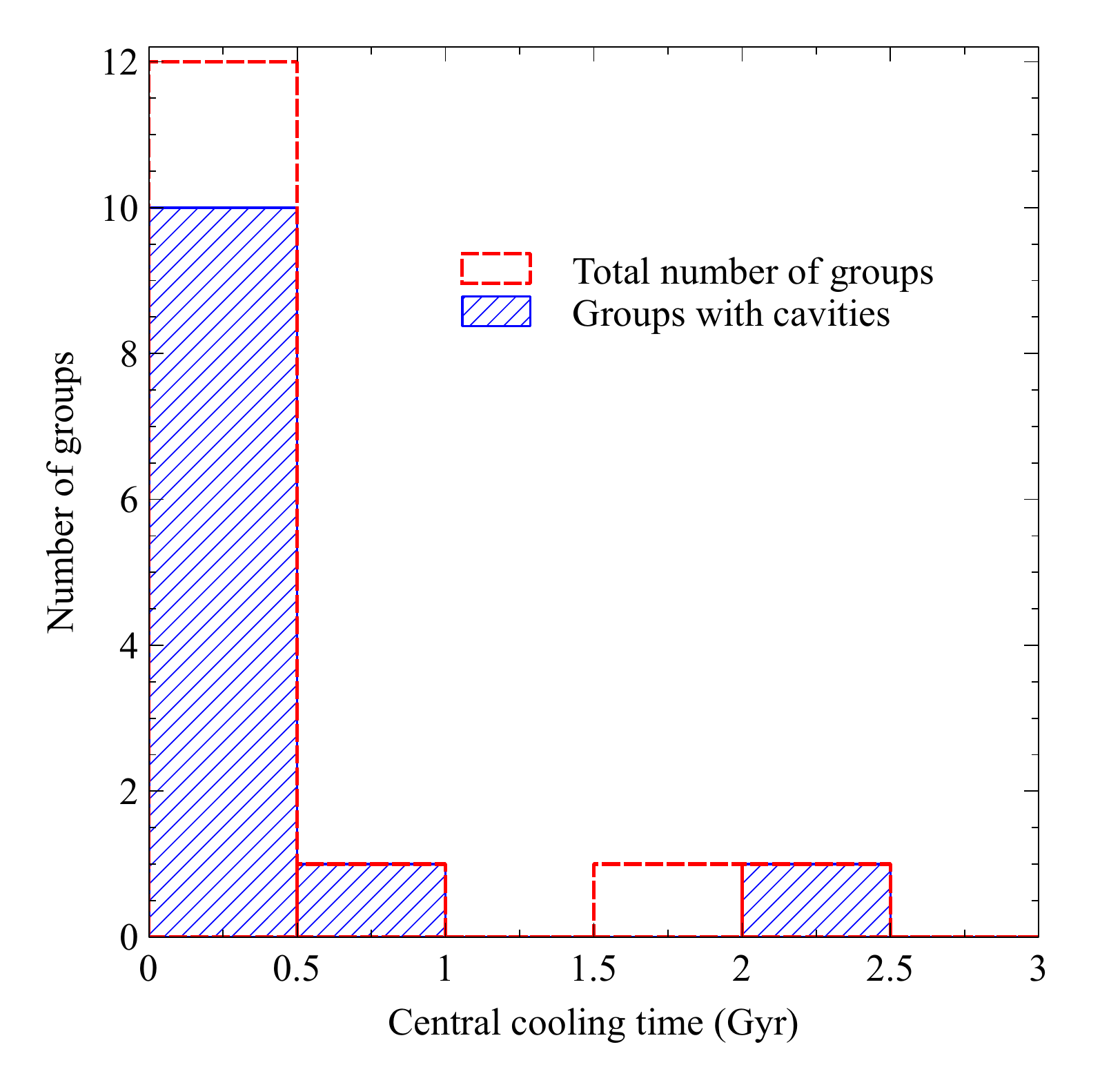}
\includegraphics[trim =  0cm 14cm 4.5cm 0cm, clip, height=3.3in, width=3.3in]{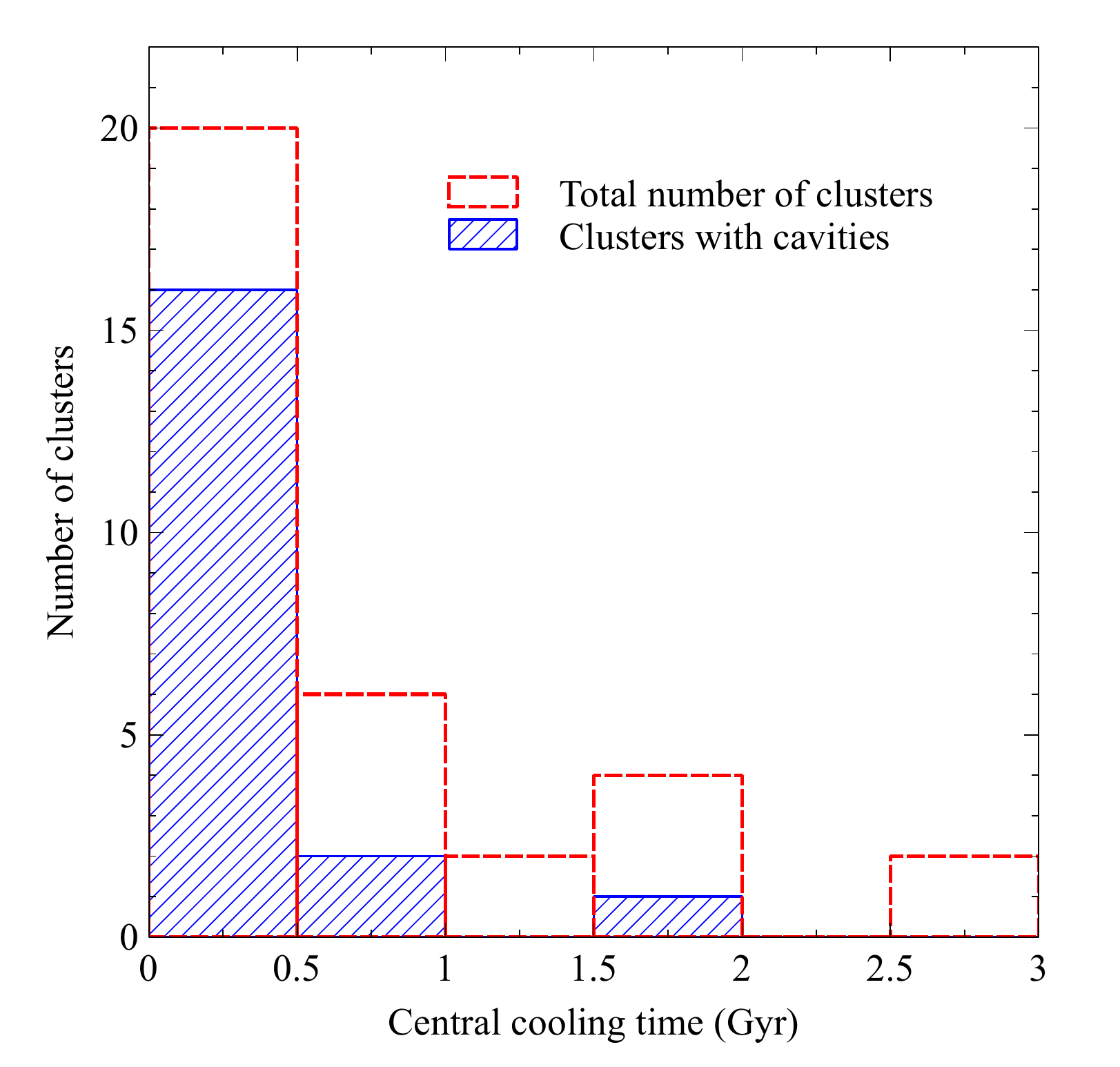}
\caption{The distribution of groups (left panel) and clusters (right panel) in our subsample, and whether they have cavities or not, with respect to their central cooling time. The dashed rectangular areas indicate the total number of groups or clusters, while the blue shaded areas represent the number of groups or clusters that do have cavities.}
\label{fig:grpandclus}
\end{center}
\end{figure*}

\section{Results}
\subsection{Clusters vs. groups}
To inspect if there is any difference in the distribution of sources with X-ray cavities with respect to their central cooling time, we examine the groups and clusters in our subsample separately. The central cooling time is defined as the cooling time calculated for the gas in the innermost spectral bin of each source. In addition, we used 0.5 Gyr bins for the central cooling time, covering the 0--3.0 Gyr range. The distributions for the groups and the clusters of our subsample are given in the left-hand and right-hand panels of Fig. \ref{fig:grpandclus}, respectively. The areas enclosed within the dashed rectangles represent the total number of groups or clusters in each bin, and the blue shaded regions indicate the number of groups or clusters that harbour X-ray cavities. We point out that all the sources from our main sample of 101 X-ray groups and clusters that have X-ray cavities, also have a central cooling time of $\leq$3 Gyr.

It is obvious from Fig. \ref{fig:grpandclus} that the vast majority of the subsample sources lie in the $\leq$0.5 Gyr bin. In fact, 32 out of 49 sources ($\sim$65 percent) of the short central cooling time subsample, are in this central cooling time bin. The number of both groups and clusters in longer central cooling time bins decreases quite rapidly. From Fig. \ref{fig:grpandclus}, we can state that the AGN duty cycle of the groups and clusters with a central cooling time of $\leq$3 Gyr is $\simeq$61 percent. However, given the added difficulty of detecting bubbles which are not visible due to projection effects, and the fact that almost all of the sources in our subsample have a detected radio source, the actual duty cycle is likely to be much higher.  

In the $\leq$0.5 Gyr central cooling time bin, 10 out of 12 groups have cavities ($\sim$83 per cent), while 16 out of 20 clusters have cavities (80 per cent). This means that the AGN duty cycle for groups and clusters in our subsample, with central cooling times $\leq$0.5 Gyr, are quite similar. Fig. \ref{fig:grpandclus} also indicates that most sources with short central cooling times harbour X-ray cavities. In other words, the majority of sources in which heating is needed to offset cooling, possess X-ray cavities. However, there are significantly fewer groups and clusters that have X-ray cavities with central cooling times longer than 0.5 Gyr, so it is not easy to extrapolate this conclusion for longer central cooling times. 


\subsection{Data quality effects}

\begin{figure}
\begin{center}
\includegraphics[trim =  0cm 14cm 4.5cm 0cm, clip, height=3.3in, width=3.3in]{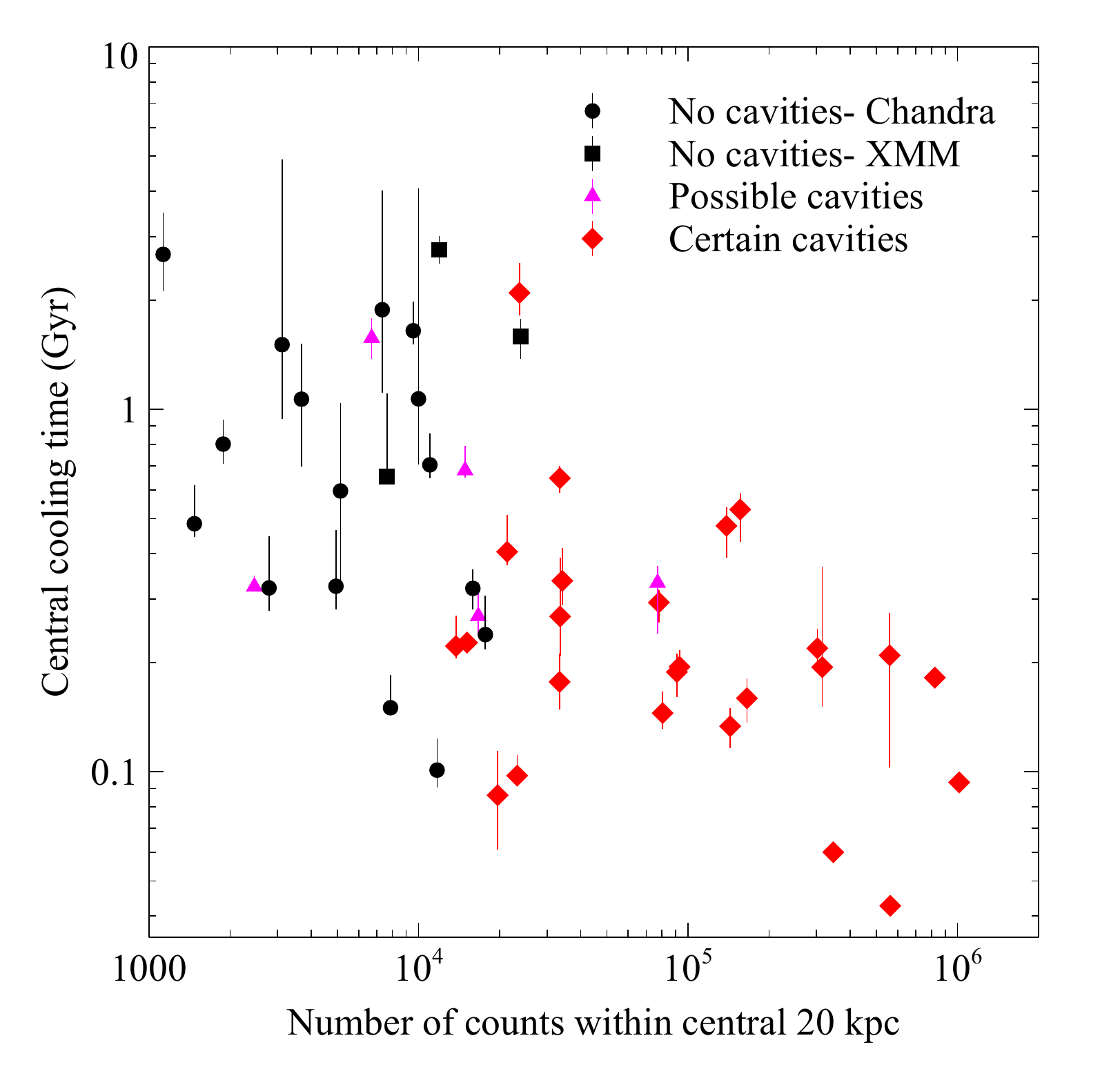}
\caption{Central cooling time vs. total number of counts within the central 20 kpc of each source in the short central cooling time subsample, adapted from Fig. \ref{fig:tcoolvscounts}. The black circles and squares represent sources with no cavities, for whose analysis {\it Chandra} and {\it XMM-Newton} data were used, respectively. The pink triangles indicate sources all of whose cavities are classified as ``possible'', while the red diamonds are the sources which have at least one ``certain'' cavity. The rightmost black square is RXCJ1840.6-7709, which has a very bright source at its core.}
\label{fig:tcoolvscounts}
\end{center}
\end{figure}
The detectability of X-ray cavities in a X-ray group or cluster could be strongly dependent on the quality of the data available on that source. To test whether the lack of data on a source impacts the detection of cavities, we calculated the number of counts within the central 20 kpc of each source in our short central cooling time sample. We use 0.5--7.0 keV images from which point sources have been removed, and that have not been exposure-corrected (in the case of {\it Chandra} data, the images have been background-subtracted). If a central point source is visible at the core of a group or cluster, we exclude it from the calculation. We point out that in the case of some of the lowest redshift sources, a circle with a radius of 20 kpc extends beyond the area of the ACIS-I and ACIS-S chips. In addition, there are three sources for which we have used {\it XMM-Newton} data, all three of which do not have X-ray cavities. These three sources will have a higher number of counts in their central 20 kpc due to the combination of detectors, and the higher sensitivity of the EPIC detectors compared to the ACIS detectors. 

We plot the central cooling time of each source against the corresponding number of total counts within 20 kpc in Fig. \ref{fig:tcoolvscounts}. This figure is similar to the left hand panel of Fig. \ref{fig:tcoolselection}, but here we have not included the ``excluded'' sources of Fig. \ref{fig:tcoolselection}. In addition, we have colour-coded the 49 remaining sources (i.e. the sources in the short central cooling time subsample), according to the detection or non-detection of an X-ray cavity. The sources with no cavities, for which we used {\it Chandra} or {\it XMM-Newton} data in the analysis, are represented by the black circles and squares, respectively. Sources whose cavities have all been classified as ``possible'', or that have at least one ``certain'' cavity, are indicated by the pink triangles and red diamonds, respectively. The rightmost black square is RXCJ1840.6-7709, which has a very bright source at its core, and as such has a high number of counts (see Appendix A for the relevant {\it XMM-Newton} image). It is evident that the sources which are found to have X-ray cavities generally have the highest number of counts and shortest cooling times, due to their cores being better resolved. In fact, with just a couple of exceptions, sources with fewer than $\sim$20000 counts within a 20 kpc radius circle from their core, do not have clearly detected X-ray cavities. 
Therefore, the values obtained from Fig. \ref{fig:grpandclus} should be regarded as lower limits, pending the availability of longer observations. Deeper observations will reveal more cavities in these sources, and raise the fraction of sources with cavities towards unity, reaffirming the fact that the incidence of cavities is underestimated.  

The inverse trend seen in Fig. \ref{fig:tcoolvscounts} is likely a signal-to-noise ratio effect. 

\subsection{Cavity size vs. temperature}

\begin{figure}
\begin{center}
\includegraphics[trim =  0cm 14cm 4.5cm 0cm, clip, height=3.3in, width=3.3in]{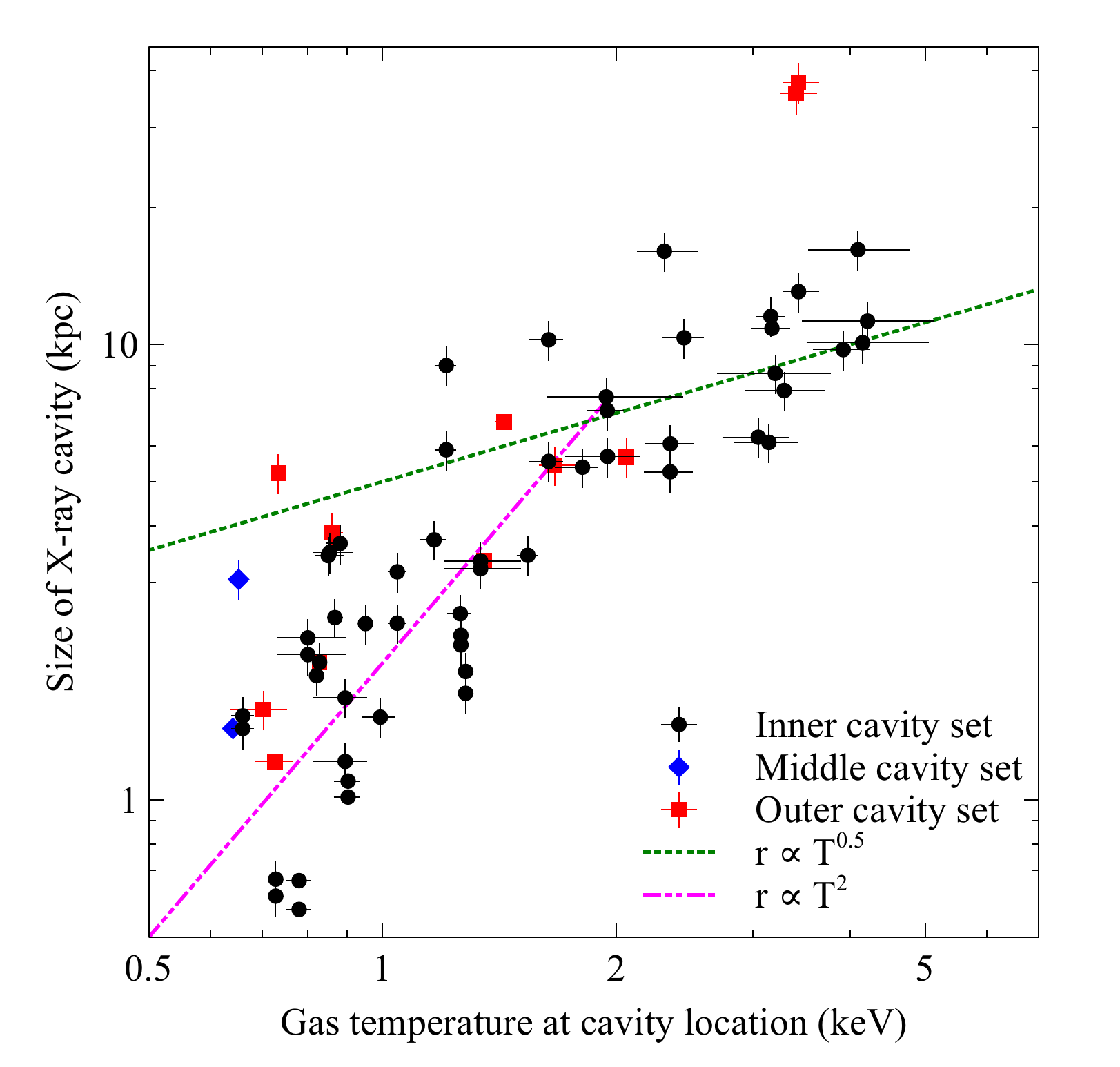}
\caption{X-ray cavity size vs. gas temperature at the cavity location. The inner (or only) set of cavities in a source, and the outer set of cavities are symbolised by the black circles and red squares, respectively. The blue diamonds represent the middle set of cavities in NGC~5813. The dotted line represents the relation $r \propto T^{0.5}$, while the dot-dashed line indicates $r \propto T^{2}$.}
\label{fig:tempvslocation}
\end{center}
\end{figure}

To examine whether there is any correlation between the size of cavities in a group or cluster, and the temperature of the host group or cluster, we plot the size of the cavity against the deprojected temperature. Specifically, we examine the correlation between cavity size and the deprojected temperature at the location of the cavity within its host group or cluster. Since we have modelled the cavities in our sample as prolate ellipsoids, we use the average of $r_{\rm {a}}$ and $r_{\rm {b}}$ as a proxy for the cavity ``radius'', $r$. The resulting plot is shown in Fig. \ref{fig:tempvslocation}. The black circles represent the inner (or only) set of cavities in a source, and the red squares the outer set of cavities. The two blue diamonds are the middle set of cavities in NGC~5813. The dotted line and dot-dashed line in the same figure indicate the relations $r \propto T^{0.5}$ and $r \propto T^{2}$, respectively, though neither is a fit to the data. As can be seen, the size of an X-ray cavity does not have a fixed dependence on the ambient ICM temperature.

We now explore the relation between the radius of a bubble, $r$, and the temperature of the ICM at its location, $T_{\rm {ICM}}$. From \cite{Churazov00}, we have
\begin{equation}
r \propto \sqrt{\frac{P_{\rm {cav}}}{p\upsilon_{\rm {K}}}} , 
\label{eqn:rbubble}
\end{equation}
where $P_{\rm {cav}}$ is the power inflating the cavity, $p$ is the pressure of the surrounding ICM (we assume that the cavity is almost in pressure equilibrium with its surrounding gas), and $\upsilon_{\rm {K}}$ is the Keplerian velocity at the location of the cavity. The Keplerian velocity is comparable to the local sound speed, $c_{\rm {S}}$, in the ICM, so
\begin{equation}
\upsilon_{\rm {K}} \simeq c_{\rm {S}} \propto T_{\rm {ICM}}^{1/2}.
\label{eqn:keplerian}
\end{equation}

For a limited range of cooling times, the Bremsstrahlung cooling time of the ICM gas, $t_{\rm {Brem}}$, is given by 
\begin{equation}
t_{\rm {Brem}} \propto \frac{T_{\rm {ICM}}^{1/2}}{n}, 
\label{eqn:tbrem}
\end{equation}
where $n$ is the number density of the ICM gas. From this relation, we have $n \propto T_{\rm {ICM}}^{1/2}$. Combining this with the ideal gas law, $p = nk_{\rm {B}}T_{\rm {ICM}}$, where $k_{\rm {B}}$ is the Boltzmann constant and $p$ is the pressure of the ICM gas, we have $p \propto T_{\rm {ICM}}^{3/2}$, for a limited range of Bremsstrahlung cooling times. The power inflating the cavity depends directly on the cooling luminosity of the cluster, $L_{\rm {cool}}$, as $P_{\rm {cav}} \propto  L_{\rm {cool}}$, which in turn depends on the bolometric luminosity, $L_{\rm {bol}}$, as $L_{\rm {cool}} \propto L_{\rm {bol}}$ \citep{Peres98}. For clusters, it is well-established that $L_{\rm {cool}} \propto T_{\rm {ICM}}^{3}$ \citep[e.g.][]{Markevitch98, Pratt09, Maughan12}. 

By substituting all the above scaling relations into Equation \ref{eqn:rbubble}, we get
\begin{equation}
r \propto T_{\rm {ICM}}^{0.5}.
\label{eqn:rbubfinal}
\end{equation}
This relation fits the general trend of the points in Fig. \ref{fig:tempvslocation} with $T_{\rm{ICM}}$ $\geq$1.5 keV reasonably well, though there is significant scatter. On the other hand, below 1.5 keV, this scaling relation no longer applies, and the scatter is much greater. This could be due to the fact that, below 1.5 keV, the relation between the X-ray luminosity and the temperature of the ICM is much less well-constrained, as has been observed by e.g. \cite{Osmond04}. The scatter in Fig. \ref{fig:tempvslocation} may be reflecting this. We have overplotted the $r \propto T_{\rm {ICM}}^{2}$ relation in the same figure to indicate this discrepancy, though this relation is not a fit to the data. 

\subsection{Cavity power vs. cooling luminosity}
\begin{figure}
\begin{center}
\includegraphics[trim =  0cm 14cm 4.5cm 0cm, clip, height=3.4in, width=3.4in]{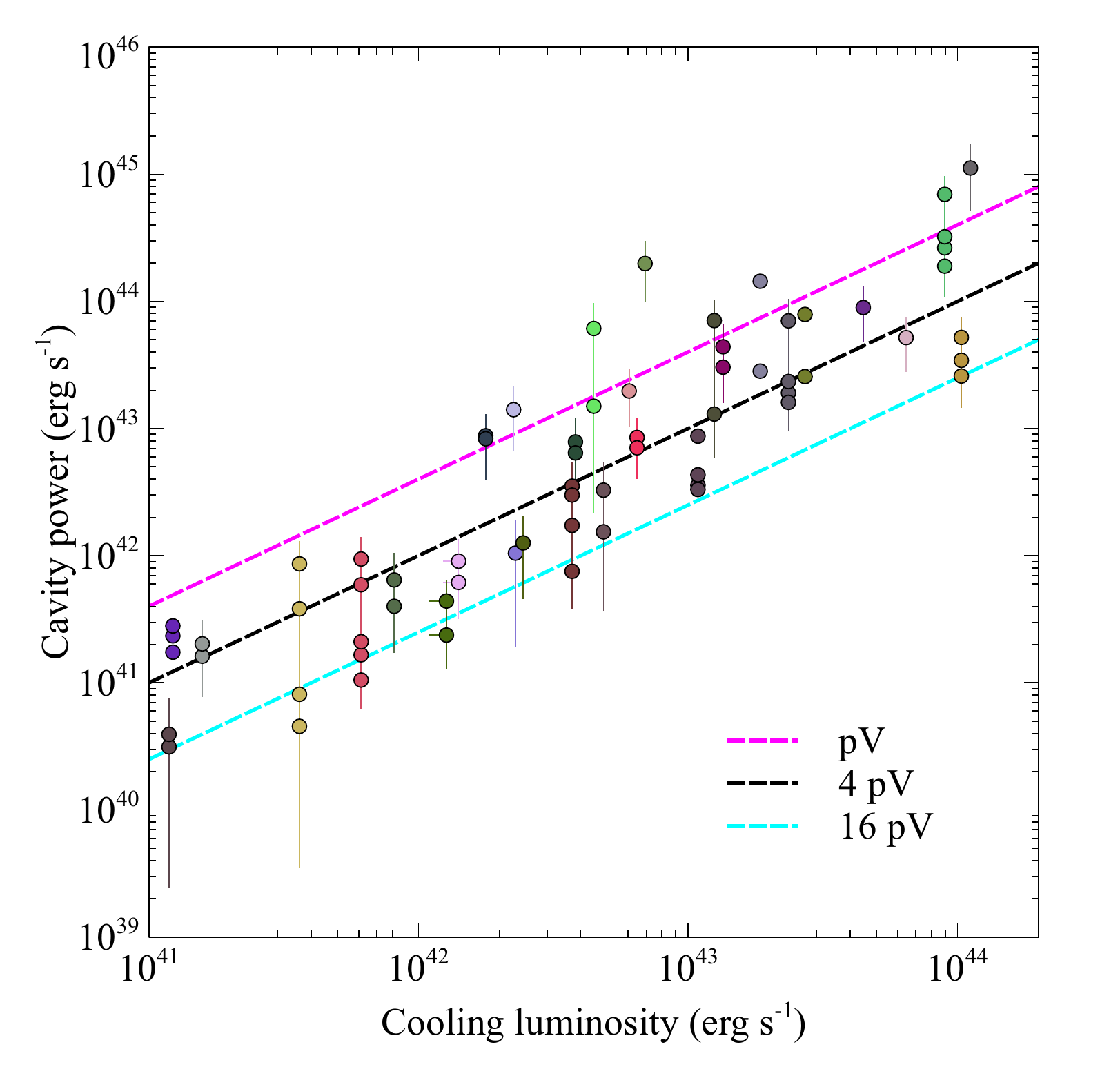}
\caption{Cavity power vs. cooling luminosity for the 42 groups and clusters in our sample. The pink, black and blue dashed lines show equality for $pV$, 4$pV$ and 16$pV$, respectively. The cavities of each source have been given the same colour.}
\label{fig:cavpower}
\end{center}
\end{figure}

As mentioned in Section 1, outbursts of energy from the brightest cluster galaxy (BCG) AGN are thought to be the main mechanism quenching cooling flows in groups and clusters of galaxies. To check whether this is the case with the groups and clusters in our subsample, we calculated the power of all the cavities in our subsample, $P_{\rm {cav}}$, and the cooling luminosity, $L_{\rm {cool}}$, of the corresponding groups and clusters. We use 
\begin{equation}
P_{\rm {cav}} = \frac{4p_{\rm {th}}V}{t_{\rm {cav}}},
\label{eq:cavpower}
\end{equation}
where $p_{\rm {th}}$ is the thermal pressure of the gas surrounding the cavity, $V$ is the volume of the cavity, and $t_{\rm {cav}}$ is the age of the cavity. As the X-ray cavities are assumed to be prolate ellipsoids, with $r_{\rm {a}}$ and $r_{\rm {b}}$ the semimajor axes along and perpendicular the jet direction respectively, the volume $V$ is
\begin{equation}
V = \frac{4}{3} {\rm \pi} r_{\rm {a}} r_{\rm {b}}^{2}.
\label{volume}
\end{equation}
Here, we have assumed that the gas in the X-ray cavities is in pressure equilibrium with the surrounding ICM, and the age of the cavity is defined as the time it would take the cavity to reach its current location at the speed of sound
\begin{equation}
t_{\rm {cav}} = \frac{R}{c_{\rm {s}}},
\label{cavage}
\end{equation}
where $R$ is the distance of the centre of the cavity from the centre of the source, and $c_{\rm {s}}$ is the local speed of sound. The latter is defined as
\begin{equation}
c_{\rm {s}} = \sqrt{\gamma \frac{k_{\rm {B}} T}{\mu m_{\rm {H}}}},
\label{soundspeed}
\end{equation} 
where $\gamma$ is the adiabatic index for the gas, $m_{\rm {H}}$ is the mass of the hydrogen atom, and $\mu$ is equal to 0.62. Here we have used $\gamma$=5/3 for a non-relativistic gas. We use this definition for the age of the cavity, as no strong shocks have been detected in groups or clusters. Therefore, the bubbles cannot be expanding at a speed much greater than the local sound speed in the ICM. We note that other studies \citep[e.g.][]{Birzan04} have also used the buoyancy rise time and cavity refill time (the amount of time it would take to refill the displaced volume as the cavity rises outwards) to calculate the age of a cavity, in addition to using the local sound speed. The resulting values for the cavity ages using these different methods do not vary significantly, and therefore our results would not affected if we used the other cavity age estimates. The cooling luminosity $L_{\rm {cool}}$ is calculated from the spectral fits, for which a {\sc wabs*apec} model was used in {\sc xspec}, and is defined as the luminosity of the gas within the cooling radius $r_{\rm {cool}}$. In turn, $r_{\rm {cool}}$ is defined as the radius within which the gas has a cooling time of $\leq$3 Gyr, in line with previous studies \citep{Dunn06, Dunn08}. As we are studying a sample of relatively nearby galaxy groups and clusters, our selection of cooling radius ensures we are able to study the innermost regions of these sample sources and take advantage of the higher spatial resolution offered, and so study AGN feedback in more detail. We have assumed a typical 20 percent error on the cavity dimensions, and the errors on $P_{\rm cav}$ were calculated using standard error propagation. The resulting plot is shown in Fig. \ref{fig:cavpower}, where the bubbles of each source have been given the same colour. The pink, black and blue dashed lines represent equality for cavity energies of $pV$, 4$pV$ and 16$pV$, respectively. 

As can be seen from Fig. \ref{fig:cavpower}, the X-ray cavities of most sources lie between the $pV$ and 16$pV$ lines, rather than above the $pV$ line or below the 16$pV$ line. This indicates that the ``bubbling mode'' of the central AGN has to be a fairly continuous process in order to offset cooling, rather than the result of an episodic outburst of AGN activity, which was triggered by an accretion event. However, there are a few sources which cannot offset the cooling flow in their cores through current AGN heating alone. These results from our analysis are in agreement with those of \cite{Birzan12}, who found that the systems with bubbles in the sample they use require a constant input of energy from the AGN (see figure 5 of the same paper). Our results also agree with recent theoretical work on AGN feedback, which favour relatively gentle and self-regulated AGN feedback, rather than feedback manifested in sudden and violent outbursts \citep[see e.g.][]{Gaspari11, Gaspari12}. Fig. \ref{fig:cavpower} is, however, in tension with the results of \cite{Nulsen07, Nulsen09}, who study a sample of nearby elliptical galaxies. The same authors find that cooling in these systems can be offset by intermittent AGN outbursts, rather than continuous bubbling activity. 

We note that \cite{Birzan12} define the cooling radius, $r_{\rm {cool}}$ as the radius within which the cooling time of the gas is 7.7 Gyr (we use 3 Gyr in our analysis). In addition, they study the correlation between cavity power and cooling luminosity for two subsamples of sources harbouring cooling flows: both subsamples require sources to have a separation $\leq$12 kpc between the optical core and the X-ray emission peak, but the first subsample additionally requires that $\eta_{\rm {min}} \leq$ 5 \citep[$\eta_{\rm {min}}$ is a measure of the gas' thermal stabillity; see][]{Voit08}, while the second requires a cooling time of 0.5--1 Gyr at 1 kpc. It is encouraging to see our results agree with those of \cite{Birzan12}, despite the differences in the subsample selection and the definition of $r_{\rm {cool}}$.

\subsection{Do smaller or larger bubbles travel faster?}
In our study, there are a number of sources that have multiple sets of bubbles, which increase in size with increasing distance from the source centre, such as Hydra~A and NGC~5813. Here we examine whether these outer, larger bubbles are a result of the merging of individual smaller bubbles, produced in separate bubbling events. 

\cite{Diehl08} showed that bubbles seen at larger distances from the cores of a sample of clusters, appear to be a lot larger than those at smaller radii. In fact, these outer bubbles are larger than expected even from simple adiabatic expansion, which is puzzling. An example of a multicavity system which has been studied in detail is Hydra~A, in which \cite{Wise07} identify 3 sets of bubbles, which show a steep increase of bubble radius with distance from the cluster core. It is therefore possible that these larger outer bubbles are due to the accumulation and merging of individual bubbles, that were created in multiple AGN cycles. Indeed, features that correspond to drops in flux in X-ray images, and could be due to the merging of many smaller individual cavities, have been seen in e.g. the Perseus cluster \citep{Fabian11} and Abell~2204 \citep{Sanders09}. After calculating the mechanical power necessary to create the outermost, and largest, set of bubbles in these clusters in a single AGN outburst, both authors conclude it is more likely that these bubbles are the result of the accumulation of many smaller bubbles produced during past AGN outbursts, rather than by a single outburst. In particular, \cite{Fabian11} conclude that bubbles must be long-lived, and that faster-moving bubbles rise and sweep up slower-rising bubbles, merging with them. It is thought that bubbles become detached from the AGN jet, rise buoyantly outwards from the cluster centre, and become ``trapped'' at some larger radius. There, it is possible that they are neutrally buoyant, either through mixing with the surrounding ICM gas, or through interactions with the local magnetic fields, which arrest their upward movement. A question which then arises is whether small bubbles catch up with previously created larger bubbles, or vice versa. In other words, do smaller or larger bubbles travel faster through the ICM?  

The terminal velocity of a bubble at its current location, assuming it is rising buoyantly through the ICM, is given by \citep{Birzan04}
\begin{equation}
\upsilon_{\rm {t}} \simeq \sqrt{\frac{2gV}{SC}},
\label{eq:termvelocity}
\end{equation}
where $g$ is the gravitational acceleration, $V$ is the volume of the bubble, $S$ is its cross section and $C$=0.75 is the drag coefficient \citep{Churazov01}. The gravitational acceleration $g$ can be calculated using the stellar velocity dispersion of the central galaxy of a group or cluster, $\sigma$, and, assuming that this galaxy is an isothermal sphere,  is defined as \citep{Binney87}
\begin{equation}
g \simeq \frac{2 \sigma^{2}}{R},
\label{eq:gravacc}
\end{equation}
where $R$ is the projected distance from the centre of the group or cluster to the centre of the cavity. Assuming the bubbles are spherical in shape, Equation \ref{eq:termvelocity} can therefore be rewritten as
\begin{equation}
\upsilon_{\rm {t}} \propto \sigma \sqrt{\frac{r}{R}}.
\label{eq:termvelocityshort}
\end{equation}
For bubbles in the same cluster, the $\sigma$ term in the above equation can be dropped, so we have $ \upsilon_{\rm {t}} \propto \sqrt{r/R}$. As a result, at a given radius in a certain group or cluster, larger bubbles will travel faster. This means that it is possible for larger bubbles to catch up and merge with smaller bubbles that were created at an earlier time. 



\subsection{Bubbling-induced metallicity gradients}
An increasing number of clusters with central abundance drops is being discovered. Some such clusters are the Centaurus cluster \citep{Panagoulia13a, Sanders02}, Abell~1644 \citep{Kirkpatrick09}, the Perseus cluster \citep{Sanders04}, and Abell~2199 \citep{Johnstone02}. There are also groups with central abundance dips, such as HCG~62 \citep{Rafferty13}. One explanation for the central abundance drops, at least in the Centaurus cluster, is that X-ray cavities drag out cool, dusty metal-enriched filaments from the cluster core, to outer regions of the cluster, where the filaments are destroyed, e.g. through sputtering in hot gas \citep[for more details, see][]{Panagoulia13a}. There, the metal-rich dust returns to the X-ray phase and mixes with the local ICM, resulting in an abundance increase. There have also been observations of radio lobes sweeping up molecular gas from cluster cores, which is the case in e.g. Abell~1835 \citep{McNamara14}. Work is underway to identify the sources that exhibit central abundance drops, as well as X-ray cavities. The results of that analysis will be the subject of a future paper. 

\section{Summary}
We searched for X-ray cavities in a volume- and $L_{\rm {X}}$-limited sample of 101 X-ray groups and clusters. We find cavities in 30 sources, all of which have a central cooling time of $\leq$3 Gyr. We then studied the subsample of 49 sources, which have a central cooling time of $\leq$3 Gyr, which, as mentioned, encompasses all sources with X-ray cavities. We then calculated cooling luminosities, $L_{\rm {cool}}$, and the power of the cavities, $P_{{\rm cav}}$, for each source with cavities. We study the dependence of the detection of X-ray cavities on the central cooling time and data quality, as well as the relation between cavity power and cooling luminosity, for the sources that harbour cavities. The main results of our analysis are:
\begin{itemize}
\item{We derive the AGN duty cycle for sources with a central cooling time of $\leq$3 Gyr, and estimate its value at $\sim$61 percent. This rises to $>$80 percent for sources with a central cooling time of $\leq$0.5 Gyr. Taking projection effects and the fact that almost all these sources have a detected central radio source into consideration, the duty cycle is probably higher. This agrees well with results of previous studies \citep[e.g.][]{Birzan04, Birzan12, Dunn06}.} 
  \item{We detect new cavities in three sources in our sample, namely NGC~3402, Abell~1644 and NGC~5846.}
    \item{We find that the ability to detect X-ray cavities, in the sources of our short central cooling time sample, strongly depends on the number of counts available in the core of the source, and hence on the data quality. Sources with fewer than 10000 counts within their central 20 kpc do not have clearly detected X-ray cavities, though all sources with $\geq$30000 counts do.}
    \item{For the groups and clusters that have  X-ray cavities, the bubbling process has to be, on average, continuous, to stop the gas from cooling and forming stars in group and cluster cores. In other words, we find that intermittent AGN outbursts are not powerful enough to offset cooling, and continous injection of energy into the ICM, in the form of bubbles, is needed. In some of our sources, the energy contained in the cavities is not enough to quench a cooling flow.}
      \item{The size of a cavity, loosely depends on the ambient ICM temperature through the relation $r \propto T^{0.5}$, down to temperatures of about 1.5 keV, below which there is much more scatter.}
        \item{The bubbles seen at larger distances from the core of a source may be the result of the merging of multiple smaller bubbles, produced in separate AGN outbursts.}
\end{itemize}

Work is underway to search for and study central abundance drops in the sources that display X-ray cavities. We aim to determine whether these abundance drops can be caused by AGN-induced bubbling activity, and if they are spatially correlated with e.g. dust emission in the infrared.

\section*{Acknowledgements}
EKP acknowledges the support of a STFC studentship. We thank the anonymous referee for helpful comments. EKP thanks Becky Canning for help with H$\alpha$ filament data. JHL is supported by NASA through the Einstein Fellowship Program, grant number PF2-130094.

The plots in this paper were created using {\sc veusz}.\footnote{http://home.gna.org/veusz/}
This research has made use of the NASA/IPAC Extragalactic Database (NED)\footnote{http://ned.ipac.caltech.edu/} which is operated by the Jet Propulsion Laboratory, California Institute of Technology, under contract with the National Aeronautics and Space Administration. 

\bibliographystyle{mn2e}
\bibliography{references}

\appendix
\section{{\it\textbf{Chandra}} and {\it\textbf{XMM-Newton}} images}
In this section, we present the corresponding 0.5--7.0 keV exposure-corrected, background-subtracted and unsharp-masked {\it Chandra} images, and the 0.5--7.0 keV background flare-cleaned and unsharp-masked {\it XMM-Newton} images for the sources in our sample. Two different sources are shown in each row, with the two left-hand and the two right-hand columns of each row containing images from two different sources. Each set of two columns shows the exposure-corrected and background-subtracted {\it Chandra} image, or the cleaned {\it XMM-Newton} image in the left-hand column, while the right-hand column shows the unsharp-masked image. Each background-subtracted, exposure-corrected or cleaned image has been smoothed with a 2-pixel Gaussian. The bar in each {\it Chandra} image is 0.5 arcmin long, while the bars in the {\it XMM-Newton} images are 3 arcmin long. The arrows indicate ``certain'' or ``possible'' cavities. The majority of the unsharp-masked images are the result of the subtraction of an 8-pixel Gaussian smoothed image from a 2-pixel Gaussian smoothed image.

\begin{figure*}
  \includegraphics[trim=1cm 11.5cm 1cm 10cm, clip, height=4.4cm, width=8.8cm]{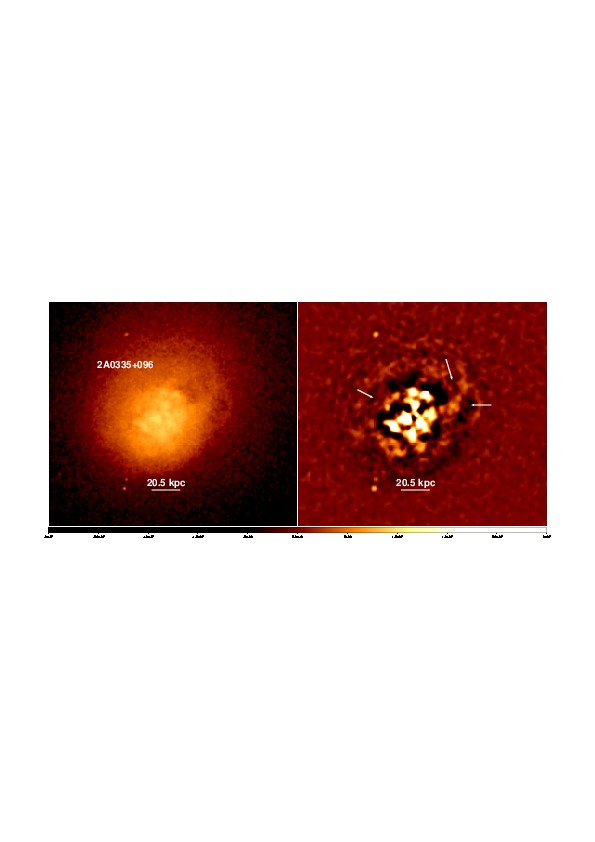}
  \includegraphics[trim=1cm 11.5cm 1cm 10cm, clip, height=4.4cm, width=8.8cm]{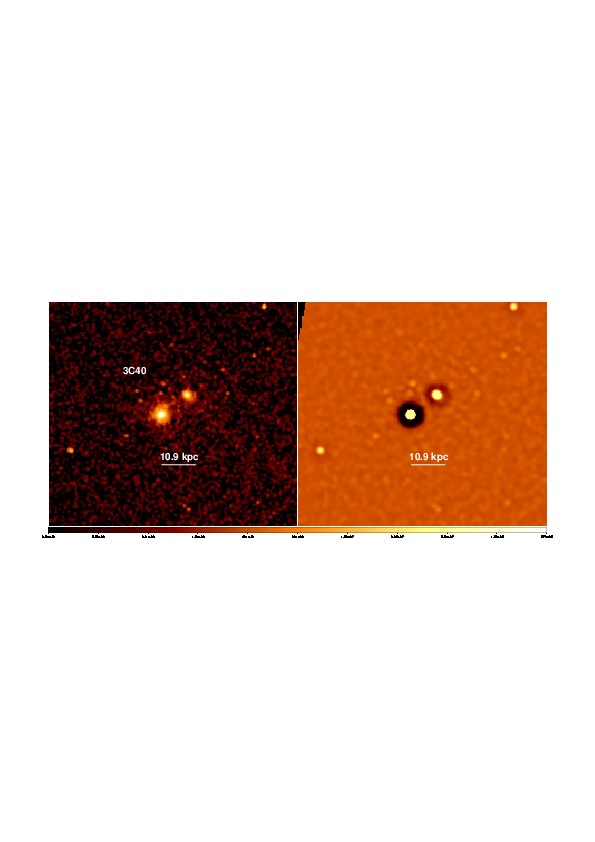}
  \includegraphics[trim=1cm 11.5cm 1cm 10cm, clip, height=4.4cm, width=8.8cm]{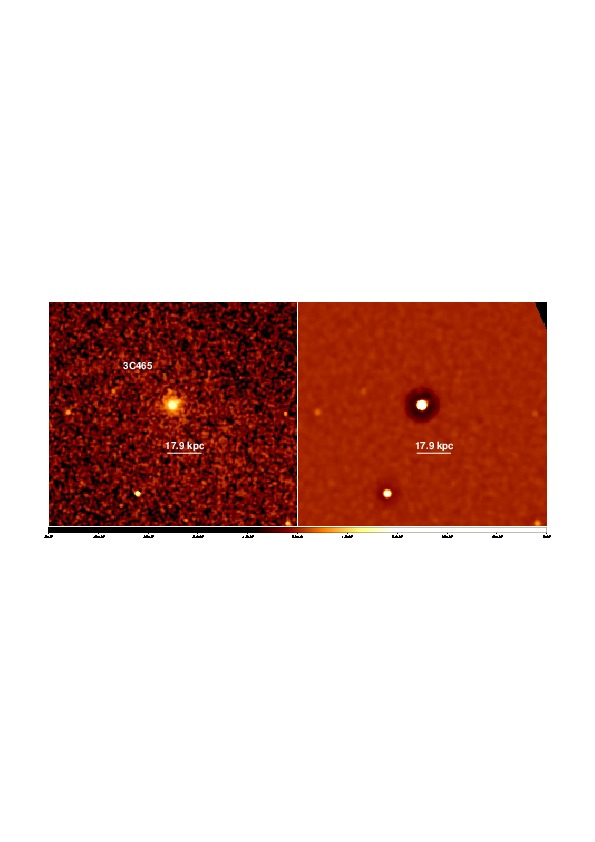}
  \includegraphics[trim=1cm 11.5cm 1cm 10cm, clip, height=4.4cm, width=8.8cm]{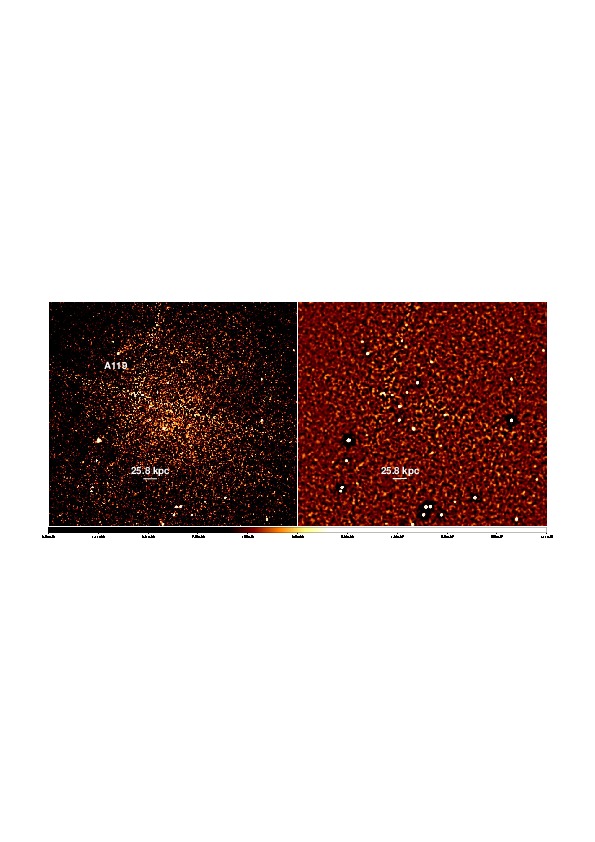}
  \includegraphics[trim=1cm 11.5cm 1cm 10cm, clip, height=4.4cm, width=8.8cm]{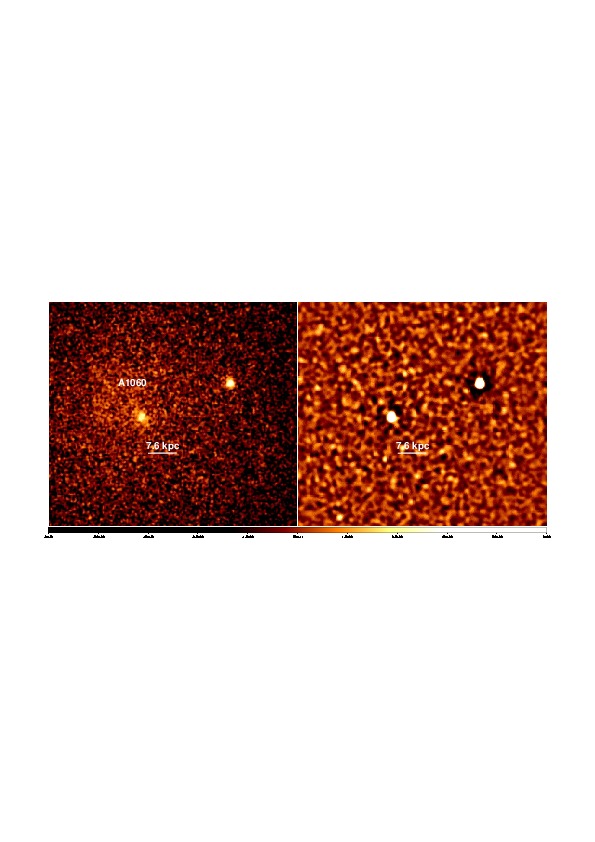}
  \includegraphics[trim=1cm 11.5cm 1cm 10cm, clip, height=4.4cm, width=8.8cm]{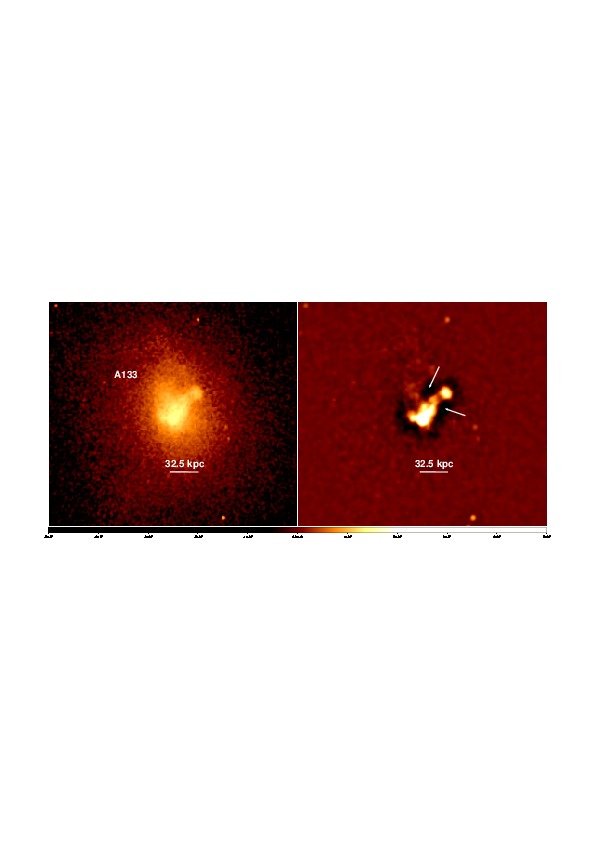}
  \includegraphics[trim=1cm 11.5cm 1cm 10cm, clip, height=4.4cm, width=8.8cm]{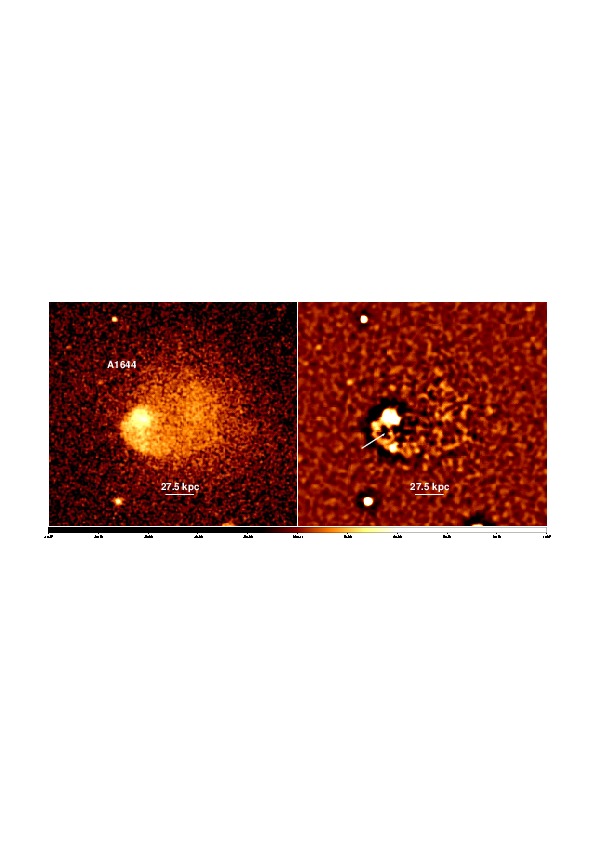}
  \includegraphics[trim=1cm 11.5cm 1cm 10cm, clip, height=4.4cm, width=8.8cm]{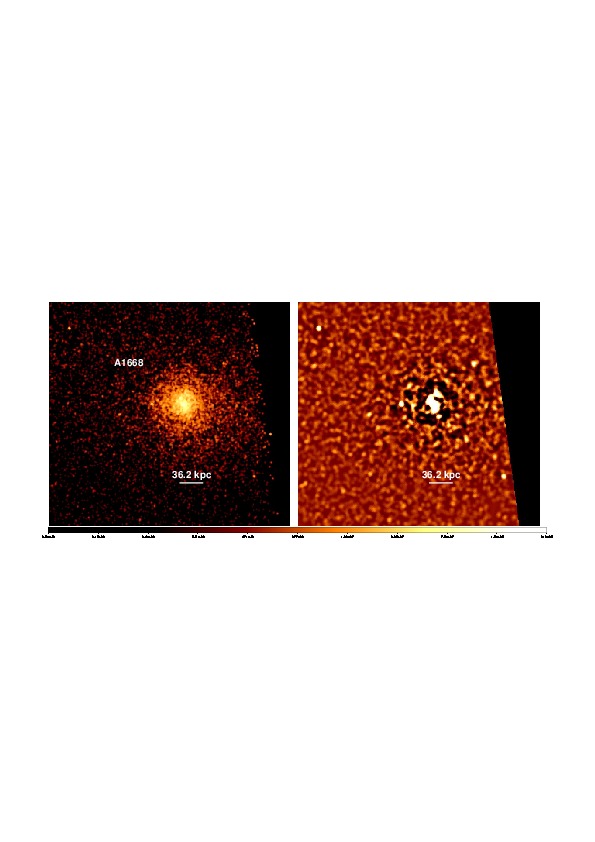}
  \includegraphics[trim=1cm 11.5cm 1cm 10cm, clip, height=4.4cm, width=8.8cm]{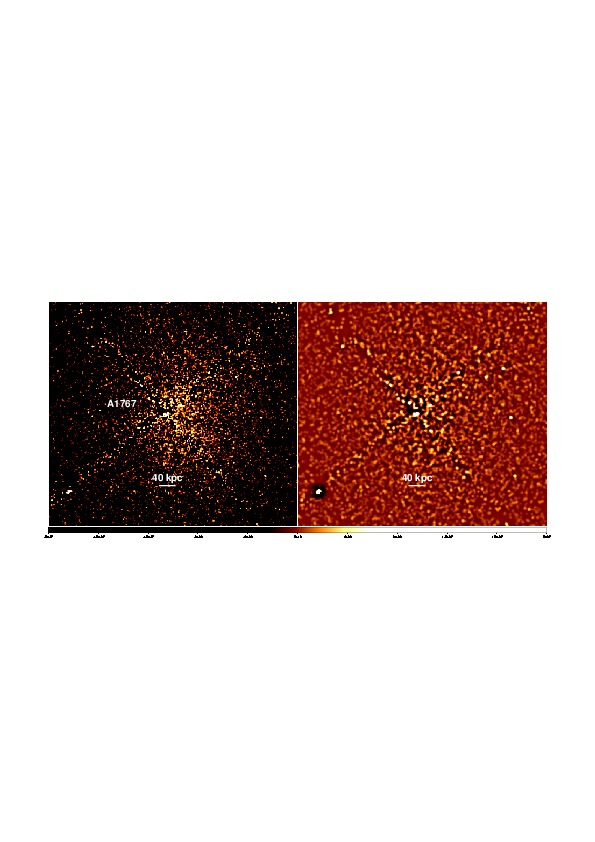}
  \includegraphics[trim=1cm 11.5cm 1cm 10cm, clip, height=4.4cm, width=8.8cm]{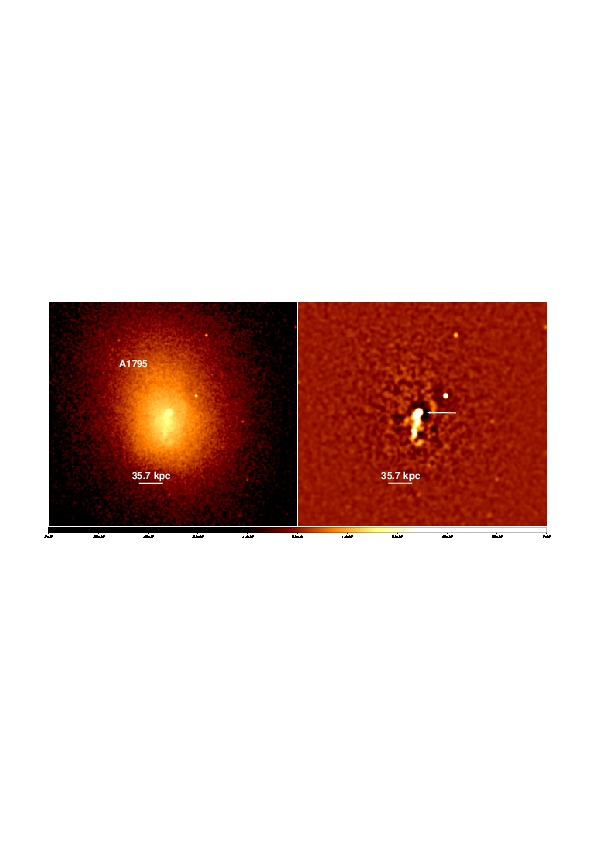}
  \caption[]{The bar is 0.5 arcmin long in all the {\it Chandra} images. All images have been created in the 0.5--7.0 keV band, and the background-subtracted, exposure-corrected images have been smoothed using a 2-pixel Gaussian. The left-hand panel for each set of two images shows the background-subtracted, exposure-corrected images, with the unsharp-masked image in the right-hand panel. The arrows indicate ``possible'' or ``certain'' cavities.}
\end{figure*}

\begin{figure*}
  \contcaption{}
  \includegraphics[trim=1cm 11.5cm 1cm 10cm, clip, height=4.4cm, width=8.8cm]{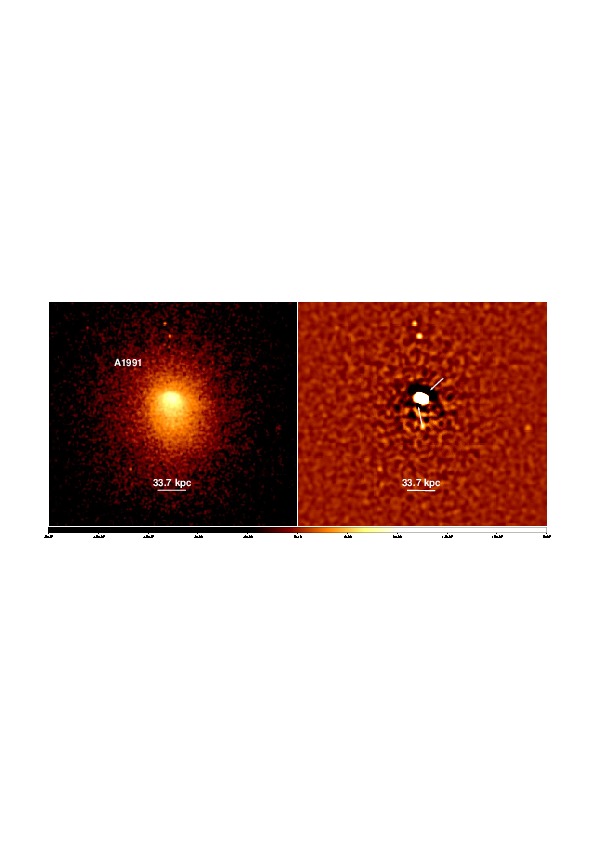}
  \includegraphics[trim=1cm 11.5cm 1cm 10cm, clip, height=4.4cm, width=8.8cm]{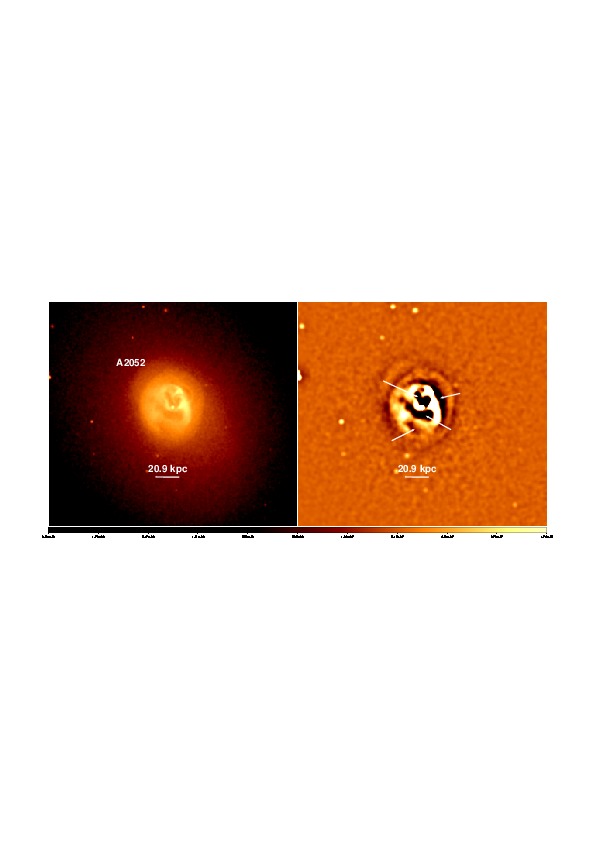}
  \includegraphics[trim=1cm 11.5cm 1cm 10cm, clip, height=4.4cm, width=8.8cm]{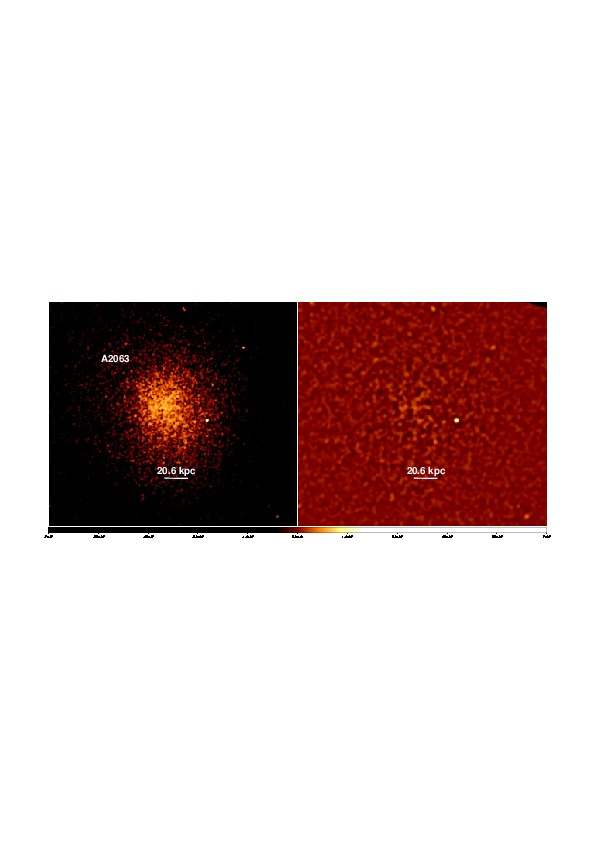}
  \includegraphics[trim=1cm 11.5cm 1cm 10cm, clip, height=4.4cm, width=8.8cm]{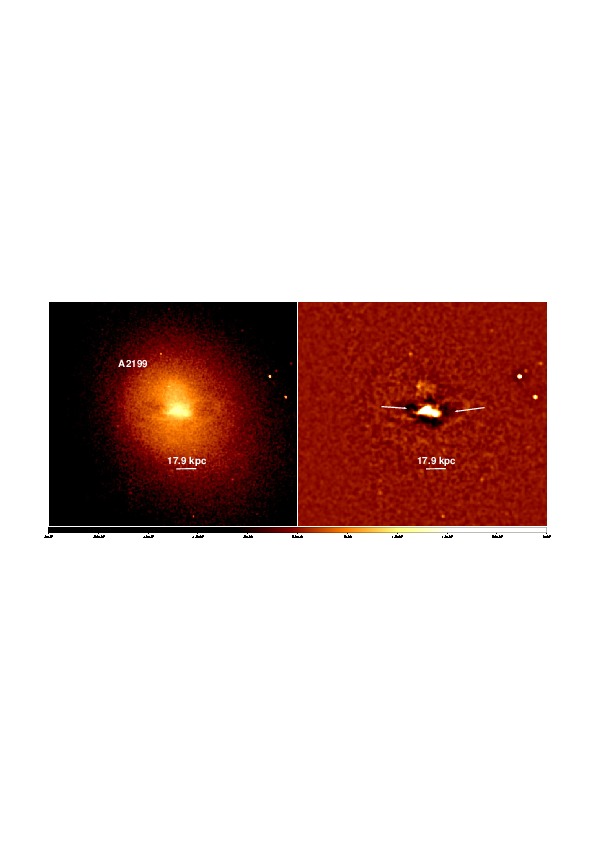}
  \includegraphics[trim=1cm 11.5cm 1cm 10cm, clip, height=4.4cm, width=8.8cm]{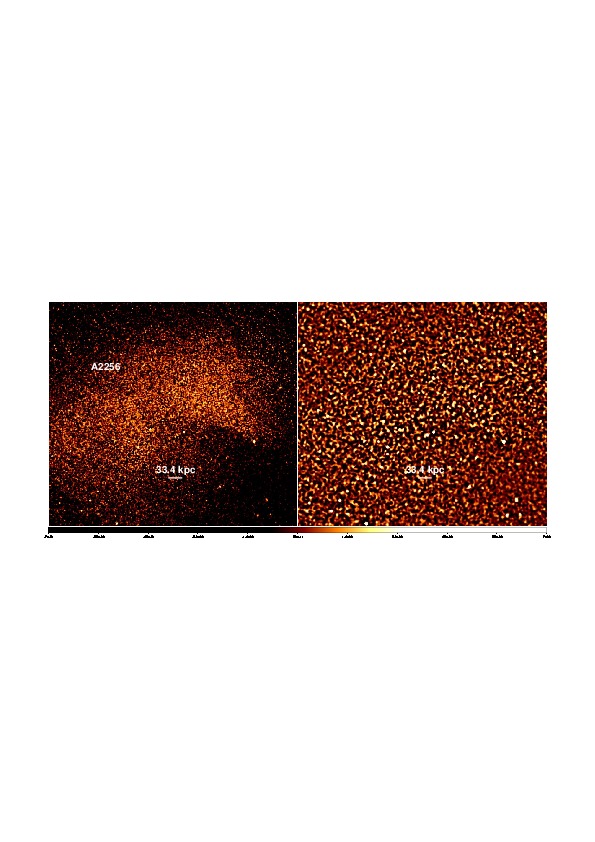}
  \includegraphics[trim=1cm 11.5cm 1cm 10cm, clip, height=4.4cm, width=8.8cm]{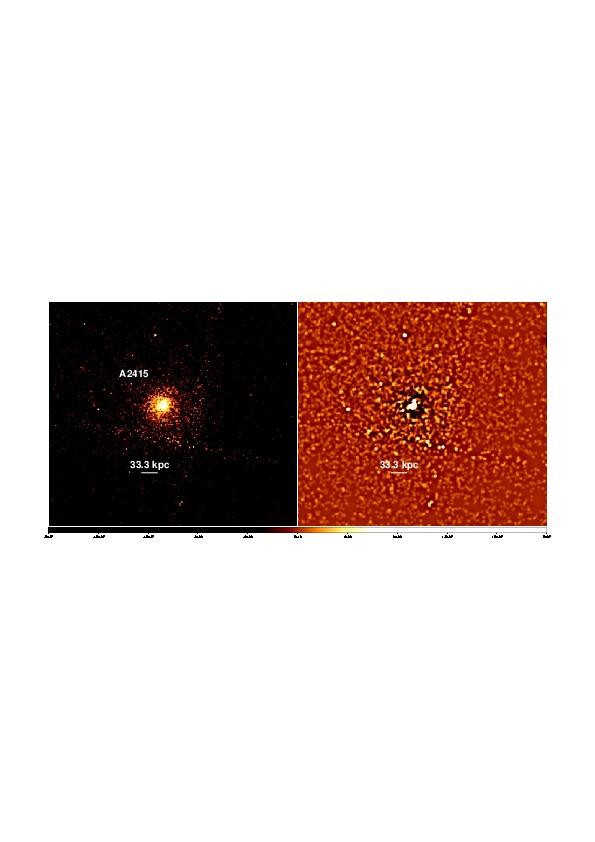}
  \includegraphics[trim=1cm 11.5cm 1cm 10cm, clip, height=4.4cm, width=8.8cm]{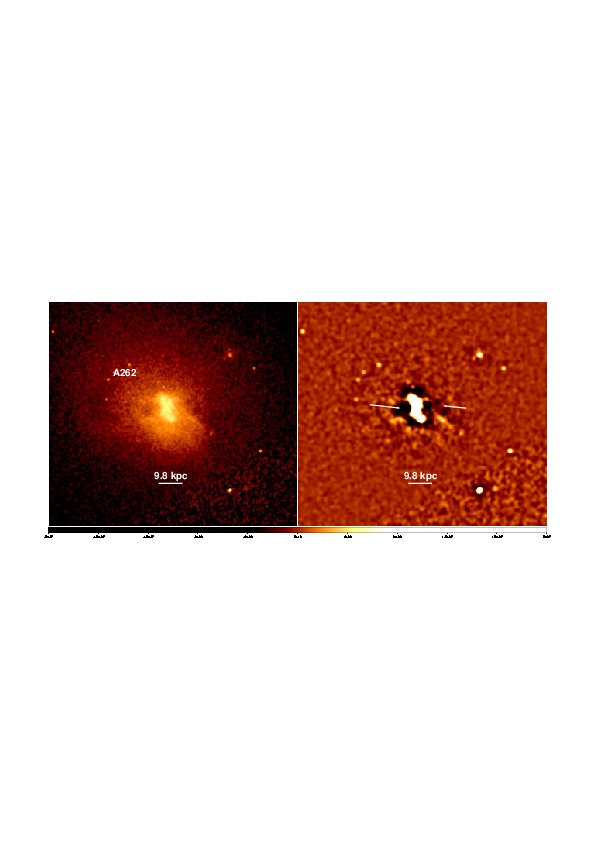}
  \includegraphics[trim=1cm 11.5cm 1cm 10cm, clip, height=4.4cm, width=8.8cm]{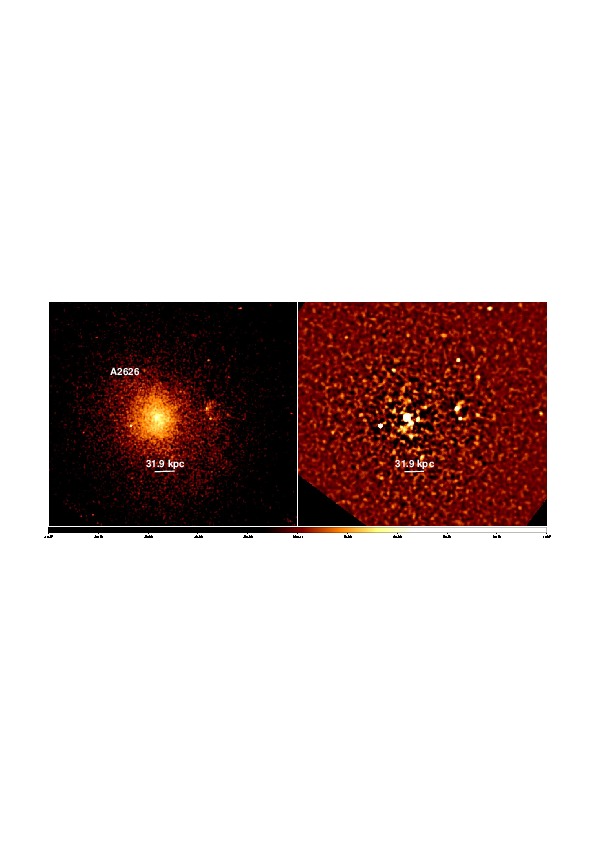}
  \includegraphics[trim=1cm 11.5cm 1cm 10cm, clip, height=4.4cm, width=8.8cm]{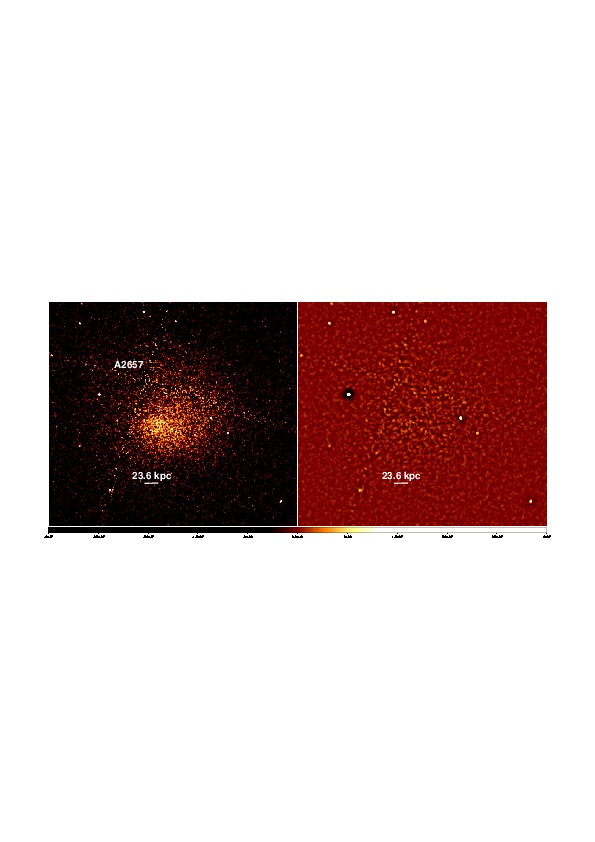}
  \includegraphics[trim=1cm 11.5cm 1cm 10cm, clip, height=4.4cm, width=8.8cm]{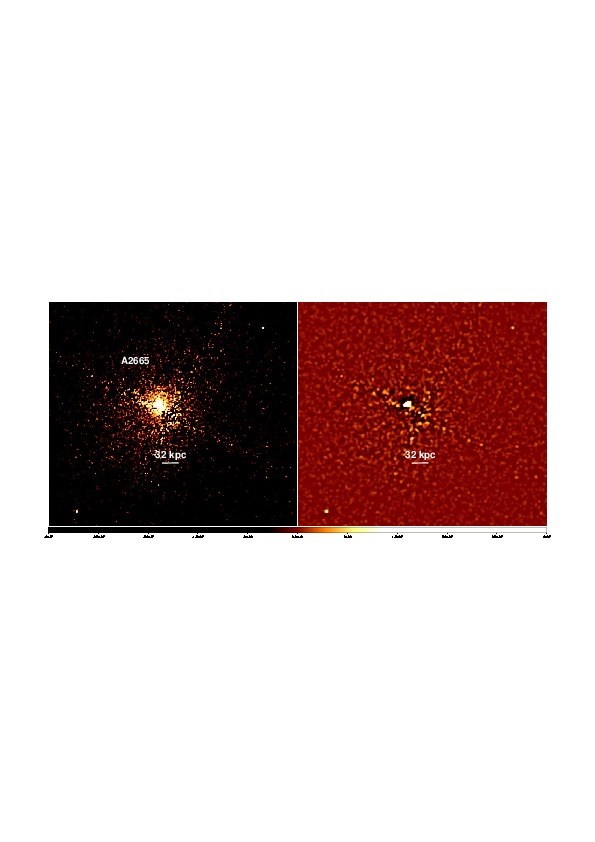}
\end{figure*}  

\begin{figure*}
\contcaption{}
  \includegraphics[trim=1cm 11.5cm 1cm 10cm, clip, height=4.4cm, width=8.8cm]{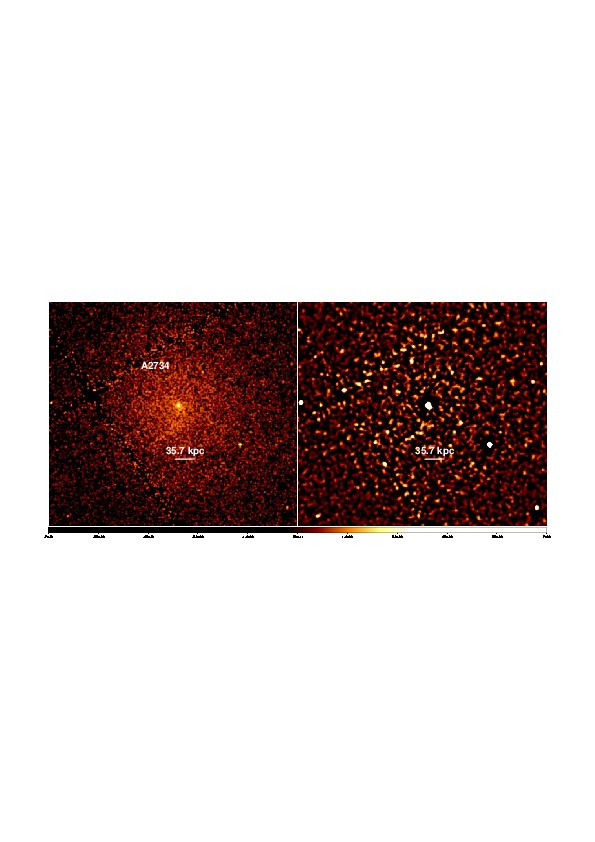}
  \includegraphics[trim=1cm 11.5cm 1cm 10cm, clip, height=4.4cm, width=8.8cm]{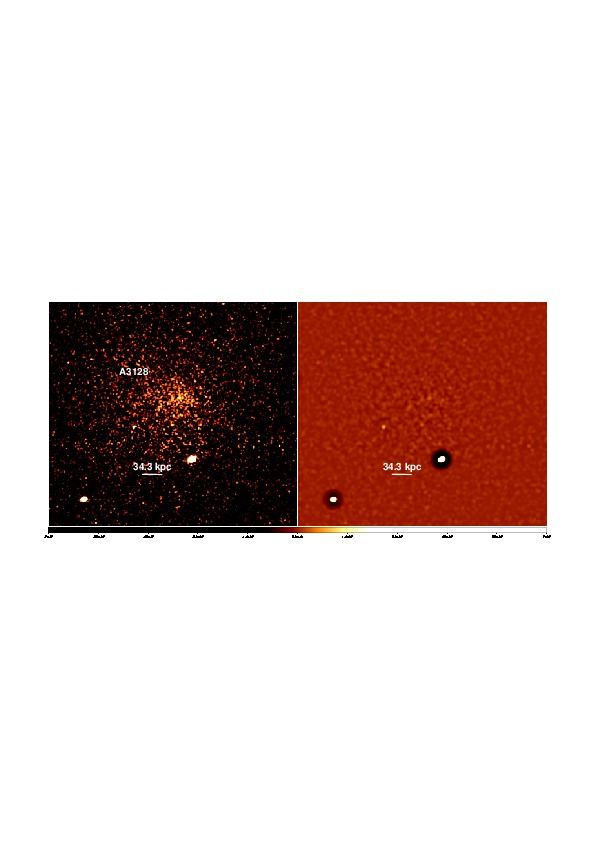}
  \includegraphics[trim=1cm 11.5cm 1cm 10cm, clip, height=4.4cm, width=8.8cm]{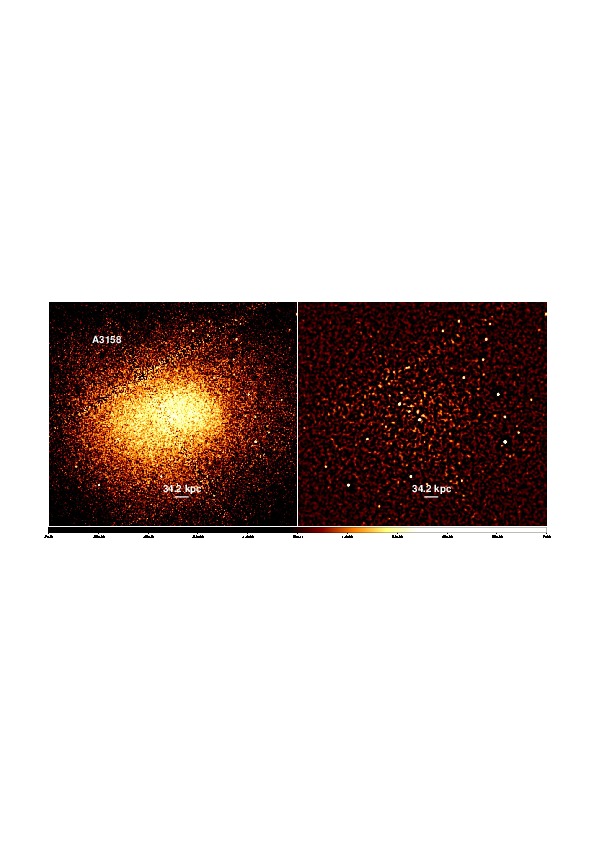}
  \includegraphics[trim=1cm 11.5cm 1cm 10cm, clip, height=4.4cm, width=8.8cm]{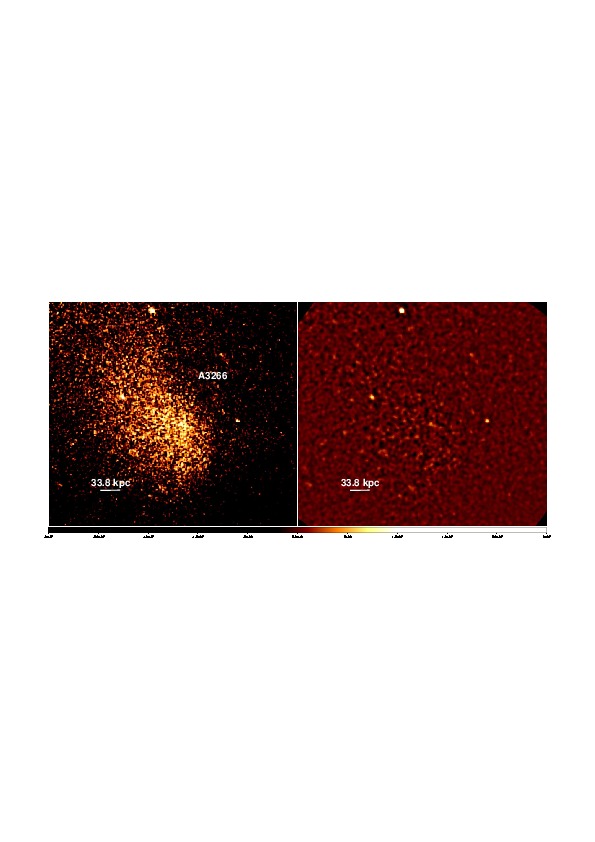}
  \includegraphics[trim=1cm 11.5cm 1cm 10cm, clip, height=4.4cm, width=8.8cm]{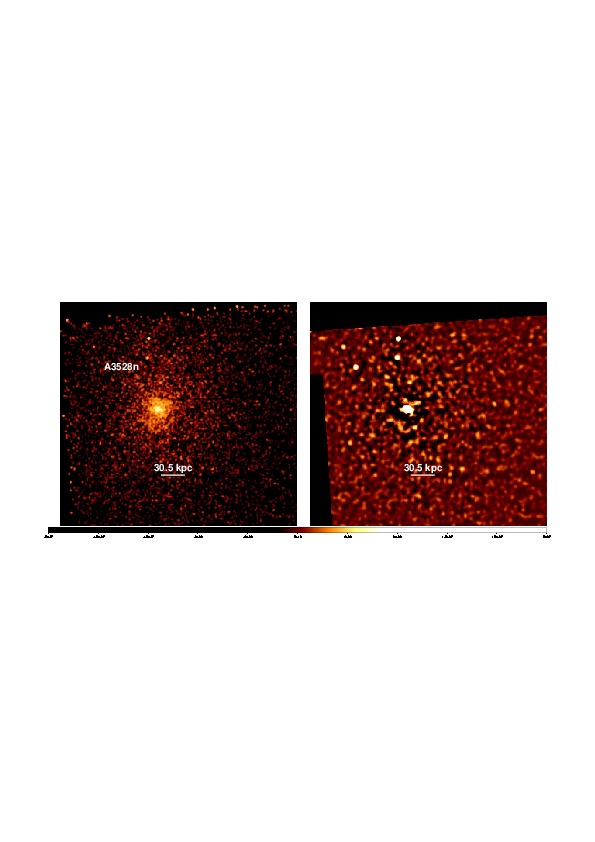}
  \includegraphics[trim=1cm 11.5cm 1cm 10cm, clip, height=4.4cm, width=8.8cm]{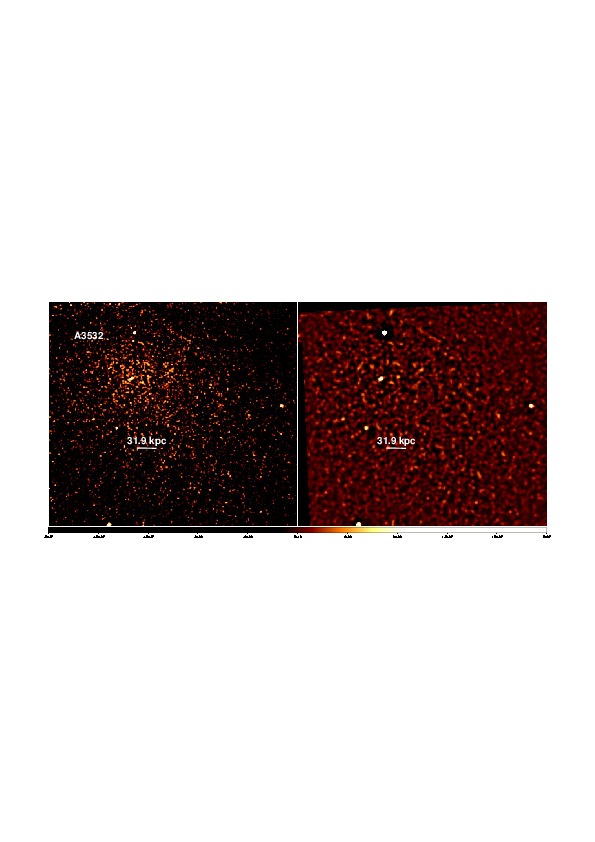}
  \includegraphics[trim=1cm 11.5cm 1cm 10cm, clip, height=4.4cm, width=8.8cm]{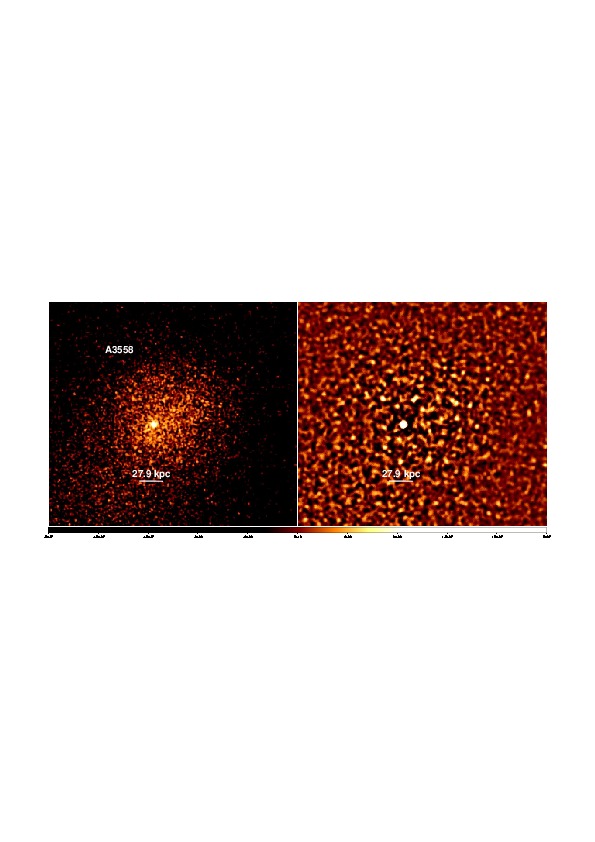}
  \includegraphics[trim=1cm 11.5cm 1cm 10cm, clip, height=4.4cm, width=8.8cm]{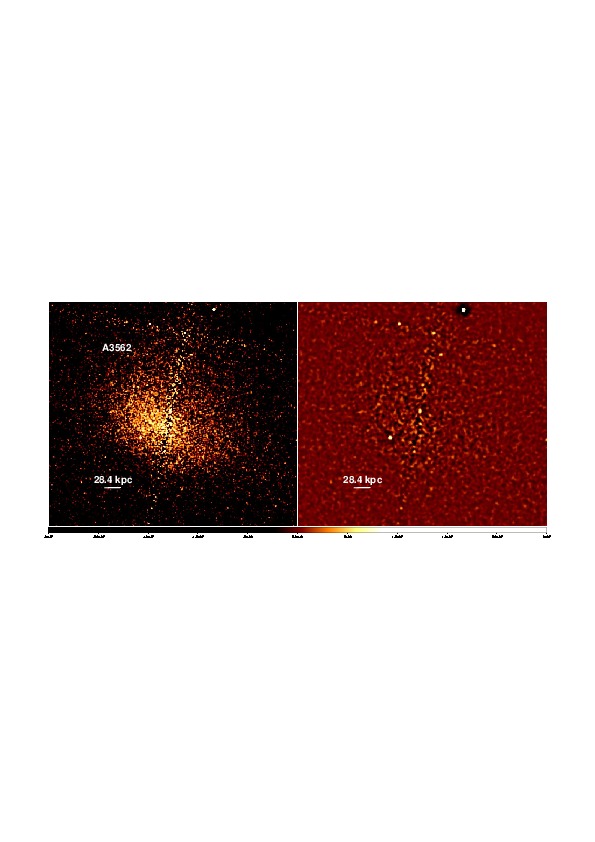}
  \includegraphics[trim=1cm 11.5cm 1cm 10cm, clip, height=4.4cm, width=8.8cm]{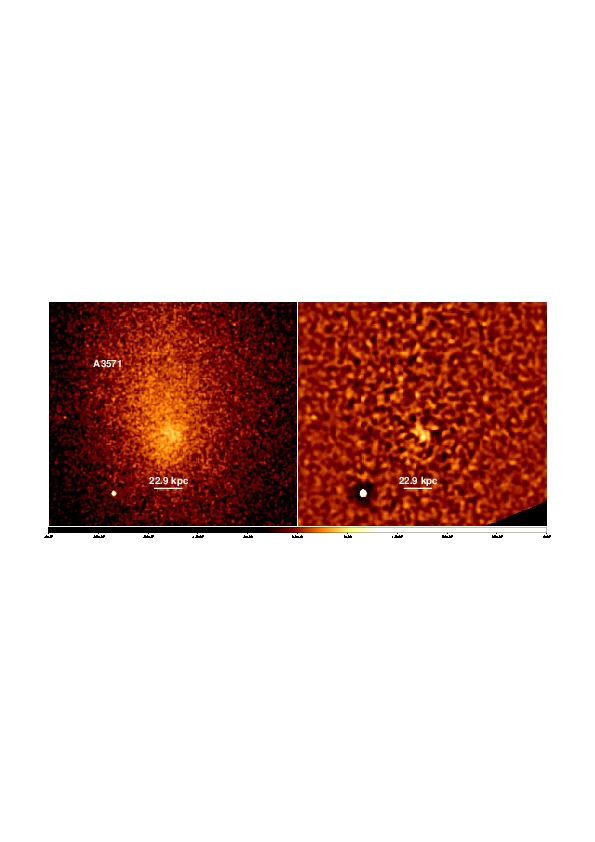}
  \includegraphics[trim=1cm 11.5cm 1cm 10cm, clip, height=4.4cm, width=8.8cm]{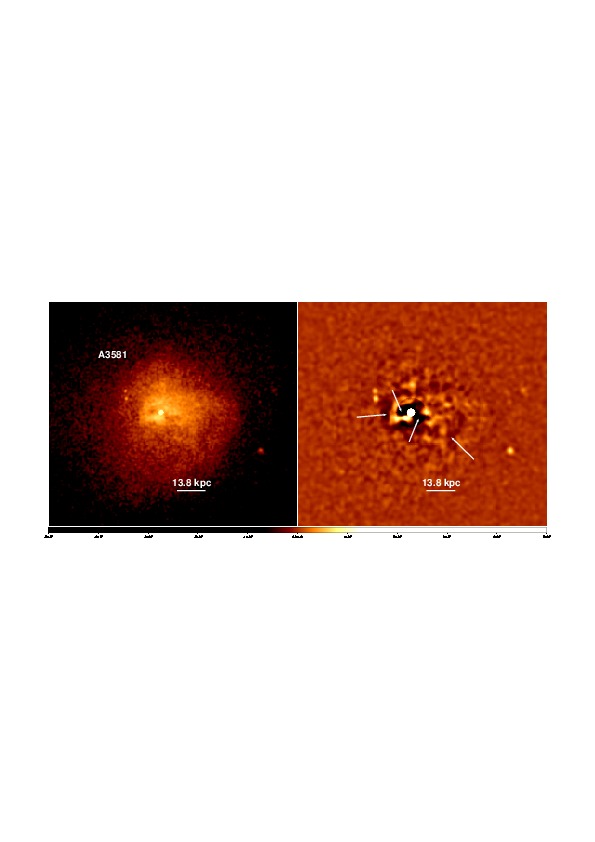}
\end{figure*}

\begin{figure*}
\contcaption{}
  \includegraphics[trim=1cm 11.5cm 1cm 10cm, clip, height=4.4cm, width=8.8cm]{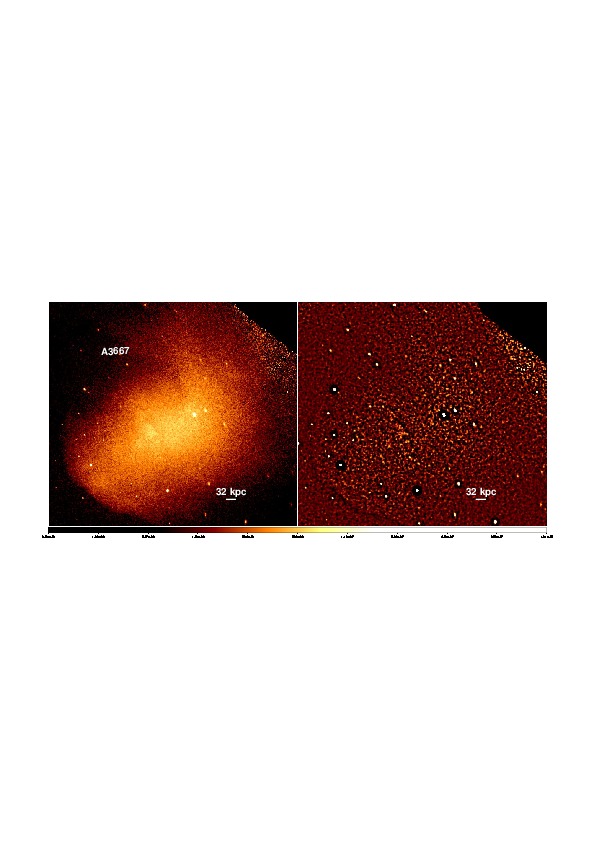}
  \includegraphics[trim=1cm 11.5cm 1cm 10cm, clip, height=4.4cm, width=8.8cm]{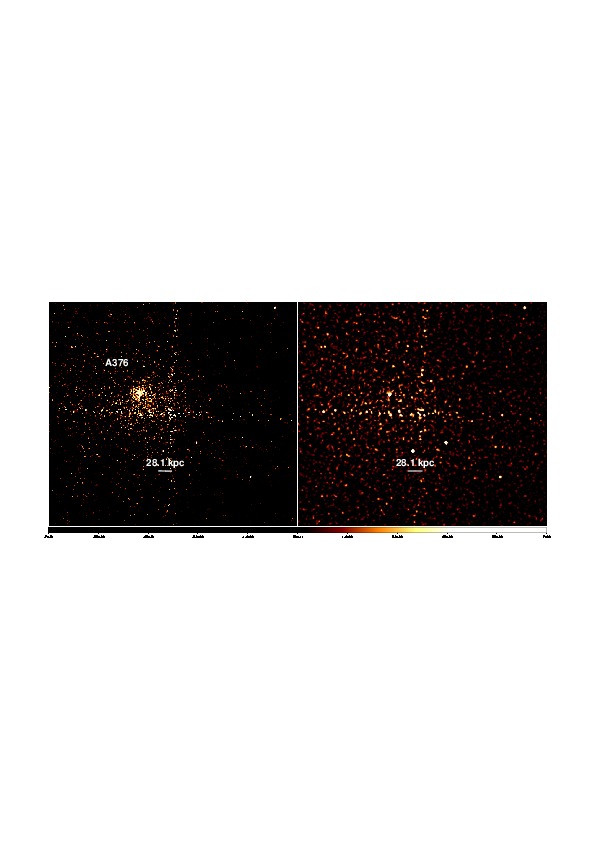}
  \includegraphics[trim=1cm 11.5cm 1cm 10cm, clip, height=4.4cm, width=8.8cm]{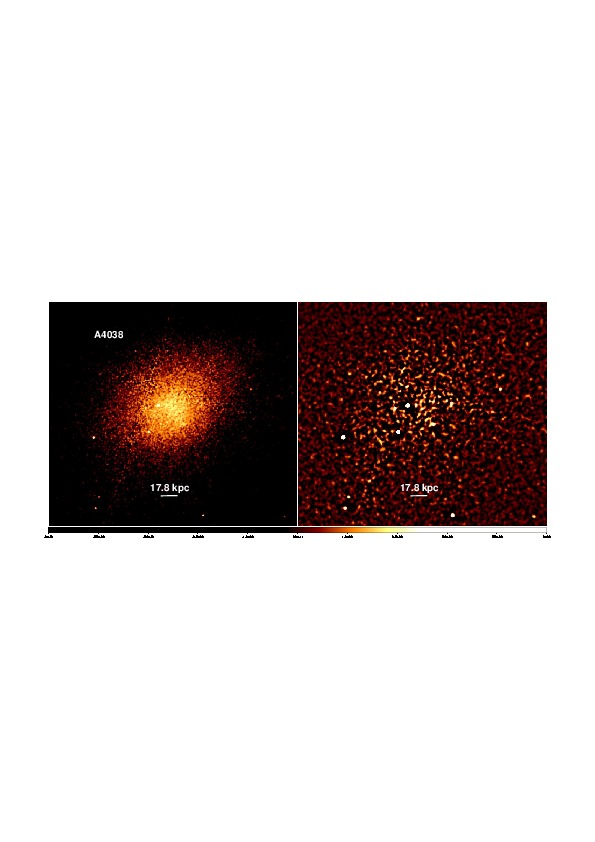}
  \includegraphics[trim=1cm 11.5cm 1cm 10cm, clip, height=4.4cm, width=8.8cm]{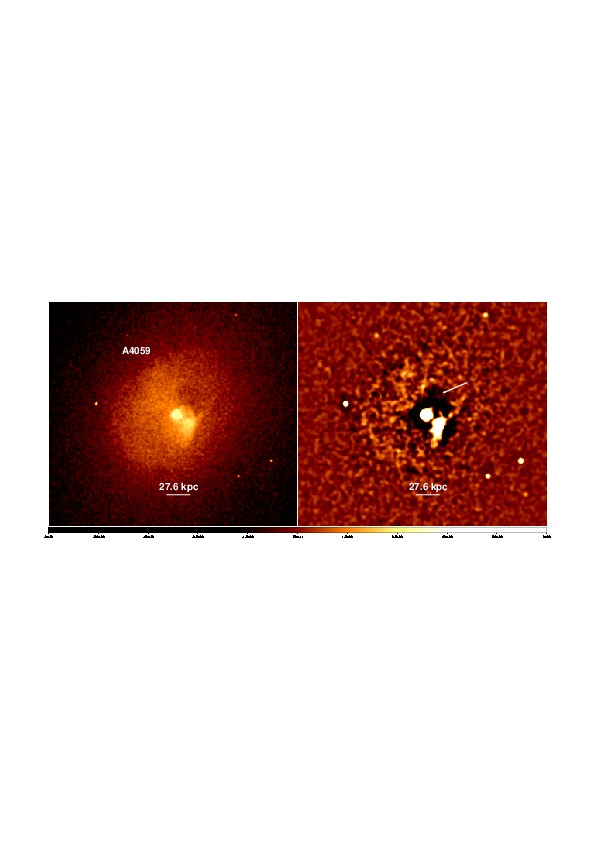}
  \includegraphics[trim=1cm 11.5cm 1cm 10cm, clip, height=4.4cm, width=8.8cm]{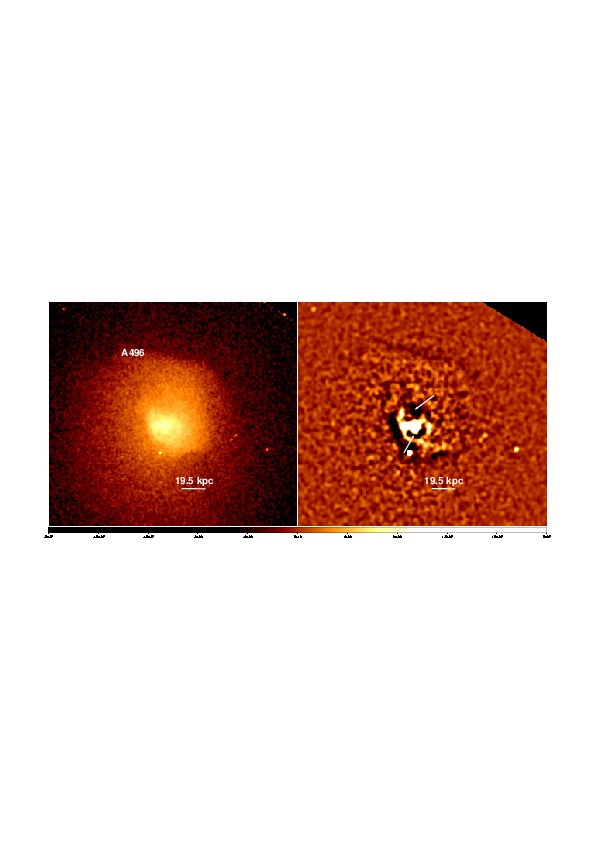}
  \includegraphics[trim=1cm 11.5cm 1cm 10cm, clip, height=4.4cm, width=8.8cm]{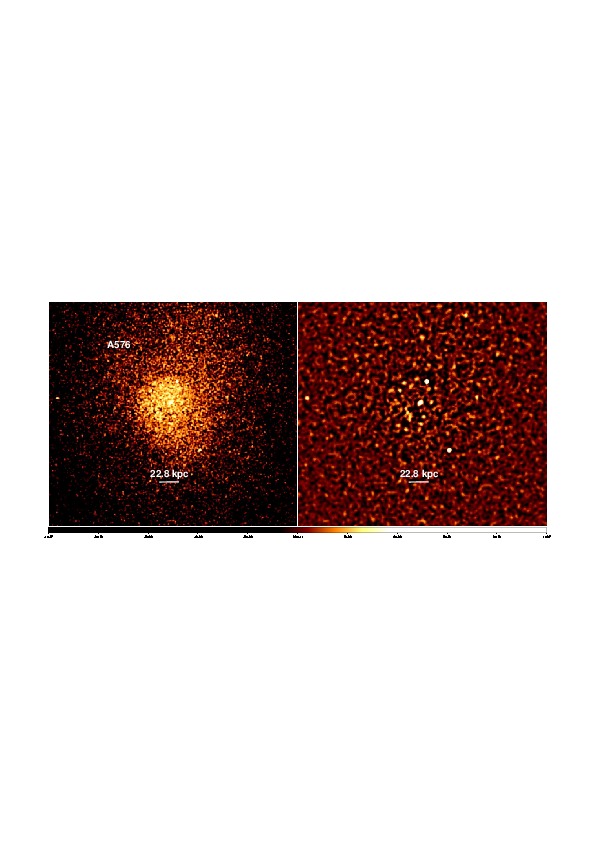}
  \includegraphics[trim=1cm 11.5cm 1cm 10cm, clip, height=4.4cm, width=8.8cm]{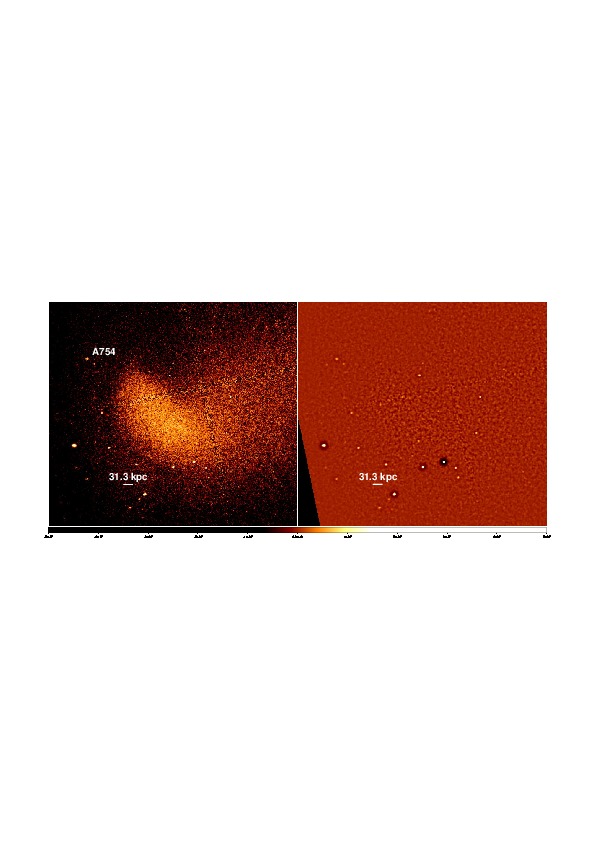}
  \includegraphics[trim=1cm 11.5cm 1cm 10cm, clip, height=4.4cm, width=8.8cm]{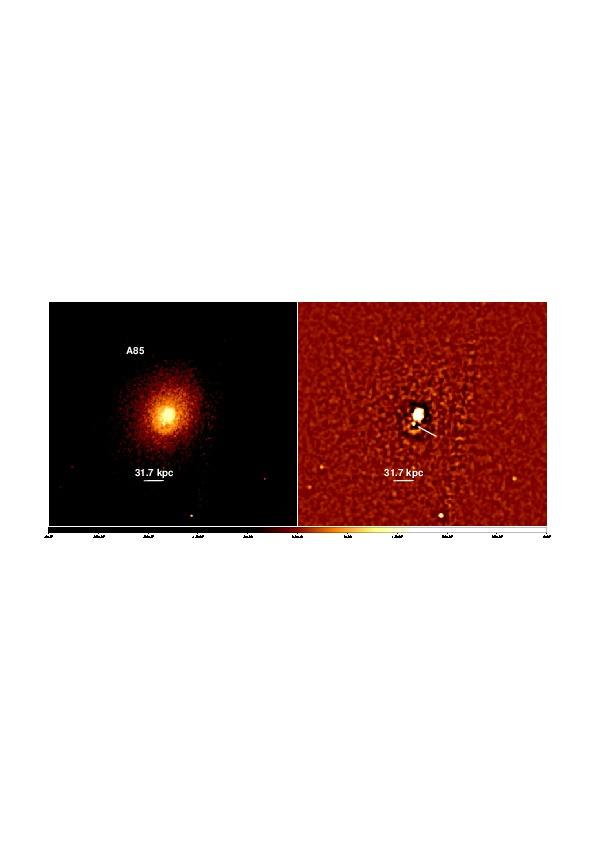}
  \includegraphics[trim=1cm 11.5cm 1cm 10cm, clip, height=4.4cm, width=8.8cm]{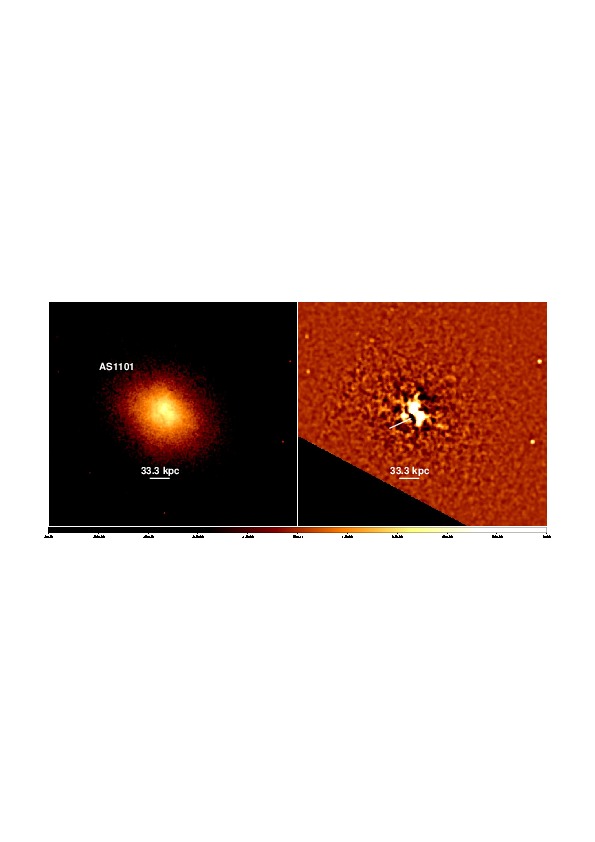}
  \includegraphics[trim=1cm 11.5cm 1cm 10cm, clip, height=4.4cm, width=8.8cm]{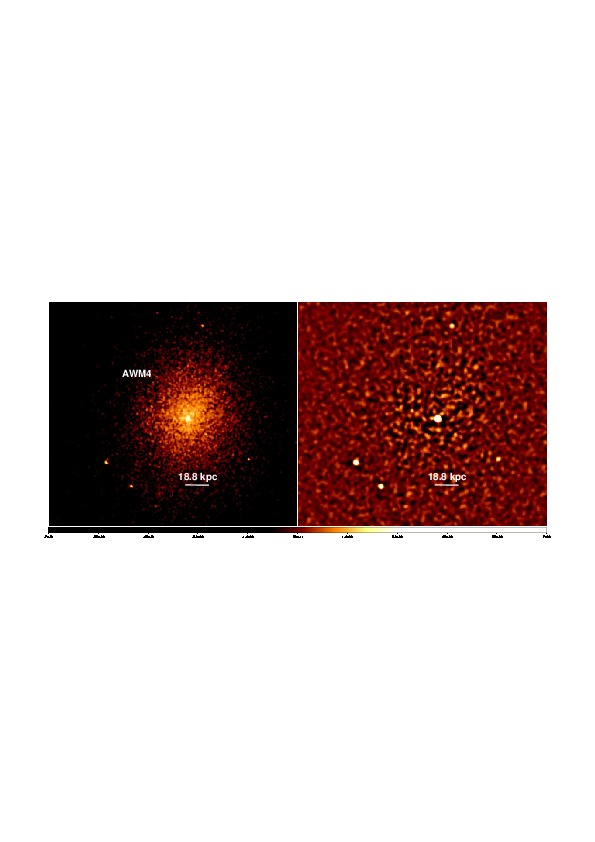}
\end{figure*}

\begin{figure*}
\contcaption{}
  \includegraphics[trim=1cm 11.5cm 1cm 10cm, clip, height=4.4cm, width=8.8cm]{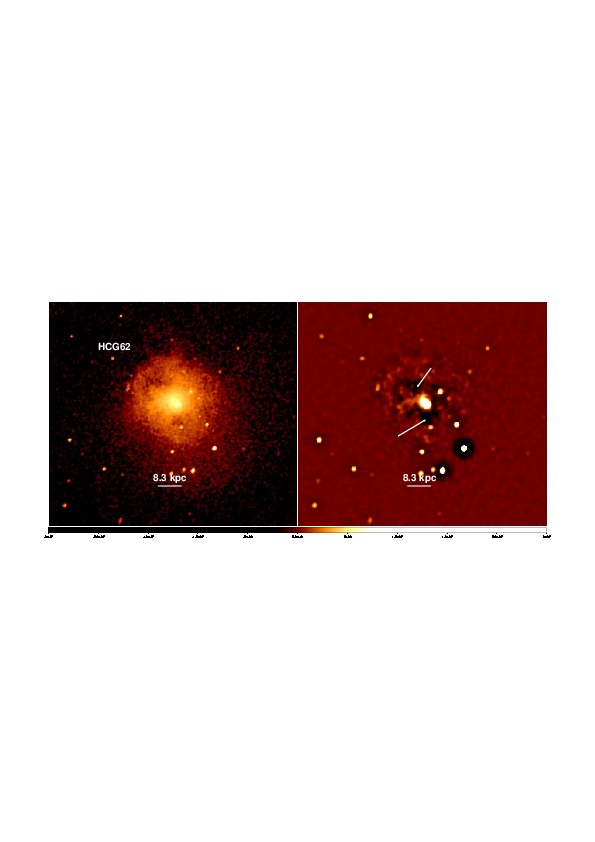}
  \includegraphics[trim=1cm 11.5cm 1cm 10cm, clip, height=4.4cm, width=8.8cm]{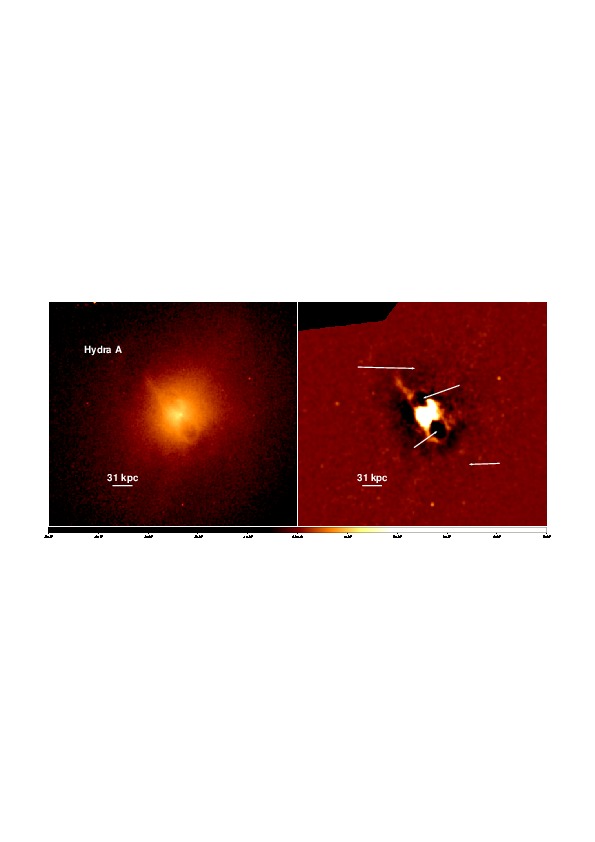}
  \includegraphics[trim=1cm 11.5cm 1cm 10cm, clip, height=4.4cm, width=8.8cm]{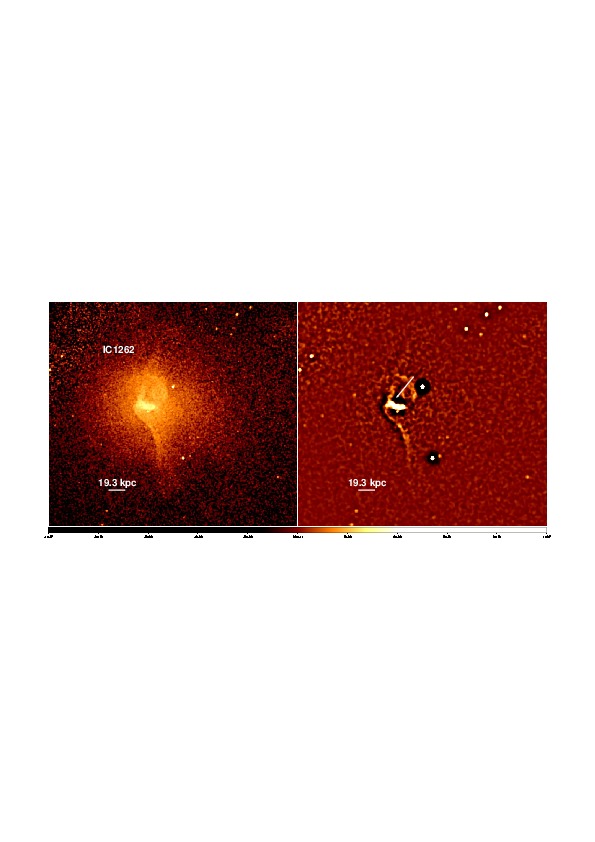}
  \includegraphics[trim=1cm 11.5cm 1cm 10cm, clip, height=4.4cm, width=8.8cm]{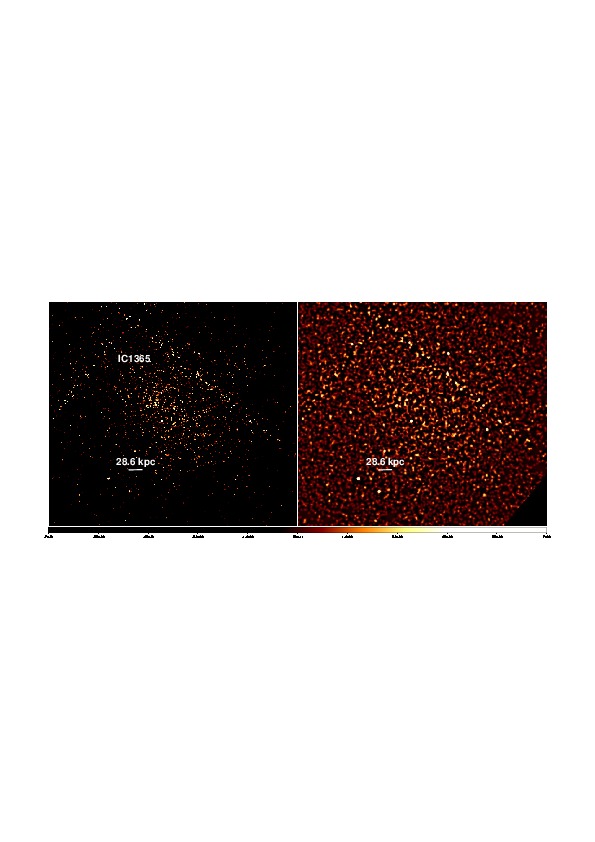}
  \includegraphics[trim=1cm 11.5cm 1cm 10cm, clip, height=4.4cm, width=8.8cm]{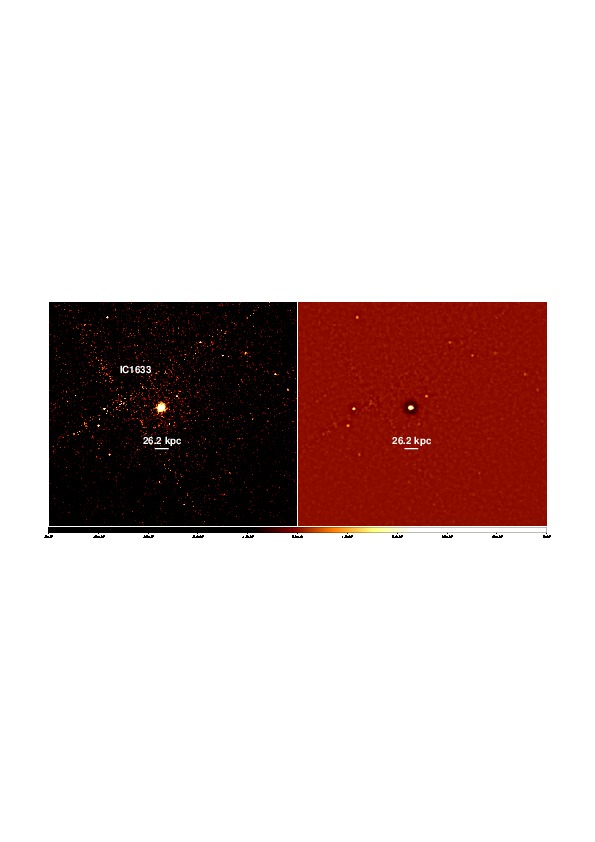}
  \includegraphics[trim=1cm 11.5cm 1cm 10cm, clip, height=4.4cm, width=8.8cm]{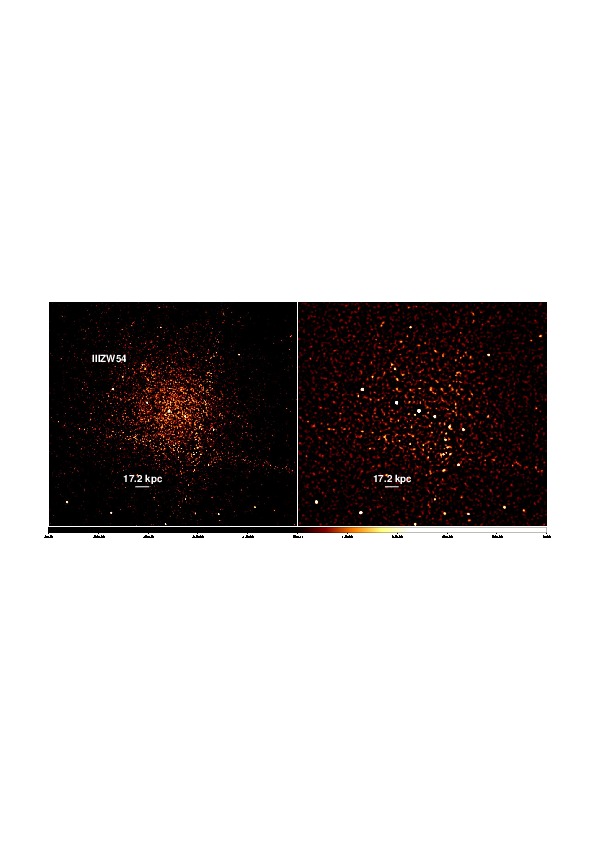}
  \includegraphics[trim=1cm 11.5cm 1cm 10cm, clip, height=4.4cm, width=8.8cm]{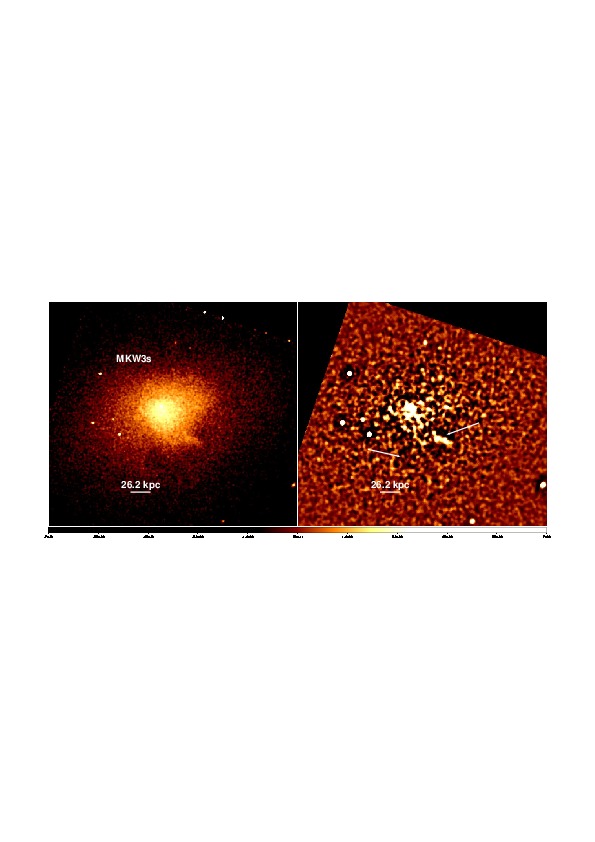}
  \includegraphics[trim=1cm 11.5cm 1cm 10cm, clip, height=4.4cm, width=8.8cm]{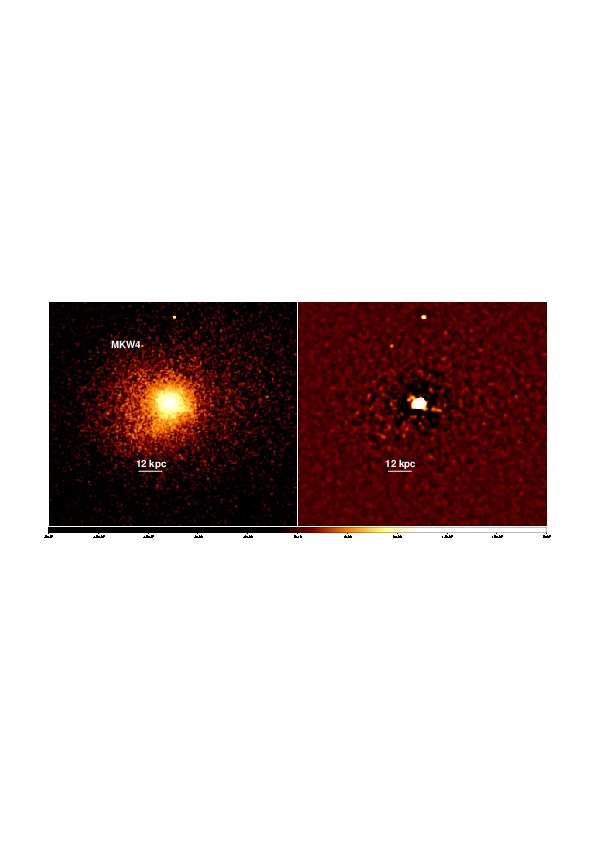}
  \includegraphics[trim=1cm 11.5cm 1cm 10cm, clip, height=4.4cm, width=8.8cm]{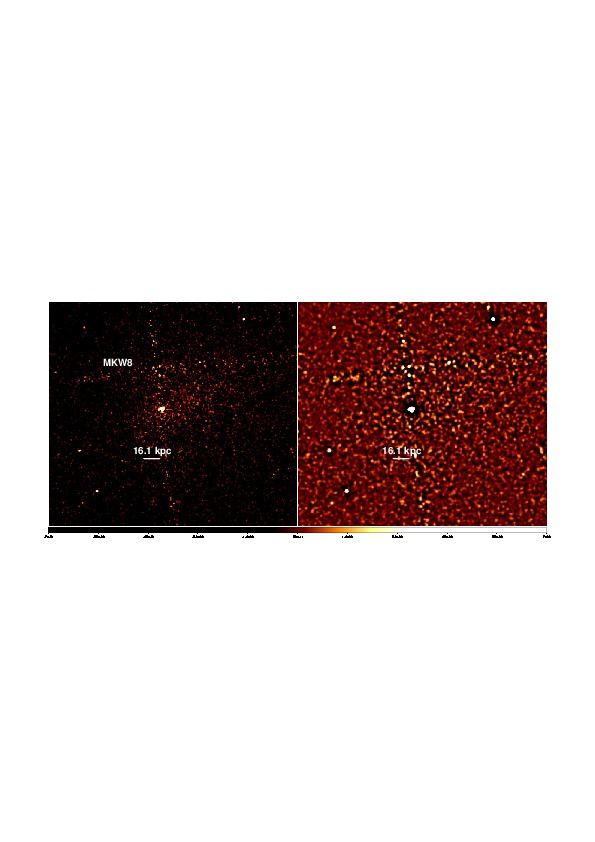}
  \includegraphics[trim=1cm 11.5cm 1cm 10cm, clip, height=4.4cm, width=8.8cm]{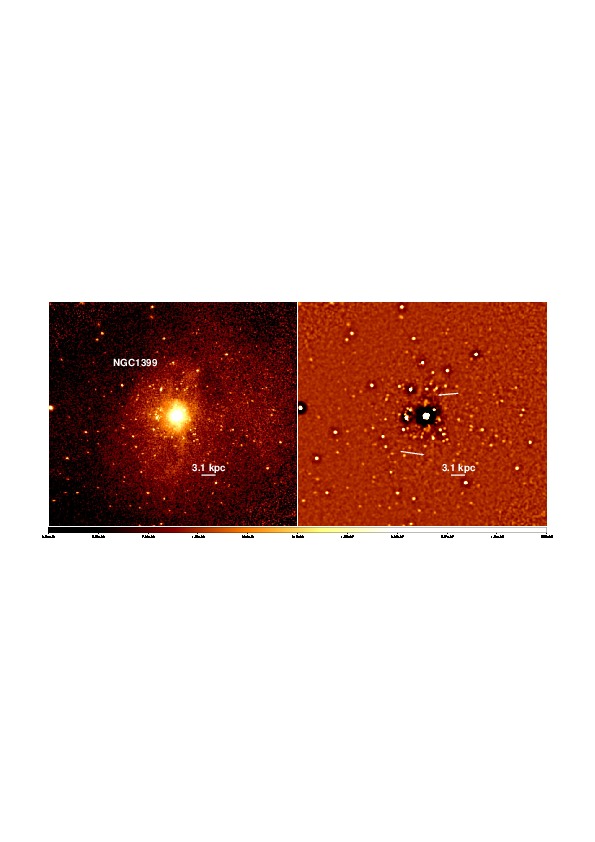}
\end{figure*}

\begin{figure*}
\contcaption{}
  \includegraphics[trim=1cm 11.5cm 1cm 10cm, clip, height=4.4cm, width=8.8cm]{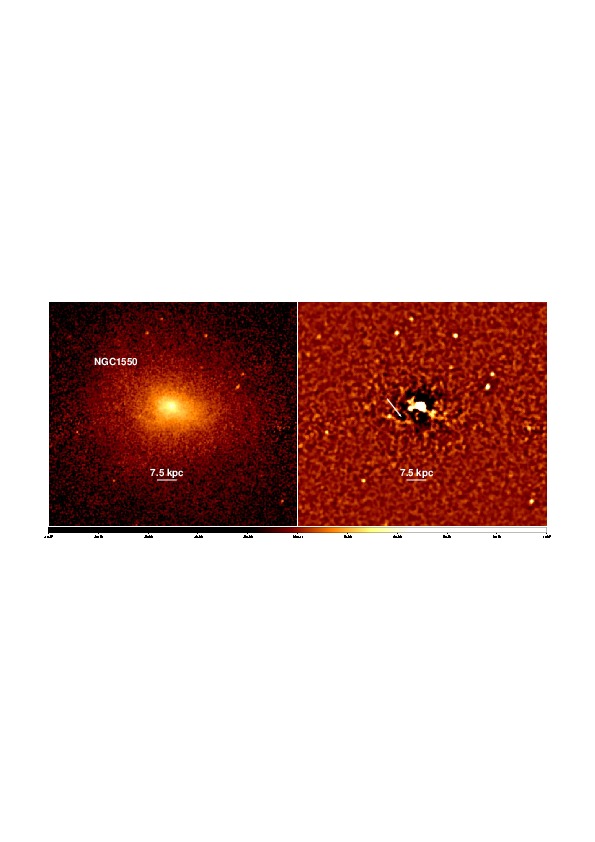}
  \includegraphics[trim=1cm 11.5cm 1cm 10cm, clip, height=4.4cm, width=8.8cm]{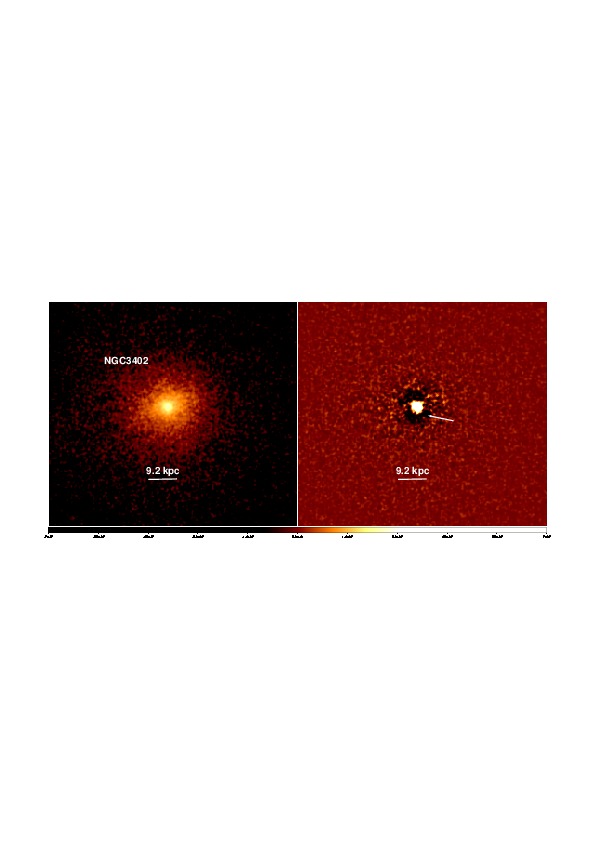}
  \includegraphics[trim=1cm 11.5cm 1cm 10cm, clip, height=4.4cm, width=8.8cm]{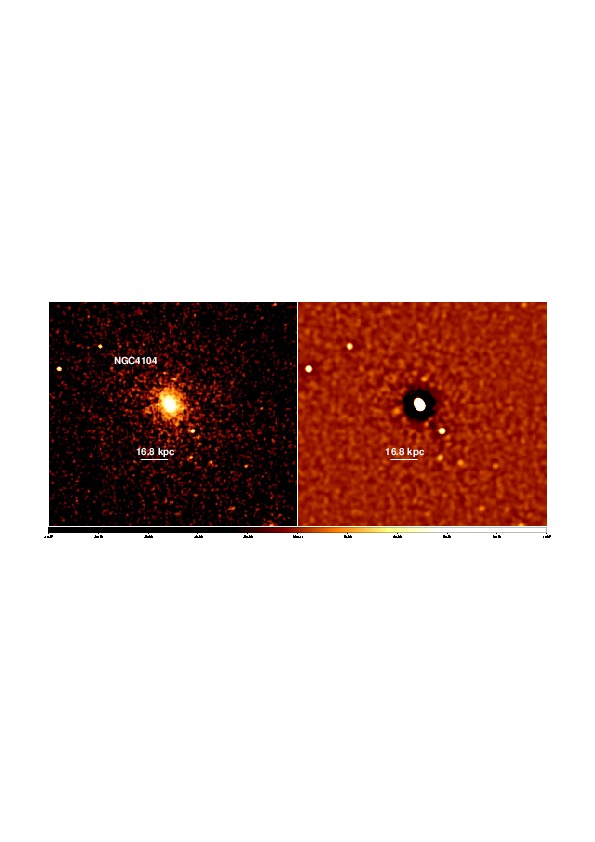}
  \includegraphics[trim=1cm 11.5cm 1cm 10cm, clip, height=4.4cm, width=8.8cm]{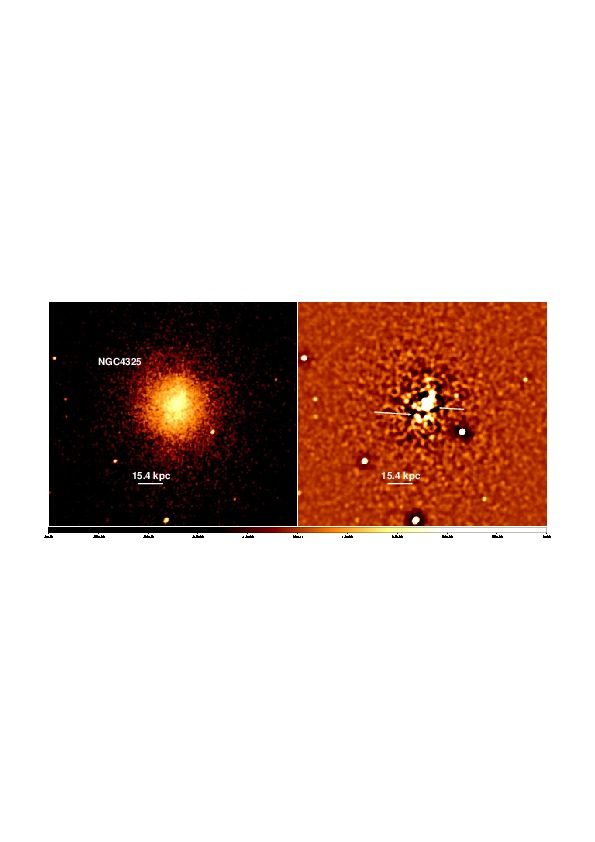}
  \includegraphics[trim=1cm 11.5cm 1cm 10cm, clip, height=4.4cm, width=8.8cm]{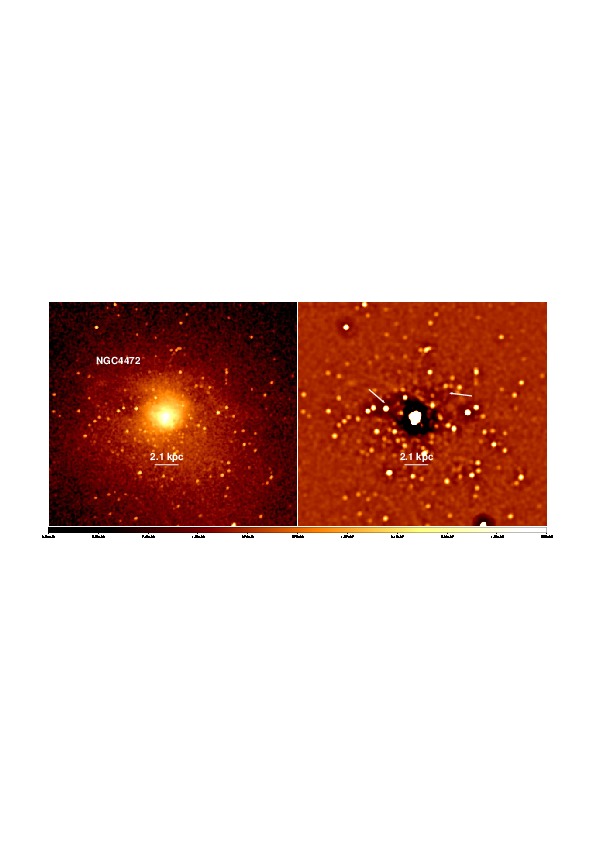}
  \includegraphics[trim=1cm 11.5cm 1cm 10cm, clip, height=4.4cm, width=8.8cm]{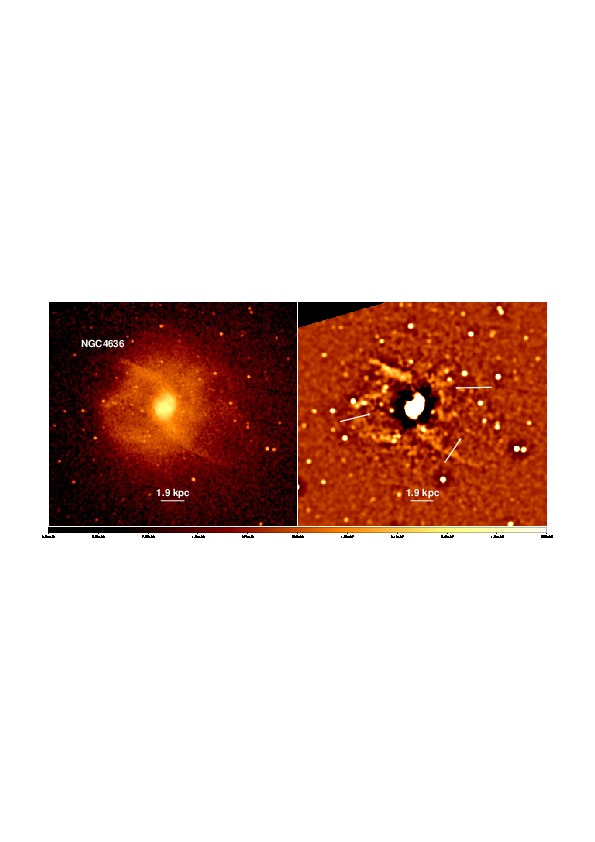}
  \includegraphics[trim=1cm 11.5cm 1cm 10cm, clip, height=4.4cm, width=8.8cm]{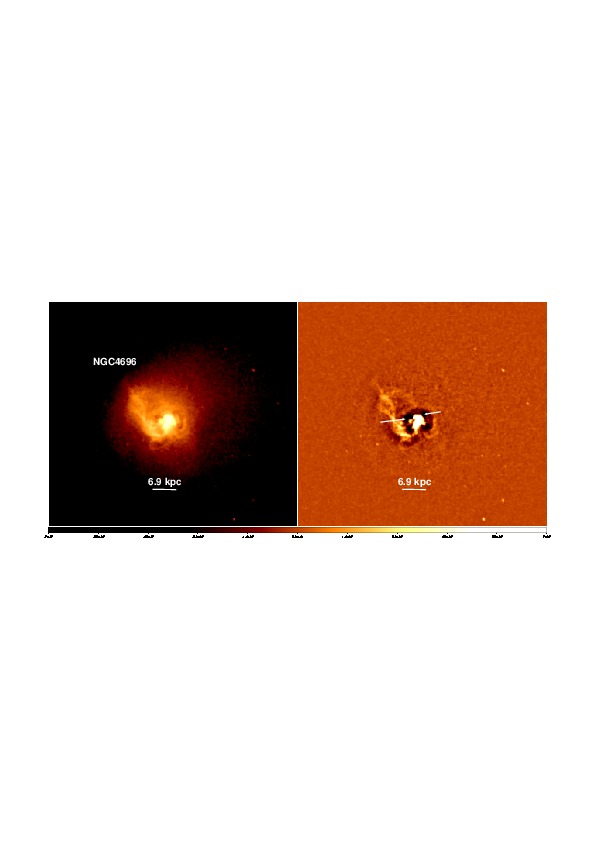}
  \includegraphics[trim=1cm 11.5cm 1cm 10cm, clip, height=4.4cm, width=8.8cm]{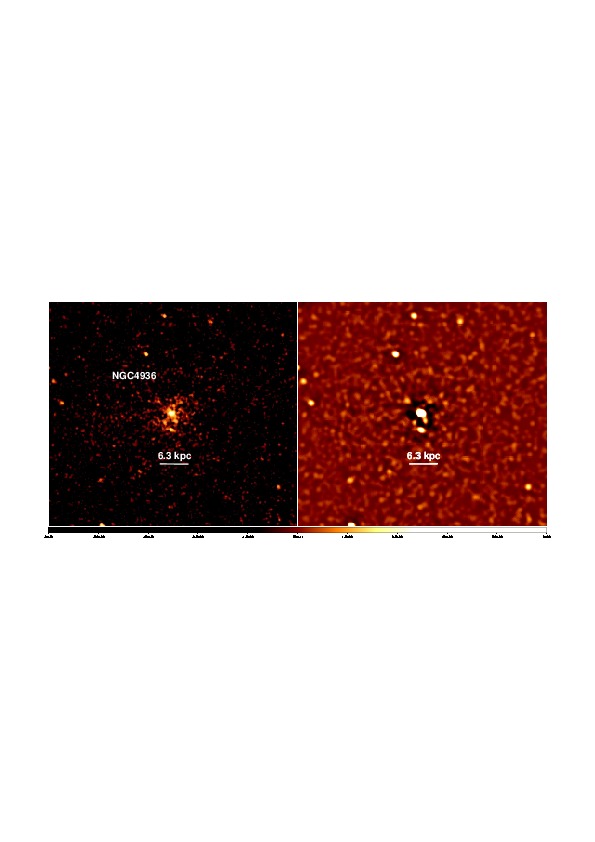}
  \includegraphics[trim=1cm 11.5cm 1cm 10cm, clip, height=4.4cm, width=8.8cm]{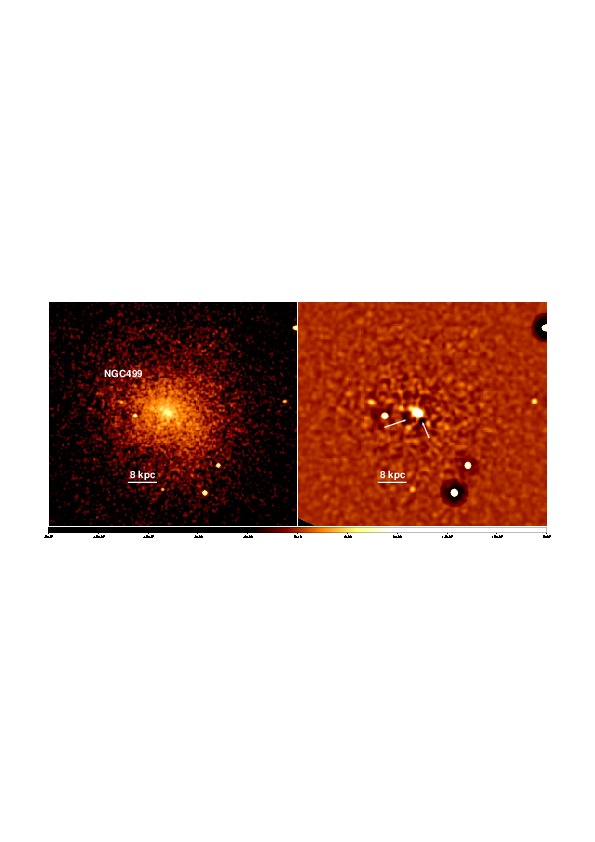}
  \includegraphics[trim=1cm 11.5cm 1cm 10cm, clip, height=4.4cm, width=8.8cm]{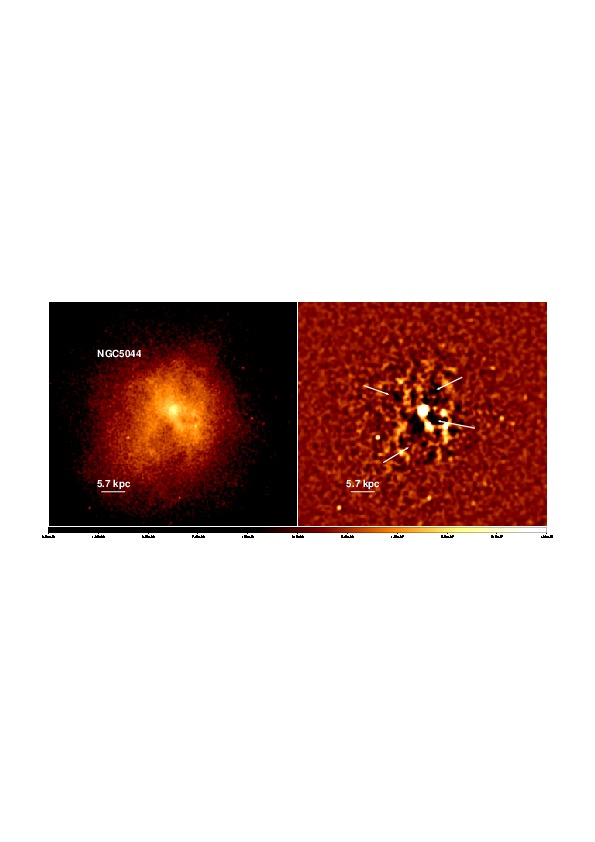}
\end{figure*}

\begin{figure*}
\contcaption{}
  \includegraphics[trim=1cm 11.5cm 1cm 10cm, clip, height=4.4cm, width=8.8cm]{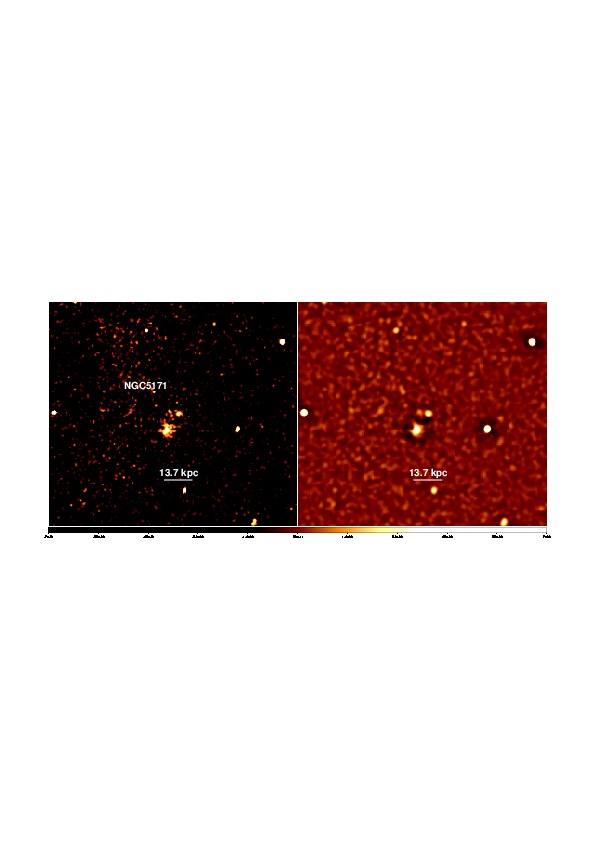}
  \includegraphics[trim=1cm 11.5cm 1cm 10cm, clip, height=4.4cm, width=8.8cm]{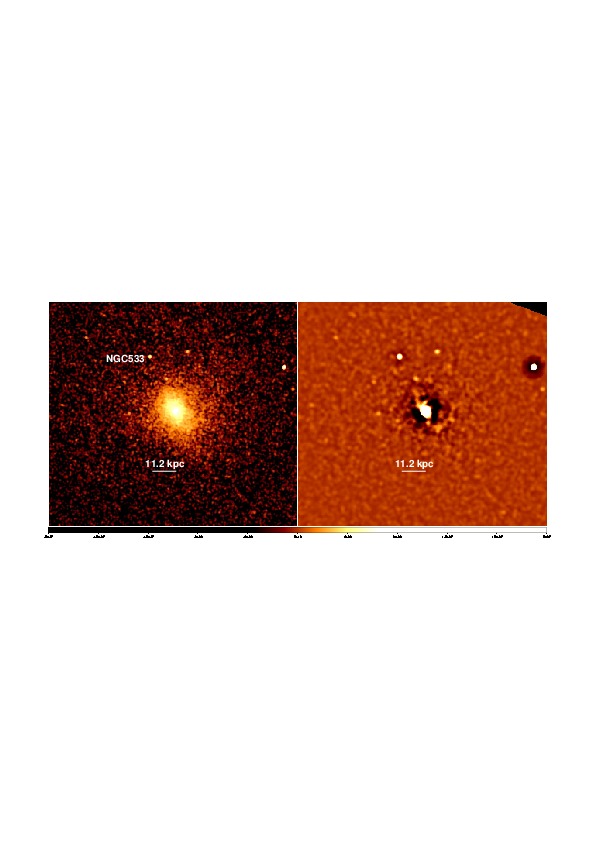}
  \includegraphics[trim=1cm 11.5cm 1cm 10cm, clip, height=4.4cm, width=8.8cm]{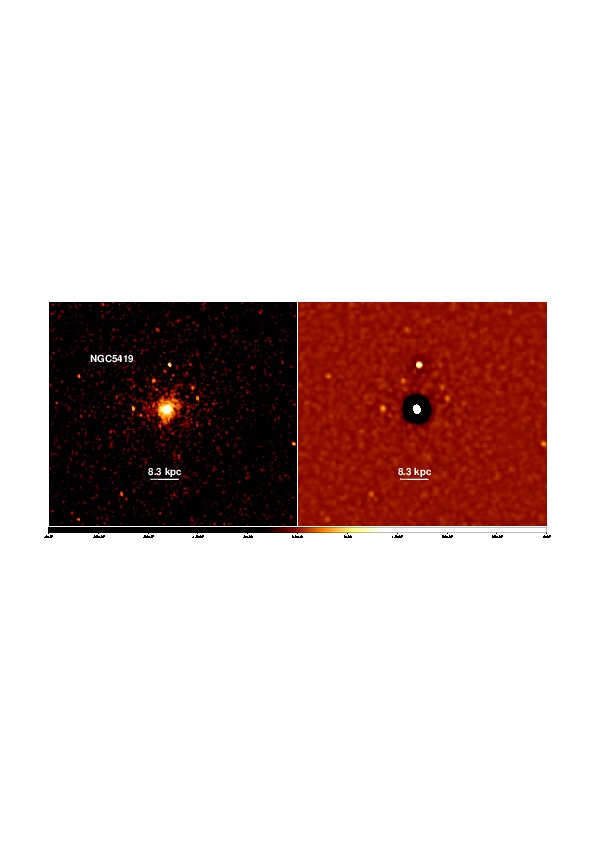}
  \includegraphics[trim=1cm 11.5cm 1cm 10cm, clip, height=4.4cm, width=8.8cm]{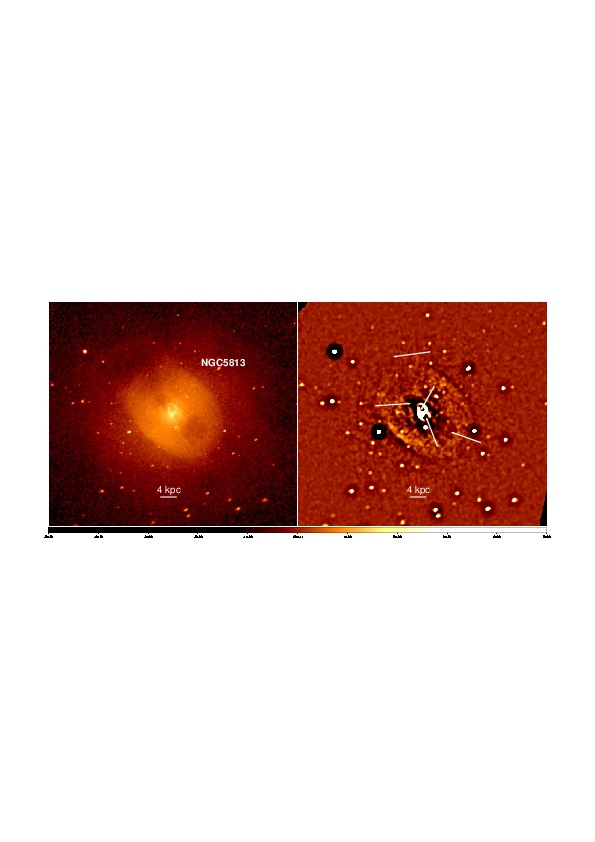}
  \includegraphics[trim=1cm 11.5cm 1cm 10cm, clip, height=4.4cm, width=8.8cm]{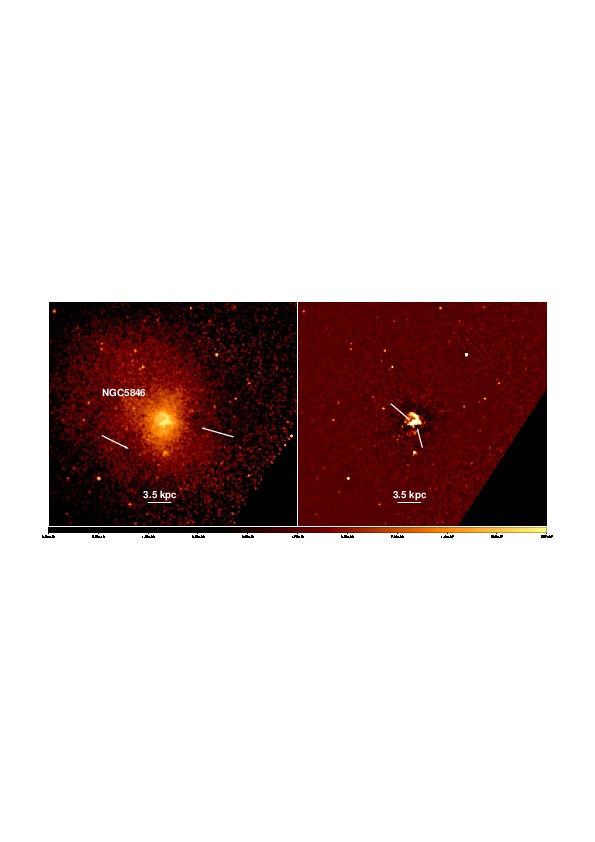}
  \includegraphics[trim=1cm 11.5cm 1cm 10cm, clip, height=4.4cm, width=8.8cm]{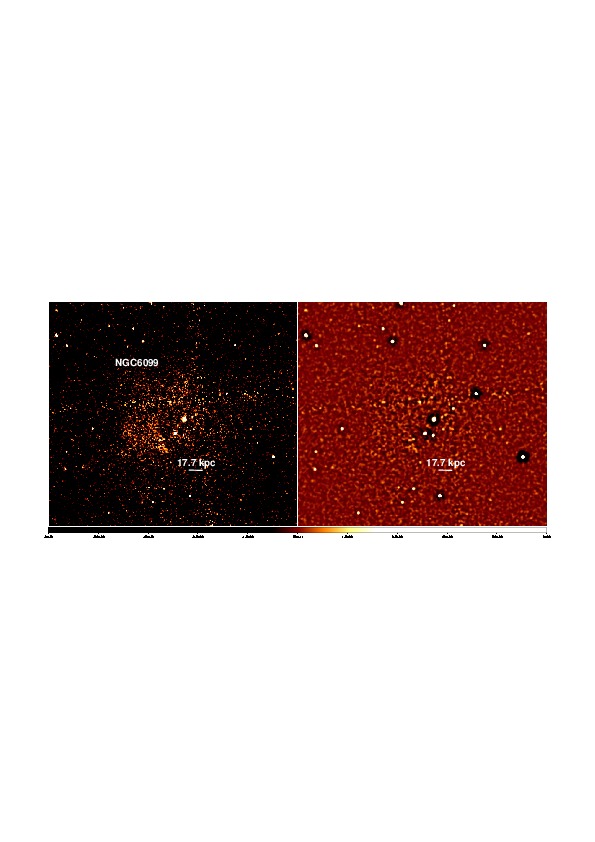}
  \includegraphics[trim=1cm 11.5cm 1cm 10cm, clip, height=4.4cm, width=8.8cm]{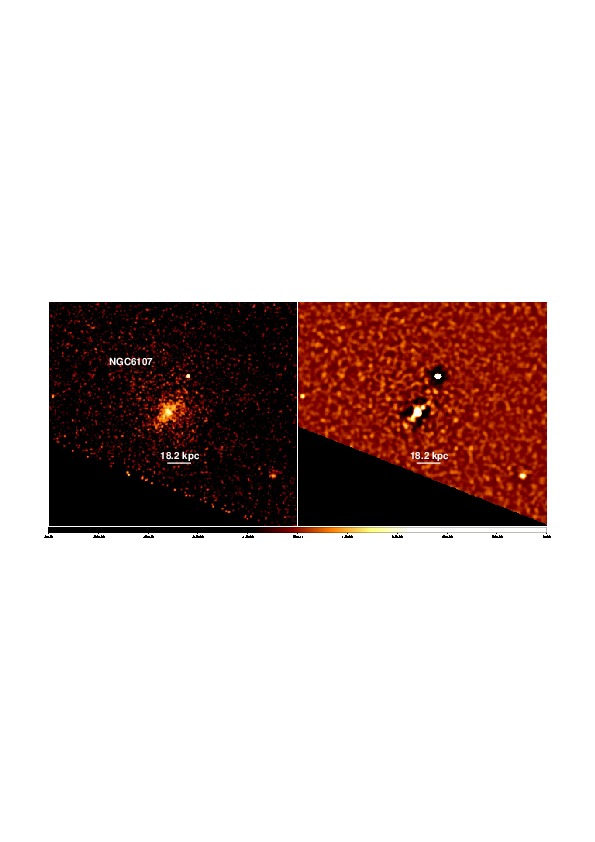}
  \includegraphics[trim=1cm 11.5cm 1cm 10cm, clip, height=4.4cm, width=8.8cm]{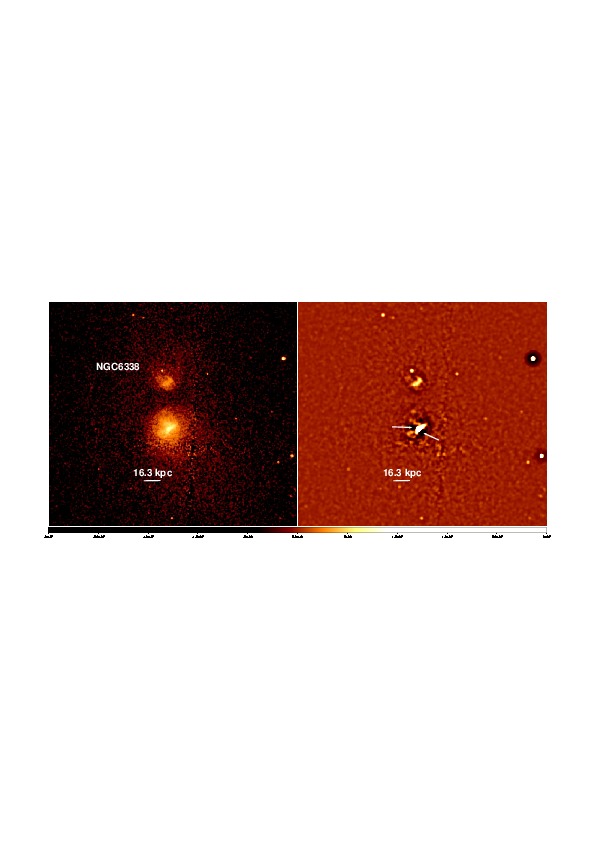}
  \includegraphics[trim=1cm 11.5cm 1cm 10cm, clip, height=4.4cm, width=8.8cm]{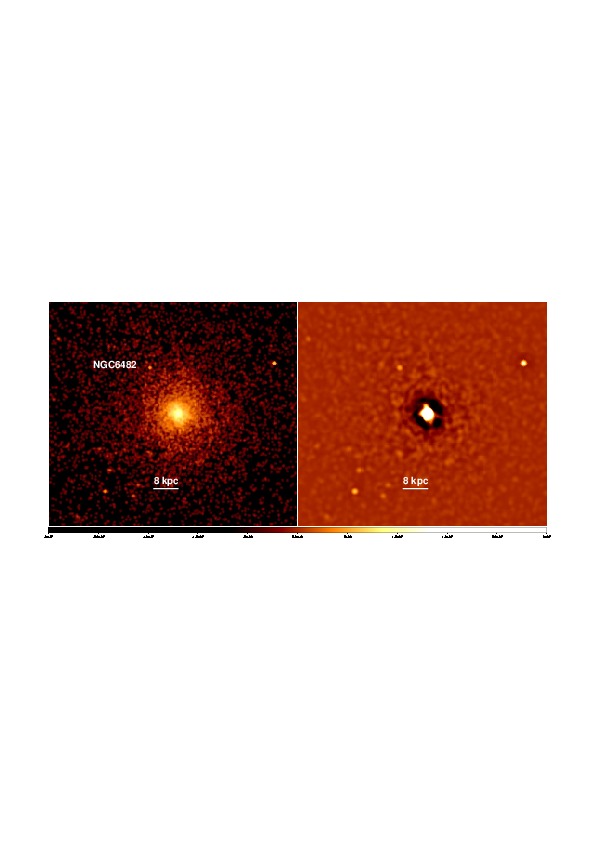}
  \includegraphics[trim=1cm 11.5cm 1cm 10cm, clip, height=4.4cm, width=8.8cm]{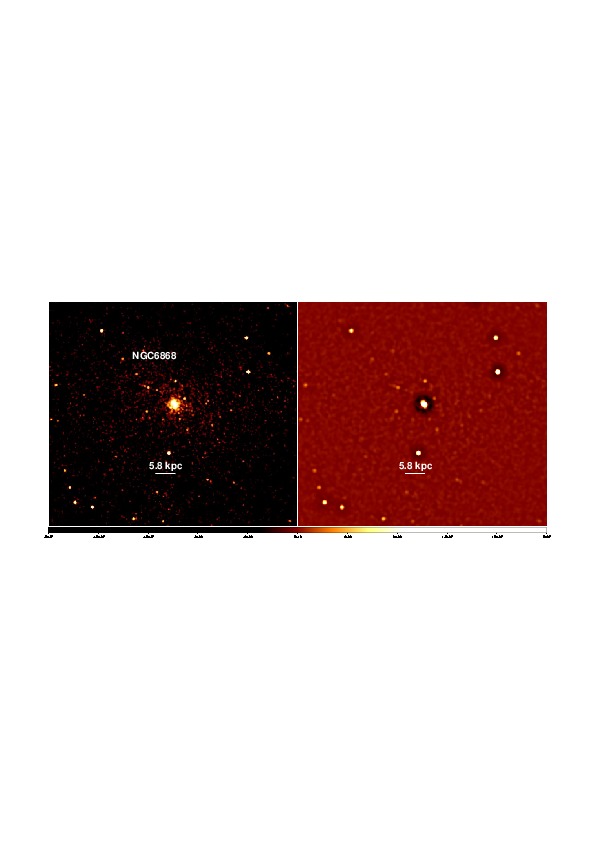}
\end{figure*}

\begin{figure*}
\contcaption{}
\includegraphics[trim=1cm 11.5cm 1cm 10cm, clip, height=4.4cm, width=8.8cm]{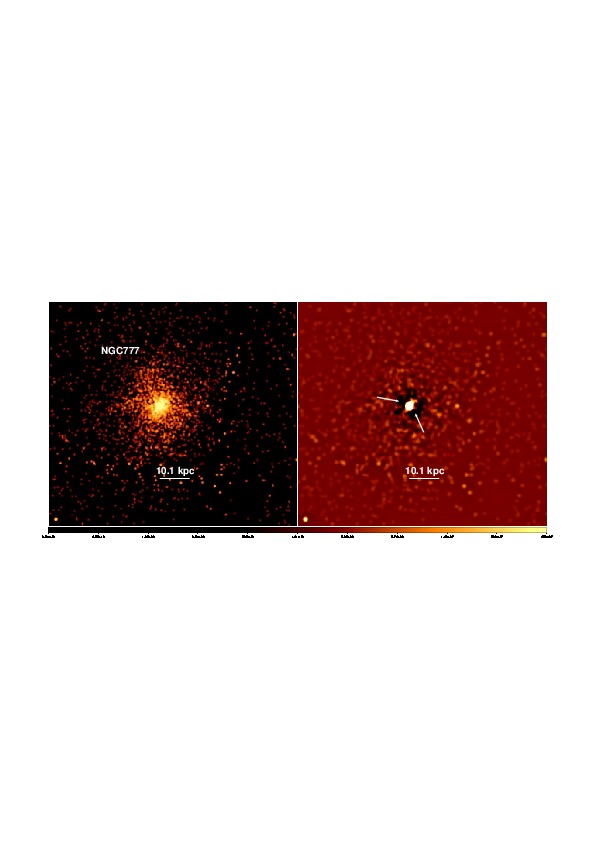}
\includegraphics[trim=1cm 11.5cm 1cm 10cm, clip, height=4.4cm, width=8.8cm]{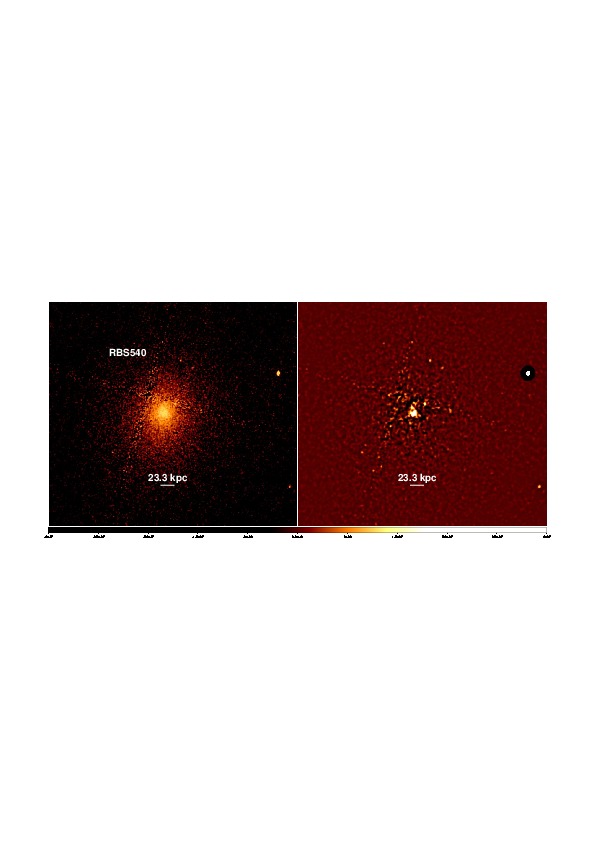}
\includegraphics[trim=1cm 11.5cm 1cm 10cm, clip, height=4.4cm, width=8.8cm]{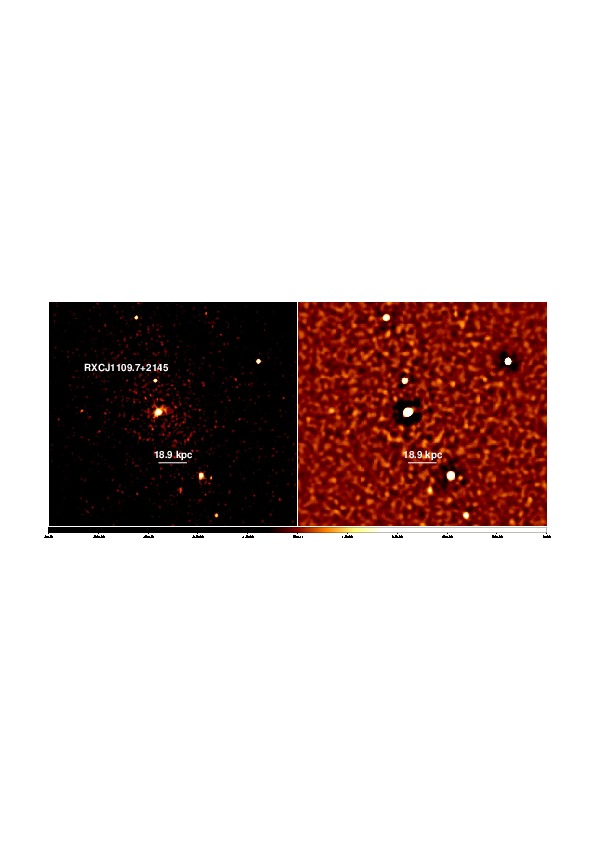}
\includegraphics[trim=1cm 11.5cm 1cm 10cm, clip, height=4.4cm, width=8.8cm]{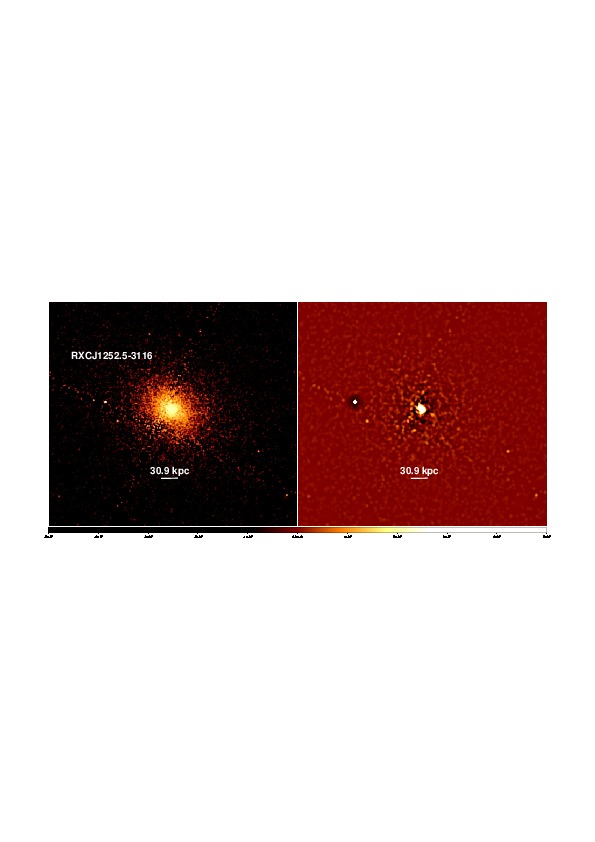}
\end{figure*}

\begin{figure*}
\includegraphics[trim=1.0cm 11.5cm 1cm 10cm, clip, height=4.2cm, width=8.6cm]{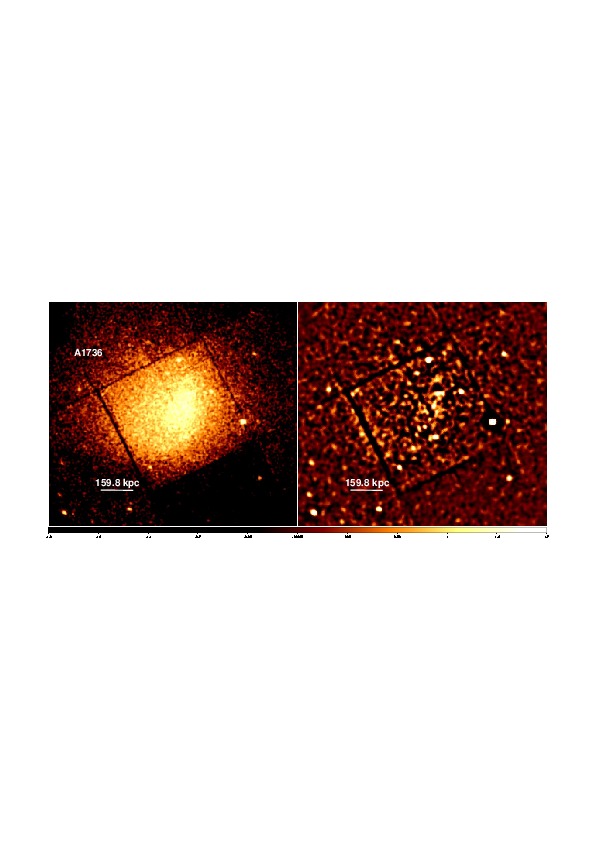}
\includegraphics[trim=1.0cm 11.5cm 1cm 10cm, clip, height=4.2cm, width=8.6cm]{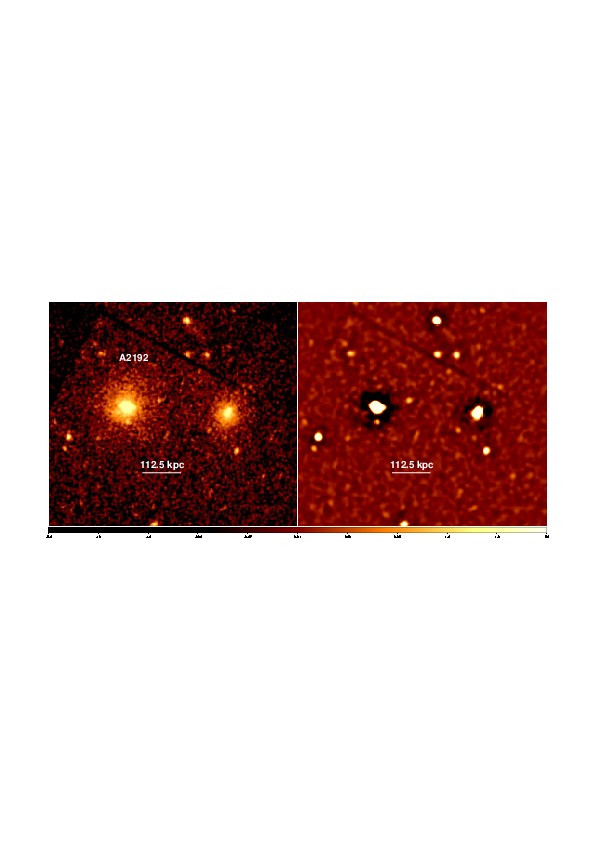}
\includegraphics[trim=1.0cm 11.5cm 1cm 10cm, clip, height=4.2cm, width=8.6cm]{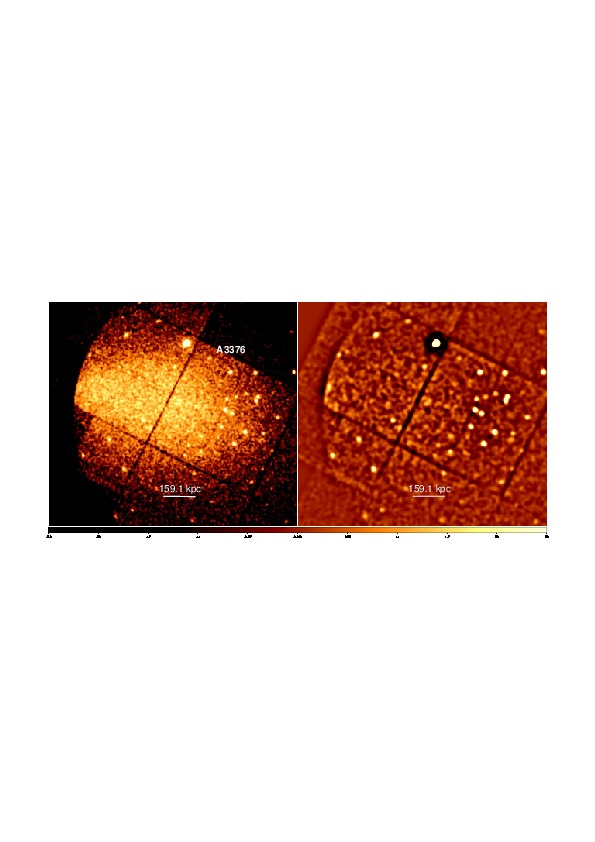}
\includegraphics[trim=1.0cm 11.5cm 1cm 10cm, clip, height=4.2cm, width=8.6cm]{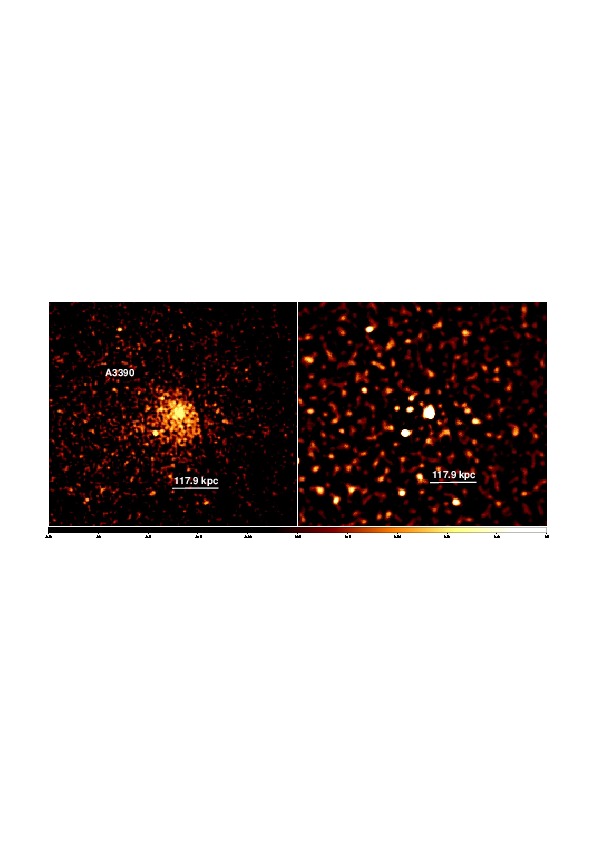}
\includegraphics[trim=1.0cm 11.5cm 1cm 10cm, clip, height=4.2cm, width=8.6cm]{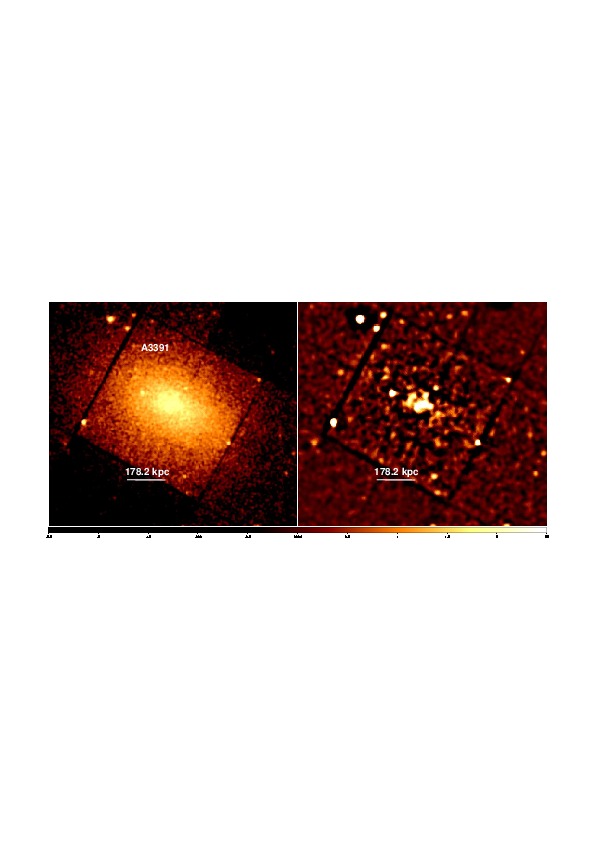}
\includegraphics[trim=1.0cm 11.5cm 1cm 10cm, clip, height=4.2cm, width=8.6cm]{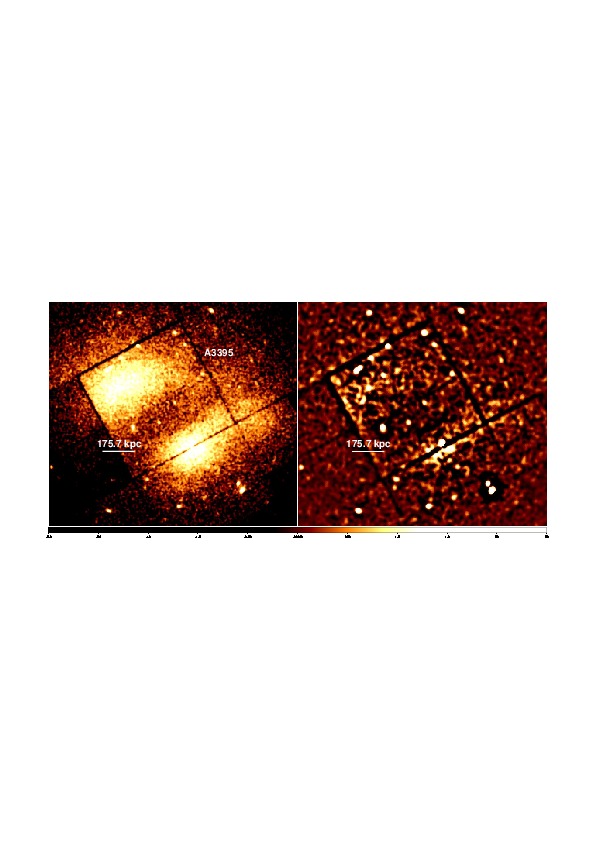}
\includegraphics[trim=1.0cm 11.5cm 1cm 10cm, clip, height=4.2cm, width=8.6cm]{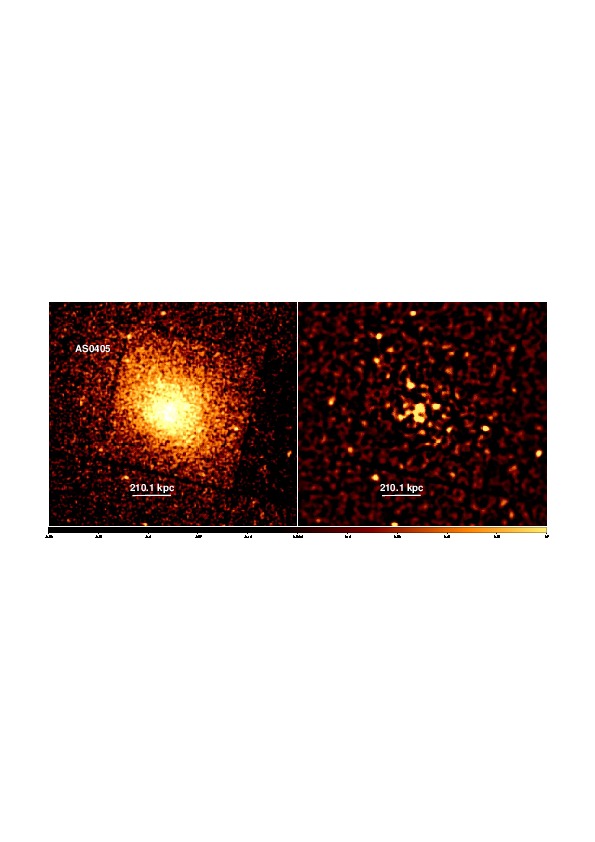}
\includegraphics[trim=1.0cm 11.5cm 1cm 10cm, clip, height=4.2cm, width=8.6cm]{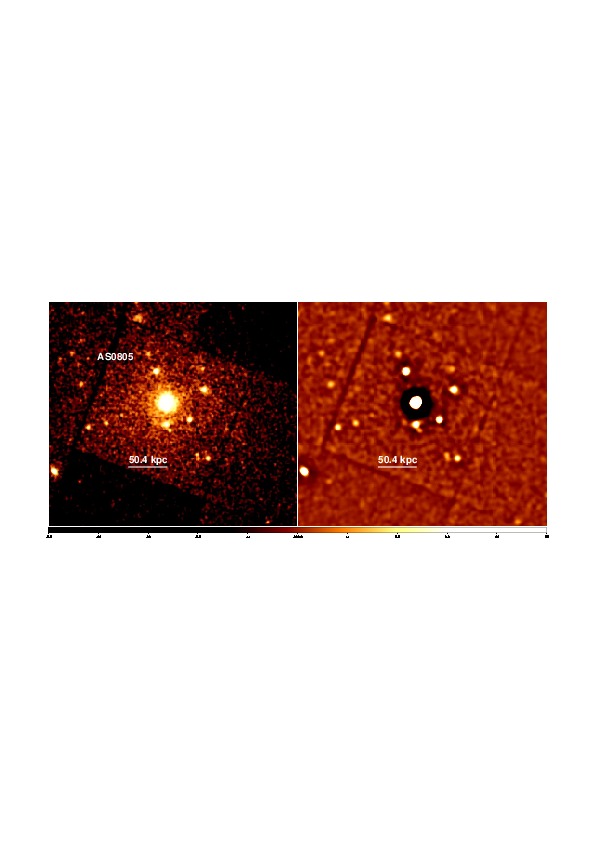}
\includegraphics[trim=1.0cm 11.5cm 1cm 10cm, clip, height=4.2cm, width=8.6cm]{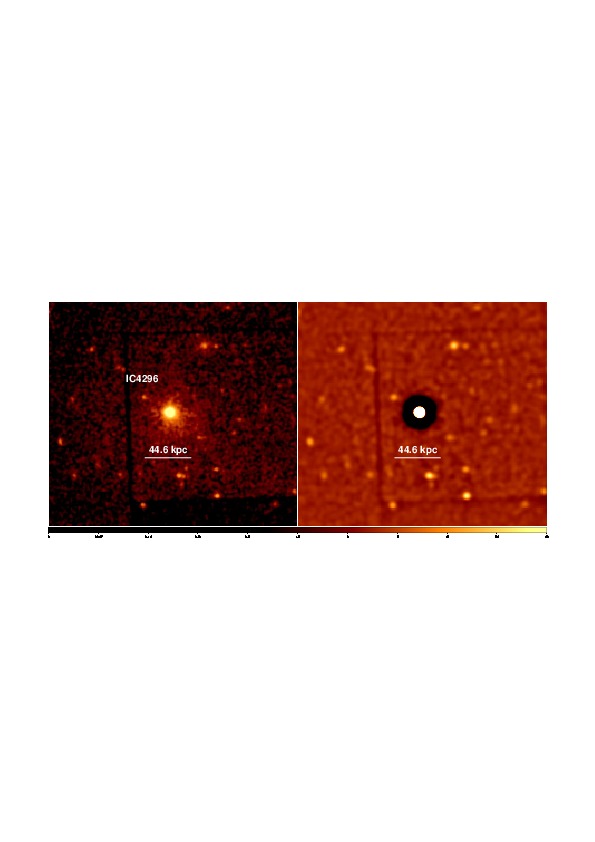}
\includegraphics[trim=1.0cm 11.5cm 1cm 10cm, clip, height=4.2cm, width=8.6cm]{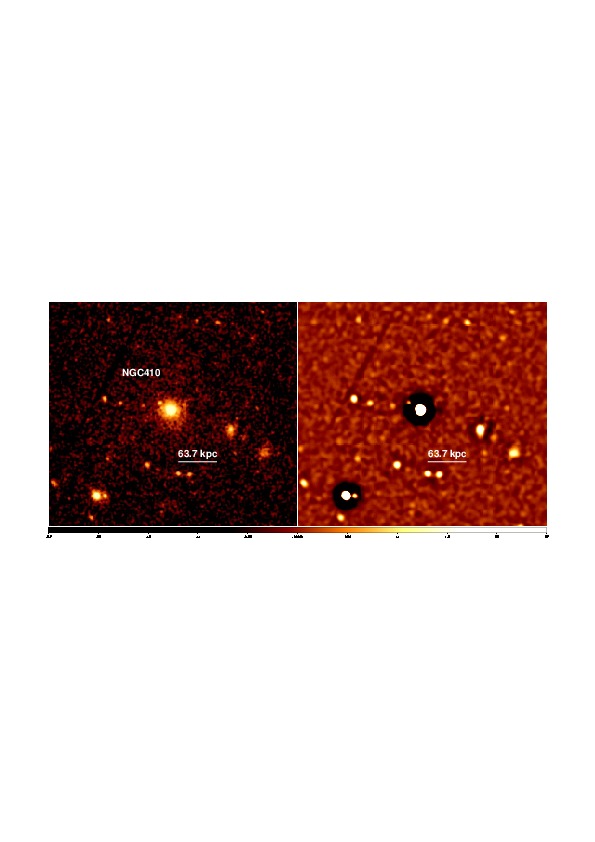}
\caption[]{The bar is 3 arcmin long in all the images. All images were generated in the 0.5--7.0 keV energy band, and the original images have been smoothed using a 2-pixel Gaussian.The left-hand panel for each set of two images shows the cleaned images, with the unsharp-masked image in the right-hand panel. The arrows indicate ``possible'' or ``certain'' cavities.}
\end{figure*}

\begin{figure*}
\contcaption{}
\includegraphics[trim=1.0cm 11.5cm 1cm 10cm, clip, height=4.2cm, width=8.6cm]{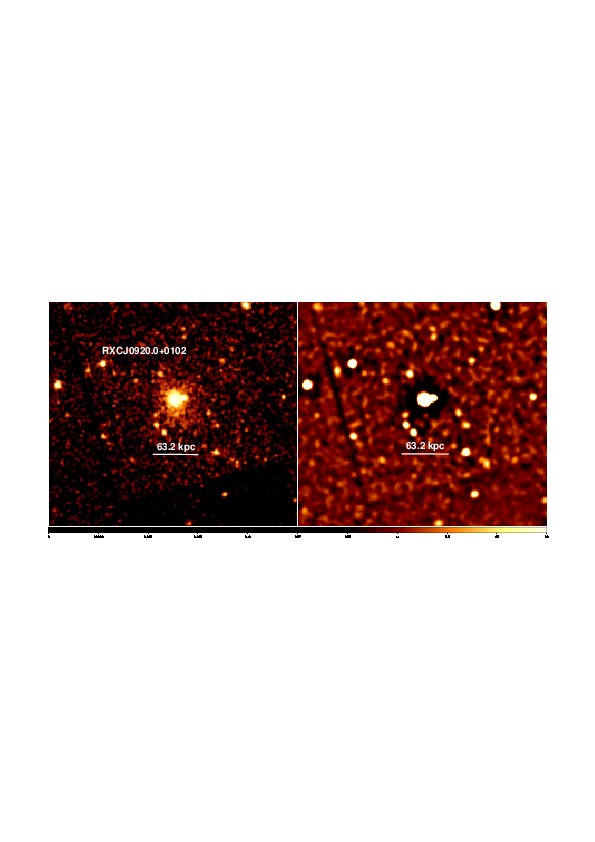}
\includegraphics[trim=1.0cm 11.5cm 1cm 10cm, clip, height=4.2cm, width=8.6cm]{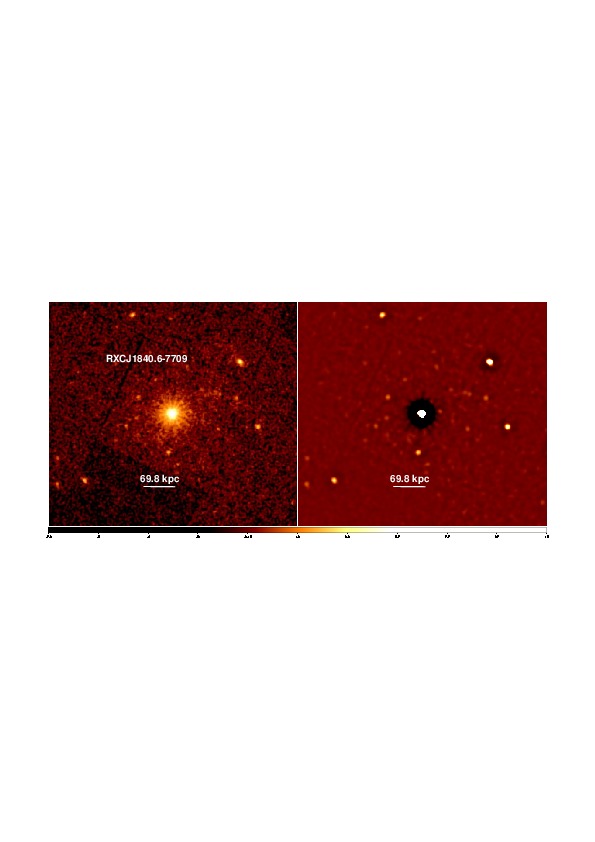}
\includegraphics[trim=1.0cm 11.5cm 1cm 10cm, clip, height=4.2cm, width=8.6cm]{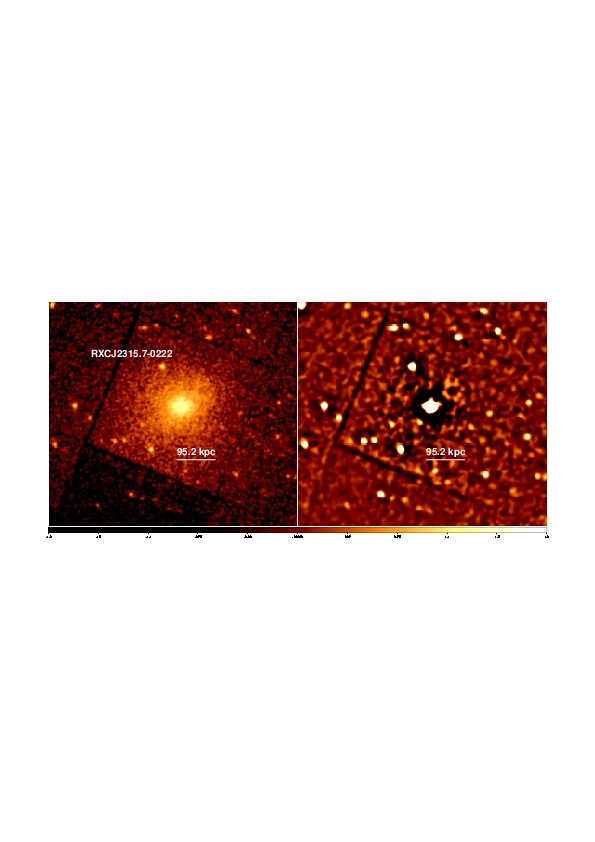}
\includegraphics[trim=1.0cm 11.5cm 1cm 10cm, clip, height=4.2cm, width=8.6cm]{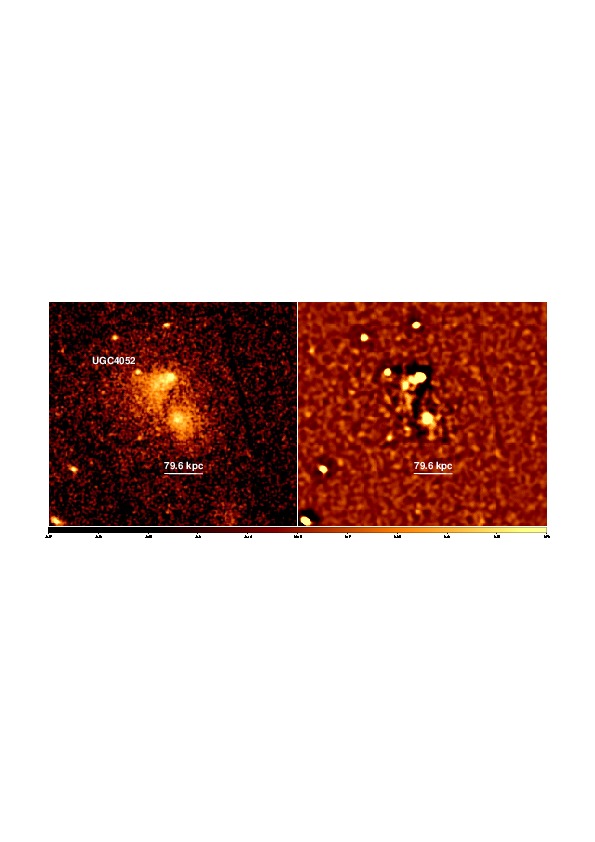}
\end{figure*}

\end{document}